\documentclass[12pt]{article}

\usepackage[T1]{fontenc}
\usepackage{natbib}
\RequirePackage[colorlinks,citecolor=blue,urlcolor=blue]{hyperref}
\usepackage[english]{babel}
\usepackage{amsthm}
\usepackage{amsmath}
\usepackage{amsfonts}
\usepackage{amssymb}
\usepackage{graphicx}
\usepackage{enumerate}
\usepackage{color}
\usepackage{array}
\usepackage{bbm}
\usepackage{multirow}
\usepackage{rotating}
\usepackage{xr}
\usepackage{setspace}
\usepackage{pdflscape}
\usepackage{tikz}
\usetikzlibrary{arrows,shapes.geometric,positioning}

\def\frac#1#2{{\textstyle{#1\over#2}}}

\DeclareSymbolFont{AMSb}{U}{msb}{m}{n}
\DeclareMathSymbol{\Natural}{\mathbin}{AMSb}{"4E}
\DeclareMathSymbol{\Integer}{\mathbin}{AMSb}{"5A}
\DeclareMathSymbol{\Real}{\mathbin}{AMSb}{"52}
\DeclareMathSymbol{\Rational}{\mathbin}{AMSb}{"51}
\DeclareMathSymbol{\Imaginary}{\mathbin}{AMSb}{"49}
\DeclareMathSymbol{\Complex}{\mathbin}{AMSb}{"43} 
\DeclareMathSymbol{\Disk}{\mathbin}{AMSb}{"44} 
\def\bi{\begin{itemize}}
\def\ei{\end{itemize}}
\def\bd{\begin{description}}
\def\ed{\end{description}}
\def\ben{\begin{enumerate}}
\def\een{\end{enumerate}}
\def\bv{\begin{verbatim}}
\def\ev{\end{verbatim}}

\def\hat#1{{\widehat{#1}}}

\def\pr{{\rm Pr}}
\def\Pr{\pr}

\def\2to{{\ {\buildrel 2\over \longrightarrow}\ }}

\def\I1ton{{$I_1,\ldots,I_n$}}
\def\X1ton{{$X_1,\ldots,X_n$}}
\def\Y1ton{{$Y_1,\ldots,Y_n$}}
\def\Z1ton{{$Z_1,\ldots,Z_n$}}
\def\R1ton{{$R_1,\ldots,R_n$}}
\def\e1ton{{$e_1,\ldots,e_n$}}
\def\t1ton{{$t_1,\ldots,t_n$}}
\def\x1ton{{$x_1,\ldots,x_n$}}
\def\y1ton{{$y_1,\ldots,y_n$}}
\def\z1ton{{$z_1,\ldots,z_n$}}

\begin{document}
\thispagestyle{empty}
\baselineskip=28pt
\vskip 5mm
\begin{center} {\Large{\bf Insights into the drivers and spatio-temporal trends of extreme Mediterranean wildfires with statistical deep-learning}}
\end{center}

\baselineskip=12pt
\vskip 5mm

\begin{center}
\large
Jordan Richards$^{1*}$, Rapha\"el Huser$^1$,\\ Emanuele Bevacqua$^2$,  Jakob Zscheischler$^{2,3}$
\end{center}

\footnotetext[1]{
\baselineskip=10pt Statistics Program, Computer, Electrical and Mathematical Sciences and Engineering (CEMSE) Division, King Abdullah University of Science and Technology (KAUST), Thuwal 23955-6900, Saudi Arabia. $^*$E-mail: jordan.richards@kaust.edu.sa}
\footnotetext[2]{
\baselineskip=10pt Department of Computational Hydrosystems, Helmholtz Centre for Environmental Research - UFZ, Leipzig, Germany}
\footnotetext[3]{
\baselineskip=10pt Technische Universit\"at Dresden, Dresden, Germany.}
\baselineskip=17pt
\vskip 4mm
\centerline{\today}
\vskip 6mm

\begin{center}
{\large{\bf Abstract}}
\end{center}
Extreme wildfires continue to be a significant cause of human death and biodiversity destruction within countries that encompass the Mediterranean Basin. Recent worrying trends in wildfire activity (i.e., occurrence and spread) suggest that wildfires are likely to be highly impacted by climate change. In order to facilitate appropriate risk mitigation, it is imperative to identify the main drivers of extreme wildfires and assess their spatio-temporal trends, with a view to understanding the impacts of the changing climate on fire activity. To this end, we analyse the monthly burnt area due to wildfires over a region encompassing most of Europe and the Mediterranean Basin from 2001 to 2020, and identify high fire activity during this period in eastern Europe, Algeria, Italy and Portugal. We build an extreme quantile regression model with a high-dimensional predictor set describing meteorological conditions, land cover usage, and orography, for the domain. To model the complex relationships between the predictor variables and wildfires, we make use of a hybrid statistical deep-learning framework that allows us to disentangle the effects of vapour-pressure deficit (VPD), air temperature, and drought on wildfire activity. Our results highlight that whilst VPD, air temperature, and drought significantly affect wildfire occurrence, only VPD affects wildfire spread. Furthermore, to gain insights into the effect of climate trends on wildfires in the near future, we focus on the extreme wildfires in August 2001 and perturb VPD and temperature according to their observed trends. We find that, on average over 
Europe, trends in temperature (median over Europe: $+0.04$K per year) lead to a relative increase of $17.1\%$ and $1.6\%$ in the expected frequency and severity, respectively, of wildfires in August 2001; similar analyses using VPD (median over Europe: $+4.82$Pa per year) give respective increases of $1.2\%$ and $3.6\%$. Our analysis finds evidence suggesting that global warming can lead to spatially non-uniform changes in wildfire activity.  
\baselineskip=16pt

\par\vfill\noindent

\pagenumbering{arabic}
\baselineskip=24pt

\newpage

\section{Introduction}
With the frequency and severity of wildfires expected to be exasperated by climate change \citep{di2019climate,dupuy2020climate, jones2020climate,ruffault2020increased,ribeiro2022}, many countries with Mediterranean climates now experience regularly-occurring and devastating wildfires. The combined effect of extensive heatwaves and droughts in 2022 lead to Europe observing its second-largest annual burnt area on record by August 4th \citep{Abnett2022}. Uncontrolled, wildfires pose a major risk to both environmental and ecological systems throughout the world; wildfires contribute significantly to global CO$_2$ emissions \citep{liu2014wildland,2021copernicus}, with worrying trends predicted under a changing climate \citep{de2013climate,knorr2016air}, and lead to destruction of biomass and biodiversity reduction amongst both plants and animals \citep{diaz2002satellite, moreira2007modelling, pausas2008wildfires,bradshaw2011little}. Moreover, wildfires carry a number of anthropogenic health risks, with hundreds of human fatalities in the past few decades being directly attributed to European wildfires \citep{san2013analysis,kron2019changes,molina2019analysis}; the indirect consequences of wildfires on human health through increased air pollution and particulates is much more difficult to quantify \citep{jimenez2020assessment,weilnhammer2021extreme}. There exists a clear need for the development of robust statistical frameworks that can be used to facilitate the prevention and risk mitigation of European wildfires, particularly those that lead to extreme fuel consumption and burnt acreage, and thus, high economic cost and pollution. To that end, here we develop a model that identifies the drivers of Mediterranean Europe wildfire occurrence and extreme spread, whilst simultaneously producing risk maps that can be used to characterise high-risk areas.
\par
{The extremal characteristics of European wildfires have previously been studied using various statistical methodologies. \cite{de2009spatial}, \cite{mendes2010spatial} and \cite{turkman2010asymptotic} model sizes of individual wildfires in Portugal using the generalised Pareto distribution (GPD), with the latter two studies using a Bayesian hierarchical framework; a similar hierarchical model was applied to French wildfire sizes by \cite{pimont2021}.  Point processes tools have been exploited for modelling occurrences of wildfires in  Spain, using hurdle models \citep{serra2014spatio}, Portugal, via empirical clustering and kernel density estimation \citep{Tonini2017} and France, using log-Gaussian Cox processes \citep{gabriel2017detecting, opitz2020point, koh2021spatiotemporal}. \cite{rios2018studying} adopt a zero-inflated semi-additive beta regression model for jointly modelling wildfire size and occurrence in Galicia. Wildfire size and impact is often characterised through measures of aggregated burnt area for spatio-temporal regions \citep{xi2019statistical}. Whilst there exists some debate over appropriate probability distributions for wildfire sizes \citep{cumming2001parametric,cui2008we,hantson2016global,pereira2019statistical}, many studies have shown that burnt area is typically heavy-tailed \citep{pereira2019statistical, koh2021spatiotemporal, richards2022}. As our focus is on modelling extreme wildfires, we employ the asymptotically-justified GPD  \citep{coles2001introduction}.}
\par
Typical approaches designed to identify the drivers of wildfire risk often rely on regression-type statistical models, see, for example, \cite{vilar2011logistic}, \cite{vilar2016modeling}, \cite{rios2018studying} and \cite{xi2019statistical}. Whilst simple linear, and additive, regression models are computationally easy to fit and facilitate fast statistical inference, they cannot capture highly complex or non-linear structure in data. Due to the complex and non-stationary nature of the climate in southern Europe and the Mediterranean basin \citep{lionello2006mediterranean}, as well as the high diversity in land-cover types, that is, both fuel abundance and relative combustibility \citep{san2012land,malinowski2020automated}, it is highly unlikely that simple regression models will be appropriate here. Recent advances in wildfire modelling have seen substantially better fitting models and predictive performance from machine learning and deep-learning approaches, see, for example, \cite{radke2019firecast}, \cite{zhang2019forest}, \cite{bergado2021predicting}, \cite{bjaanes2021deep}, \cite{cisneros2021combined}, and \cite{koh2021gradient}, as these are significantly better than simple regression models at capturing complex structure in data, and scale well to high-dimensional data. {A more complex model is not necessarily guaranteed to provide a better fit to wildfire data, but we believe that this is a safe assumption to make as \cite{richards2022} showcase large gains in predictive power for a neural network-based model of extreme US wildfire spread, relative to classical regression models. As their data shares similar complexity with ours, we adapt aspects of their methodology to model European wildfires. }

Modelling frameworks for fitting GPDs with deep-learning exist \citep{rietsch2013network, carreau2007hybrid, carreau2011stochastic, ceresetti2012evaluation,pasche2022neural,wilson2022deepgpd}, but typically these models cannot be used to identify the drivers of risk; standard neural networks often lack interpretability due to their large number of trainable parameters. Recently, \cite{richards2022} proposed the partially-interpretable neural network (PINN) framework for semi-parametric regression, with the influence of a subset of predictors modelled using ``interpretable" parametric, or semi-parametric, functions and the influence of the rest of the predictors modelled using non-parametric neural networks (see Section~\ref{pinnrepsec}); they used this framework to fit an extreme-value point-process model to U.S. wildfire data, and we extend their approach to create a bespoke GPD model for extreme European and Mediterranean wildfires. Whilst \cite{richards2022} focused on identifying drivers and estimating extreme burnt area quantiles, we here also study spatio-temporal trends and climate change impacts. \par
The paper is outlined as followed. In Section~\ref{Data_sec}, we introduce the data used in our study of extreme European wildfires. Section~\ref{model_sec} details the extreme-value deep-learning model used to perform our analyses, with details of the GPD and PINN frameworks provided in Subsections~\ref{POTsec} and \ref{DL_sec}, respectively. Our analyses are presented in Section~\ref{application_sec} with separate consideration given for the interpretable results, wildfire risk assessment, and climate change impacts, in Subsections~\ref{interp_results_sec}--\ref{climate_change_sec}. We conclude the paper in Section~\ref{conclusion_sec}. 
\section{Data}
\label{Data_sec}
As the impact of wildfires is not directly observable, we quantify it through a measure of burnt area, which is a useful proxy for both fuel consumption and emissions \citep{koh2021spatiotemporal}. Let $\{Y(s,t):s \in \mathcal{S},t \in \mathcal{T}\}$ be the aggregated burnt area (BA), measured in km$^2$, for a spatio-temporal gridbox $(s,t)$ indexed by a spatial domain $\mathcal{S}\subset\mathbb{R}^2$ and temporal domain $\mathcal{T} \subset \mathbb{R}_+$, and let $\{\mathbf{X}(s,t):s\in\mathcal{S},t\in\mathcal{T}\}$ denote a $d$-dimensional space-time process of chosen predictor variables where $\mathbf{X}(s,t)=(X_1(s,t),\dots,X_d(s,t))^T$ for all $(s,t)\in \mathcal{S}\times\mathcal{T}$ and for $d\in\mathbb{N}$. We denote observations of $\mathbf{X}(s,t)$ and $Y(s,t)$ by $\mathbf{x}(s,t)$ and $y(s,t)$, respectively, and note that typically observations are recorded at a finite collection of space-time locations; we denote these space-time observation locations by $\boldsymbol{\omega}\subset\mathcal{S}\times\mathcal{T}$.\par
\begin{figure}[t]
\centering
\includegraphics[width=0.9\linewidth]{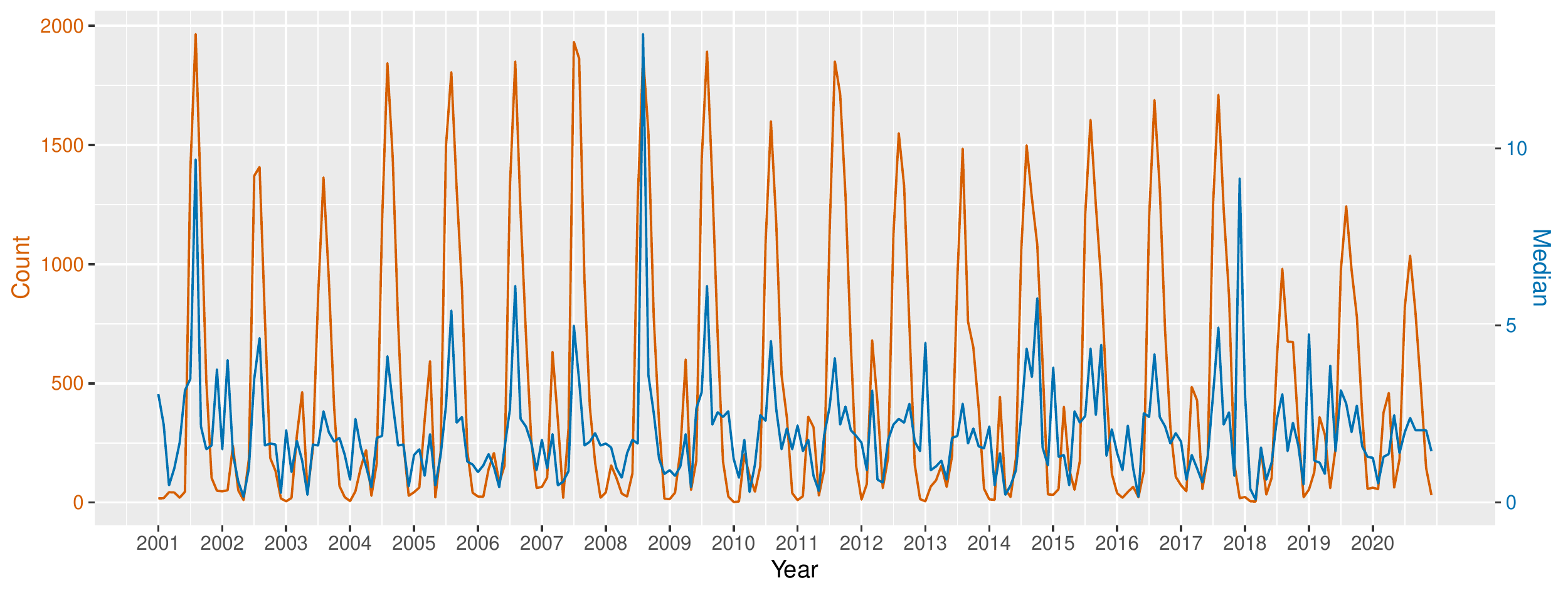} 
\caption{{Monthly (brown) counts, and (blue) median, of non-zero BA across the entire spatial domain [km$^2$]. The time series spans all months in 2001 to 2020, inclusive.}}
\label{obs_ts}
\end{figure}
\begin{figure}[t]
\centering
\begin{minipage}{0.49\linewidth}
\centering
\includegraphics[width=\linewidth]{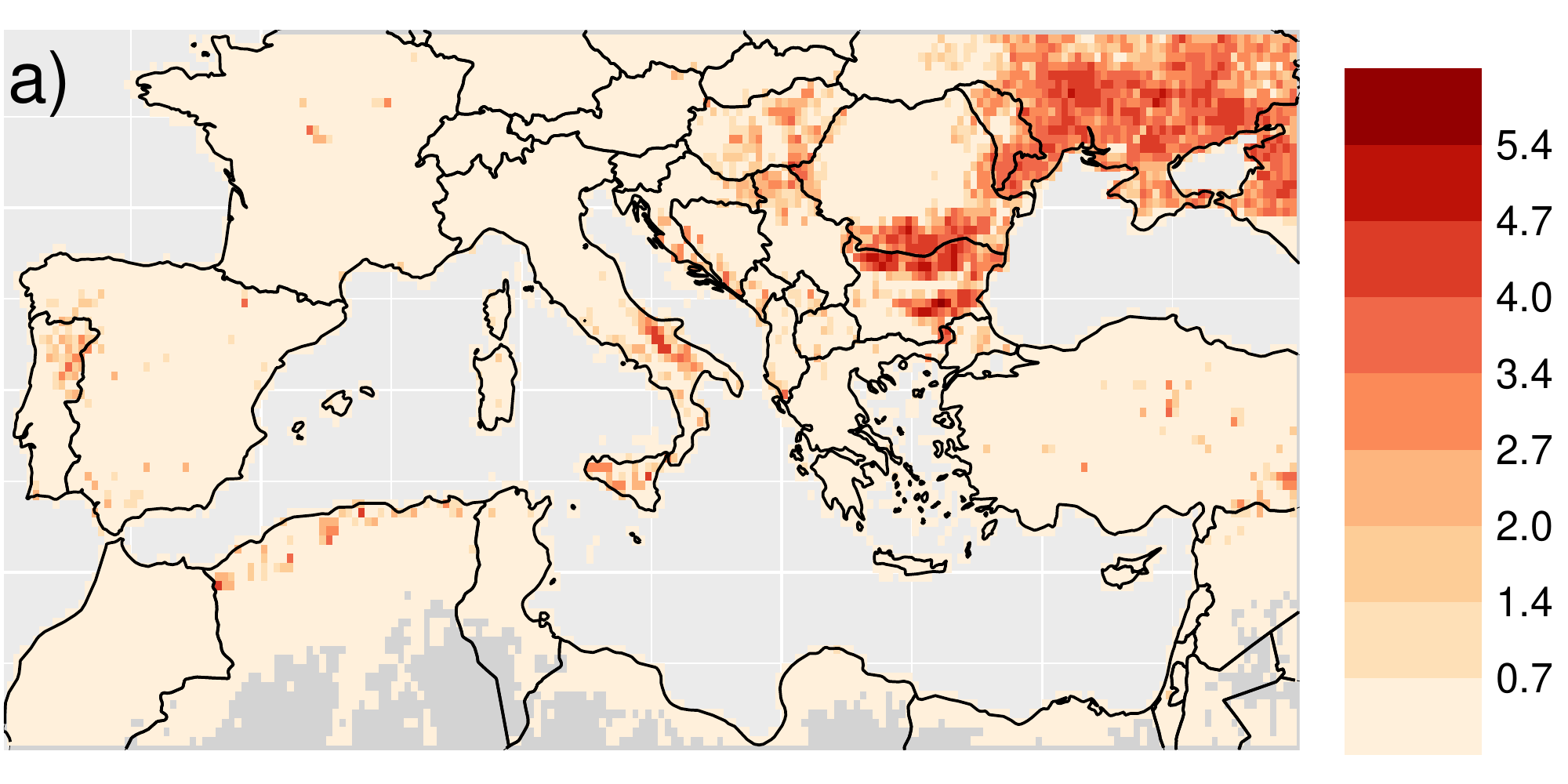} 
\end{minipage}
\hfill
\begin{minipage}{0.49\linewidth}
\centering
\includegraphics[width=\linewidth]{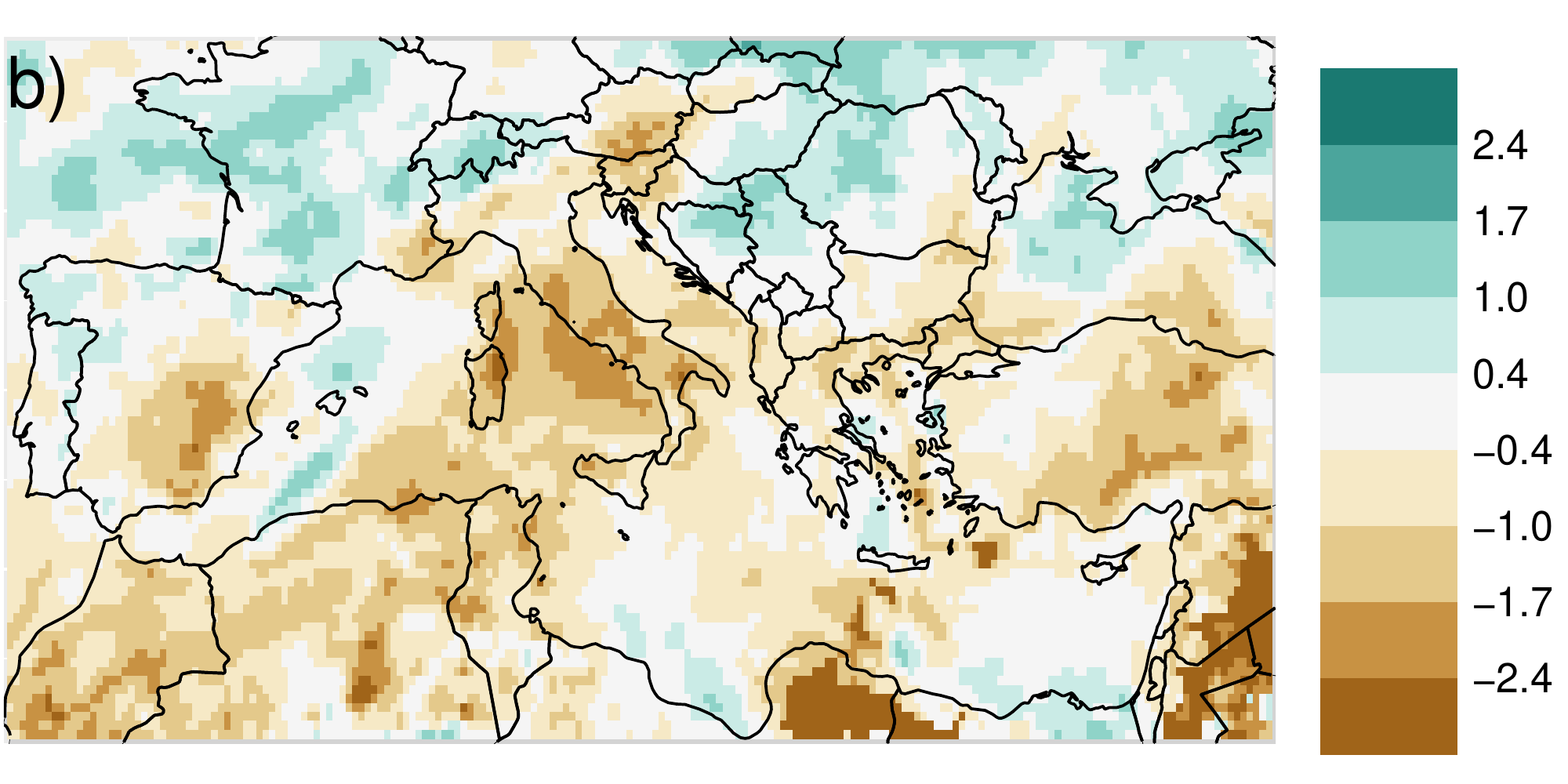} 
\end{minipage}
\centering
\begin{minipage}{0.49\linewidth}
\centering
\includegraphics[width=\linewidth]{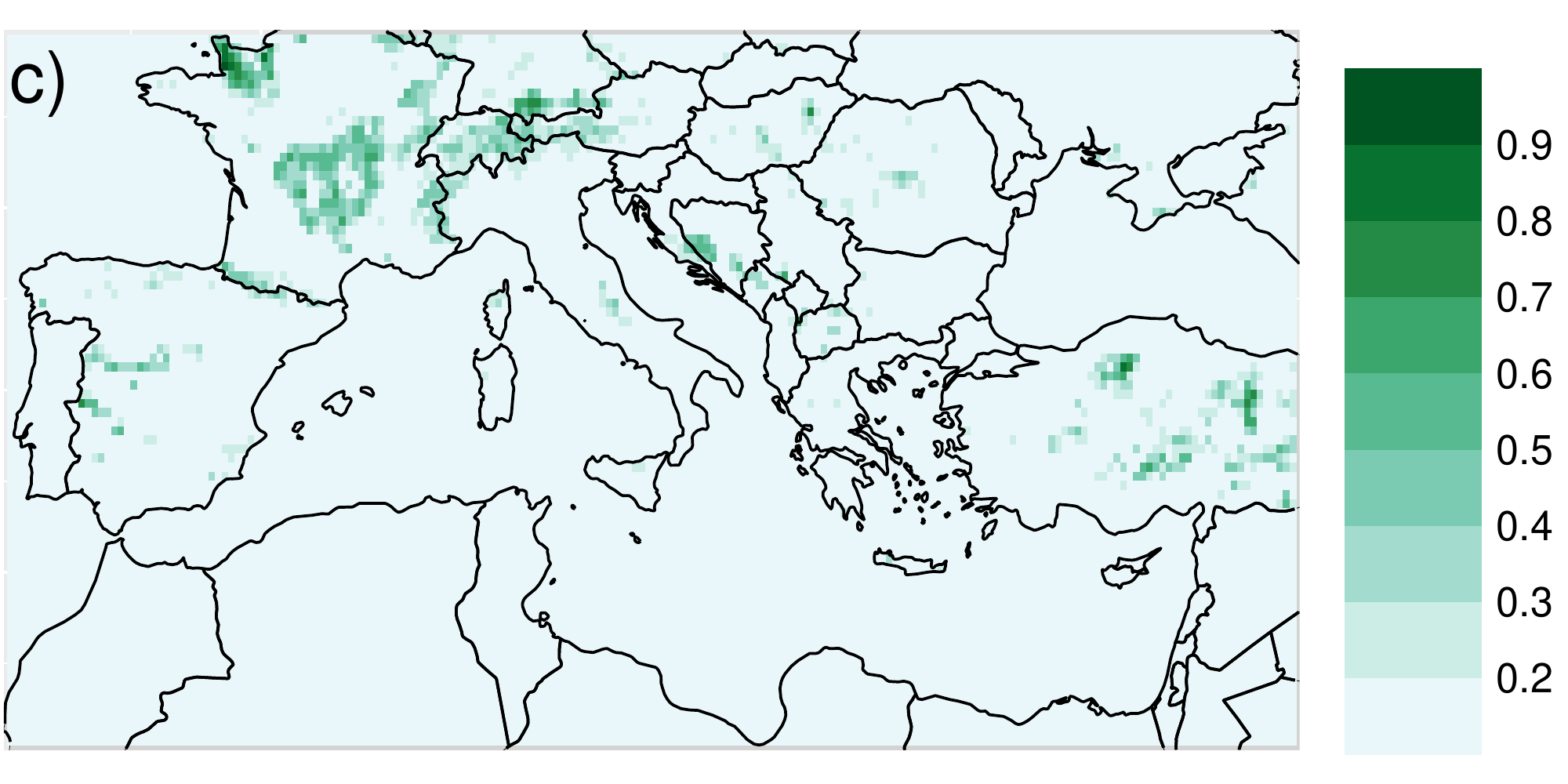} 
\end{minipage}
\hfill
\begin{minipage}{0.49\linewidth}
\centering
\includegraphics[width=\linewidth]{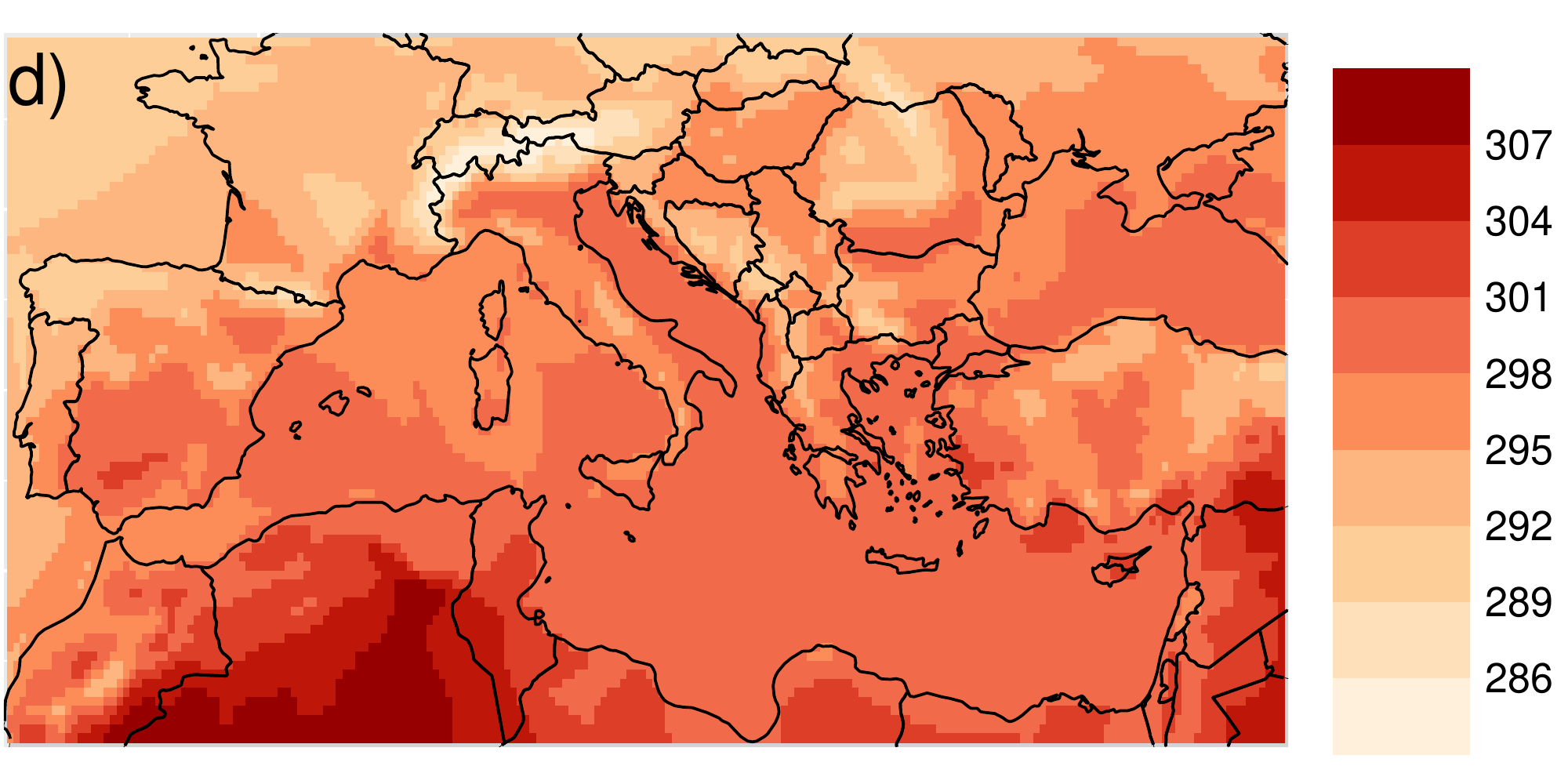} 
\end{minipage}
\begin{minipage}{0.49\linewidth}
\centering
\includegraphics[width=\linewidth]{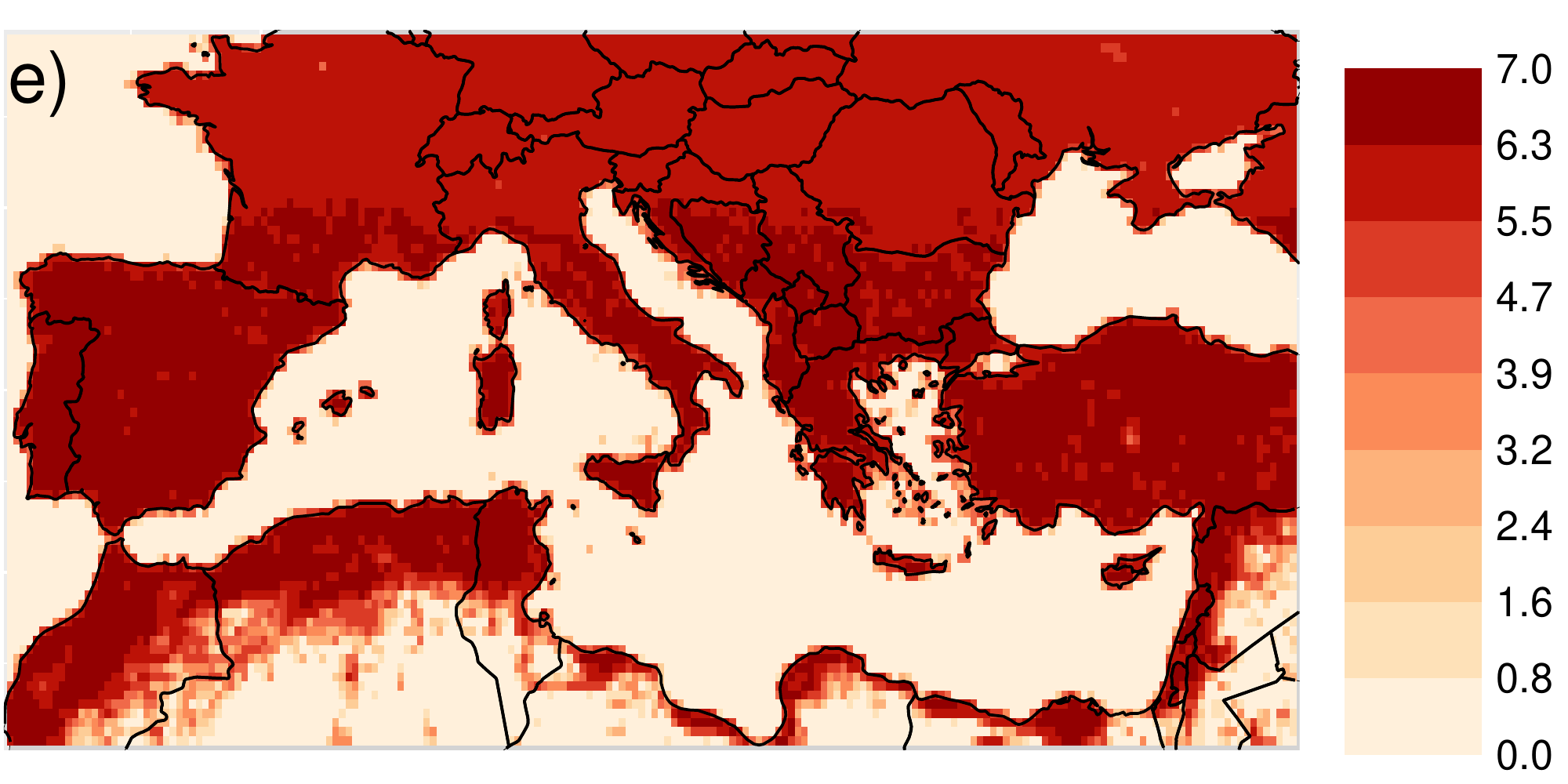} 
\end{minipage}
\hfill
\begin{minipage}{0.49\linewidth}
\centering
\includegraphics[width=\linewidth]{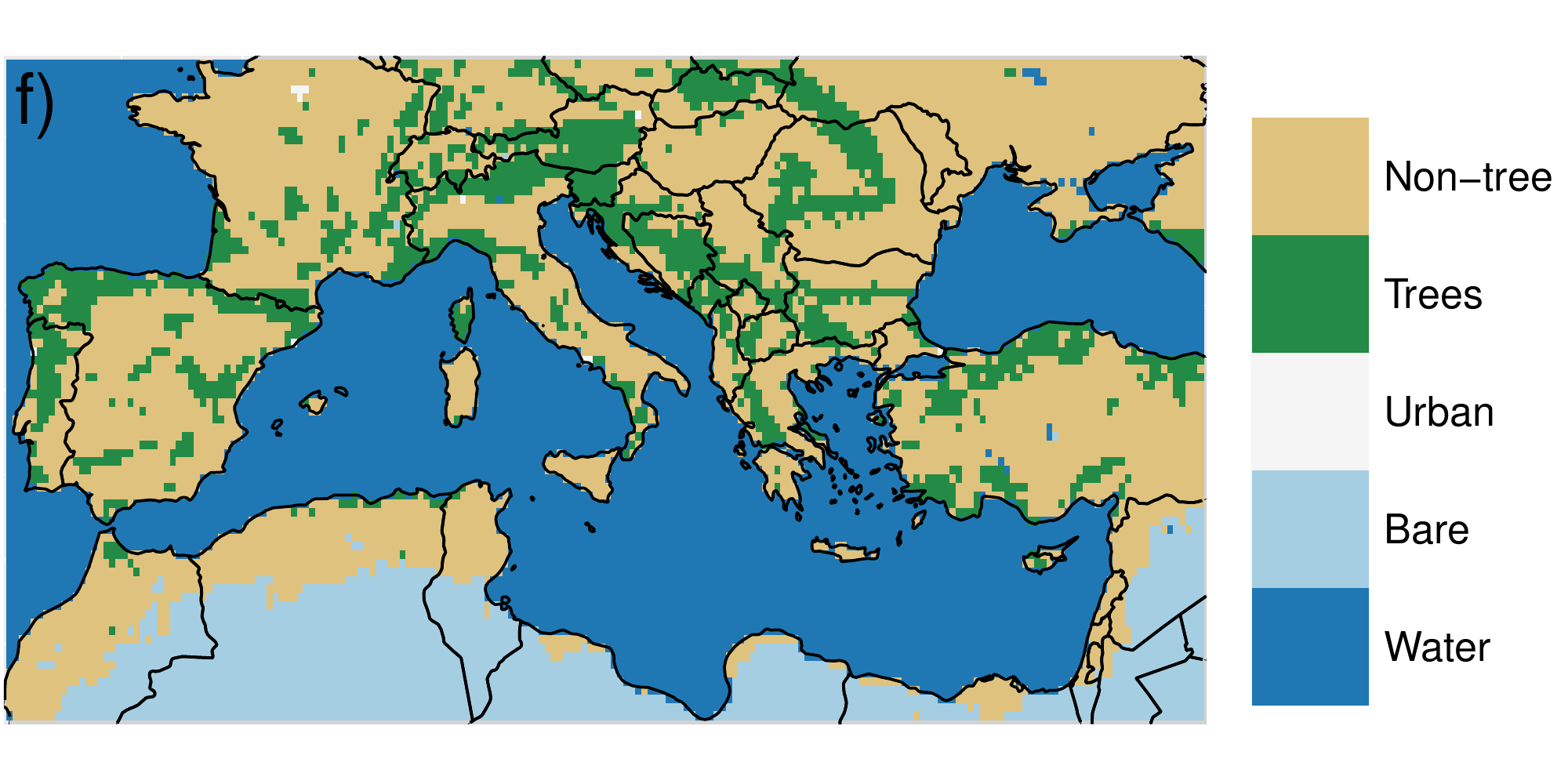} 
\end{minipage}
\caption{Maps of a) observed $\log\{1+{Y}(s,t)\}$ [burnt area; $\log(\mbox{km}^2)$], b) 3-month SPI [unitless], c) proportion of grassland coverage [unitless], d) 2m air temperature [K], {e) $\log\{1+\lambda(s,t)\}$ [burnable area; $\log(\mbox{km}^2)$] and f) dominant land cover class} for August 2001. Note that spurious values of SPI subceeding $-3$ have been truncated. }
\label{obs_maps}
\end{figure}

Our burnt area data are derived from version 5.1 of the Fire Climate Change Initiative (FireCCI) dataset 
 \citep{LIZUNDIALOIOLA2020111493}, which is generated by Moderate Resolution Imaging Spectroradiometer (MODIS) 250m reflectance data and {guided by active fire detection on a grid of pixels with spatial resolution 1km$^2$} \citep{fire4040074}. Our data have spatial resolution $0.25^\circ \times 0.25^\circ$ with temporal resolution of one month, and {the observation period covers 2001 to 2020. Figure~\ref{obs_ts} gives the monthly count and median area of non-zero BA values across the study region $\mathcal{S}$, taken to be a regular grid of pixels over large areas of southern Europe and countries encompassing the Mediterranean Basin (see Figure~\ref{obs_maps}). We choose to model only the months of June to November inclusive, which is the period when most wildfires occur. Figure~\ref{obs_ts} highlights differences in the temporal structure in wildfire occurrences and spread, and motivates us to model them separately (see Subsection~\ref{training_sec}).}  \par
We build a regression model with $d=38$ predictors of three classes: meteorological, orographical and land coverage. Thirteen meteorological variables are provided by the monthly ERA5-reanalysis on single levels \citep{hersbach2019era5}, available through the Copernicus Climate Data Service, which is given as monthly averages on a $0.25^{\circ} \times 0.25^{\circ}$ grid. Eleven variables are provided directly from ERA5: both eastern and northern components of wind velocity at 10m above ground level (m/s), temperature at 2m above ground level (K; see Figure~\ref{obs_maps}d), potential evaporation (m), evaporation (m of water equivalent), surface pressure (Pa), surface net solar, and thermal, radiation (J/m$^2$), snowmelt, snowfall and snow evaporation (all three; m of water equivalent). Monthly total precipitation (m) is used to derive a three-month standardized precipitation index (SPI, unitless), which is illustrated in Figure~\ref{obs_maps}b; SPI was derived using the $\texttt{SPEI}$ package in $\texttt{R}$ under the assumption that the data follow a gamma distribution\footnote{The data includes a small percentage ($< 0.24\%$) of unrealistically small values of SPI ($<-3$) at certain locations in arid climates with very little rainfall, where the gamma assumption used to derive SPI is not appropriate. We found that these spurious values did not have a significant impact on the model fits.}. Hourly temperature and dewpoint temperature at 2m above ground level (K) are used to derive monthly vapour-pressure deficit (VPD, measured in Pa), which refers to the difference (deficit) between the amount of moisture in the air and how much moisture the air can hold when it is saturated. Note that 2m dewpoint temperature and total precipitation are not included in the model to reduce colinearity amongst the predictors.\par
The four orographical predictors are latitude and longitude coordinates, and the mean and standard deviation of the elevation (m), for each grid-cell. Elevation estimates are derived using a densely-sampled gridded output from the $\mathtt{R}$ package ``elevatr'' \citep{hollister2017elevatr}, which accesses Amazon Web Services Terrain Tiles (\url{https://registry.opendata.aws/terrain-tiles/}); the standard deviation of the elevation is here used as a proxy for terrain roughness. \par
We also used land cover variables that describe the proportion of a grid-cell composed of one of 21 different types, including water, tree species, urban areas and grassland (see Figure~\ref{obs_maps}c); for a full list of labels, see \cite{bontemps2015multi}. We derive these predictors using a gridded land cover map, of spatial resolution 300m, that is also produced by Copernicus and is available through their Climate Data Service. For all $0.25^{\circ} \times 0.25^{\circ}$ grid-cells, the proportion of land cover types is derived from the high-resolution land cover product by counting the number of 300m$\times$300m cells of each type that fall within the boundaries of the larger grid-cell. {These predictors are dynamic and updated at the start of every year.}\par
Alongside values of BA, FireCCI provides the ``fraction of burnable area" for $(s,t)$, that is, the fraction of a spatio-temporal gridbox composed of burnable land-cover types\footnote{Non-burnable land cover types include permanent bodies of water, snow and ice, as well as urban and bare areas. All other land cover types are burnable.}; we use this to derive a measure of burnable-area, denoted $\{\lambda(s,t)\}$, by {counting the number of ``burnable'' cells from the high resolution, 300m$\times$300m, product.} Note that $\lambda(s,t)$ is not constant over time $t$ and the number of spatial locations with $\lambda(s,t)>0$ decreases monotonically from $10083$ to $10075$ with $t$. We assume that wildfires cannot form at any space-time locations with $\lambda(s,t)=0$, {as no fuel is present}, and so we treat observations $y(s,t)$ at these locations as missing\footnote{Handling of missing observations is described in Subsection~\ref{training_sec}.} for all analyses (see Figure~\ref{obs_maps}); this leaves $1,209,066$ observations. We further note that $Y(s,t)$ must satisfy $Y(s,t)\leq {\lambda}(s,t)$ for all $(s,t)\in\mathcal{S}\times\mathcal{T}$.\par
{In Figure~\ref{obs_maps}, we illustrate observations of BA, $\lambda(s,t)$ and selected predictors for August 2001. We focus on this month as i) it is exhibits one of the largest total burnt area values across the entire observation period (see Figure~\ref{obs_ts}) and ii) it serves as a reference period when we investigate the impacts of climate change on the wildfire distribution in Subsection~\ref{climate_change_sec}. Figure~\ref{obs_maps}f provides the dominant land cover class for each grid-cell. Land cover types are allocated to one of five classes: water, bare, urban areas, with the remaining vegetation types classified as either tree or non-tree; for each grid-cell, we then plot the land cover class which provides the largest proportion of its land cover. We observe that Europe is mostly dominated by non-tree land cover types, with forests typically being dominant in areas with lower average temperatures (see Figure~\ref{obs_maps}d). }
\section{Model}
\label{model_sec}
\subsection{Overview}
\label{mod_overview_sec}
As our aim is to gain insight into the drivers of both wildfire occurrence and extreme spread, we propose the following two-stage model; we first estimate the probability of wildfire occurrence $p_0(s,t):=\Pr\{Y(s,t)>0\mid\mathbf{X}(s,t)=\mathbf{x}(s,t)\}$ using a {logistic partially-interpretable neural network (see Subsection~\ref{pinnrepsec})}. We then independently model the extremal behavior of {wildfire spread, that is, the distribution of BA given that a fire has occurred}; this is given by $Y(s,t)\mid\{Y(s,t)>0, \mathbf{X}(s,t)=\mathbf{x}(s,t)\}$, which we model using the so-called peaks-over-threshold approach, detailed in Subsection~\ref{POTsec}. {Our choice to model these two components separately is motivated by observable differences in their spatial and temporal patterns (see Figures~\ref{obs_ts} and \ref{clim_maps}).} Estimates from the two models can then be combined to do inference on the full distribution of burnt area, $Y(s,t)\mid \mathbf{X}(s,t)$, by noting that $\Pr\{Y(s,t) \leq y\mid\mathbf{X}(s,t)=\mathbf{x}(s,t)\}$ equals 
\begin{equation}
\label{eq:total_prob}
1-p_0(s,t)+p_0(s,t)\Pr\{Y(s,t) \leq y\mid Y(s,t)>0,\mathbf{X}(s,t)=\mathbf{x}(s,t)\}.
\end{equation}
{Note that, in this way, our full model for BA combines separate models for the occurrence probability and for the spread.}
\subsection{Peaks-over-threshold model}
\label{POTsec}
The peaks-over-threshold (POT) approach is a widely applied framework for modelling the upper-tails of a random variable, see, for example, \cite{Pickands1975}, \cite{davison1990models} and \citet{coles2001introduction}. For a random variable $Y$, we first assume that there exists some high threshold $u$ such that the distribution of $(Y-u)\;|\;( Y > u)$ is characterised by the generalised Pareto distribution, denoted GPD$(\sigma_u,\xi)$, $\sigma_u>0,\xi\in\mathbb{R}$, with distribution function $H(y)=(1+\xi y/\sigma_u)^{-1/\xi}$ and support $y\geq 0$ for $\xi \geq 0$ and $0 \leq y \leq -\sigma_u/\xi$ for $\xi < 0$. Note that the scale parameter $\sigma_u$ is dependent on $u$, that is, it changes as we condition upon $Y$ exceeding a higher threshold; this makes it difficult to interpret estimates of $\sigma_u$ if $u$ varies over space-time locations $(s,t)$. To that end, we exploit the threshold-stability law \citep{coles2001introduction}  of the GPD and re-parameterise the distribution in terms of a scale parameter $\sigma=\sigma_u-\xi u>0$, which is independent of $u$.
\par 
{The shape parameter $\xi$ controls the limit of the upper-tail of $Y$: for $\xi <0$ and $\xi \geq 0$, we have that $Y$ has a bounded, and infinite, upper-tail, respectively. If $\xi \geq 1$, then $Y$ is very heavy-tailed with infinite expected value. This property is considered to be inappropriate for environmental applications and so we constrain $\xi <1$ throughout. A number of studies have shown that wildfire burnt areas are heavy-tailed (with $\xi > 0$; see \cite{pereira2019statistical, koh2021spatiotemporal, richards2022}), but our data satisfy the natural physical constraint that burnt areas are bounded above by the available burnable area (i.e., $y(s,t)<\lambda(s,t)$ for all $(s,t)\in\mathcal{S}\times\mathcal{T}$; see Figure~\ref{hists} in Appendix~B). This suggests that a model with bounded upper-tail (i.e., $\xi <0$) would, in principle, be more appropriate. However, it can be computationally problematic to fit such a model due to issues regarding numerical instability \citep{smith1985maximum} and parameter-dependent support of the distribution, which makes inference with neural networks particularly difficult \citep{richards2022}; we thus further constrain $\xi >0$ throughout. We note that, in theory, our model is inconsistent with the physical characteristics of the data generating process, but, in practice, the upper-tail behaviour of the data is approximated sufficiently well provided that one does not extrapolate too far into the upper-tail. At observable levels, we find that a model with $\xi>0$ fits our data better than a model with $\xi<0$ and excellent model fits are obtained with $\xi > 0$, see Subsection~\ref{risk_assess_sec}; the threshold $u$ would need to be excessively high to see the benefit of imposing $\xi<0$.}\par
We model the extremal behaviour of {non-zero spread} $Y(s,t)\;|\;\{Y(s,t)>0, \mathbf{X}(s,t)=\mathbf{x}(s,t)\}$ using the POT approach, with a spatio-temporally varying exceedance threshold $u(s,t)>0$ and scale parameter $\sigma(s,t)>0$, such that they are both functions of predictor set $\mathbf{X}(s,t)$ described in Section~\ref{Data_sec}; similarly to \cite{koh2021spatiotemporal}, we fix $\xi$ over all space and time for computational efficiency. Hence, we have $\{Y(s,t)-u(s,t)\}\;|\; \{Y(s,t)>u(s,t),\mathbf{X}(s,t)=\mathbf{x}(s,t)\}\sim\mbox{GPD}\{\sigma(s,t),\xi\}$; models for $u(s,t)$ and $\sigma(s,t)$ are detailed in Subsection~\ref{pinnrepsec}. {Whilst $u(s,t)$ is difficult to interpret, the scale $\sigma(s,t)$ can be interpreted as a measure of the conditional severity of extreme wildfires spread, should a fire occur at site $s$ in month $t$}. The distribution function of $Y(s,t)\;|\; \{Y(s,t)>0, \mathbf{X}(s,t)=\mathbf{x}(s,t)\}$, denoted by $F_{(s,t),0}(y)$ for all $(s,t)\in\mathcal{S}\times\mathcal{T}$ is given by
\begin{equation}
\label{MargTransform}
F_{(s,t),0}(y)=1-p_u(s,t)\left[1+\frac{\xi\{y-u(s,t)\}}{\sigma(s,t)+\xi u(s,t)}\right]^{-1/\xi}\;\;\;\; \text{if}\;\;y > u(s,t),
\end{equation}
where $p_u(s,t):=\Pr\{Y(s,t) > u(s,t) \;|\; Y(s,t)>0,\mathbf{X}(s,t)=\mathbf{x}(s,t)\}\in[0,1]$. As our interest lies only in the extremes of $Y(s,t)$, for $y\leq u(s,t)$ we estimate $F_{(s,t),0}(y)$ empirically; here we use the estimator 
\begin{equation}
\label{emp_eq}
\hat{F}_{(s,t),0}(y)=\{1-\hat{p}_u(s,t)\}\left[\sum_{(s,t)\in\boldsymbol{\omega}}\mathbbm{1}\{y(s,t)\leq y\}\right]\left[\sum_{(s,t)\in\boldsymbol{\omega}}\mathbbm{1}\{y(s,t)\leq u(s,t)\}\right]^{-1},
\end{equation} 
where $\mathbbm{1}$ denotes the indicator function, $\hat{p}_u(s,t)$ denotes an estimate of $p_u(s,t)$, and with observations pooled across all space-time observation locations $\boldsymbol{\omega}$. An alternative approach to an empirical estimator could be to use a parametric distribution for the bulk and lower-tail of $Y(s,t)\;|\;Y(s,t)>0$, see, for example, \cite{carreau2007hybrid} and \cite{opitz2018inla}. Similar zero-inflated models with empirical bulk and GPD upper-tails have been employed by \cite{richards2022modelling} and \cite{d2021flexible} for modelling extreme precipitation and U.S. wildfire burnt areas, respectively.
\par
Typically, we would estimate the threshold $u(s,t)$ as some high $\tau$-quantile of {non-zero spread} $Y(s,t)\;|\;\{Y(s,t)>0\}$, for $\tau \in (0,1)$. If $u(s,t)$ is assumed known, then it follows that $p_u(s,t)=1-\tau$ for all $(s,t)$; however, given the highly non-stationary model that we wish to fit and the complexity of the data, it may be inappropriate to assume that $u(s,t)$ is exactly the required $\tau$-quantile. Instead, we use a neural network to model $p_u(s,t)$ (see Subsection~\ref{pinnrepsec}), which leads to improved model fits through increased flexibility of \eqref{MargTransform}. {To illustrate the components of \eqref{MargTransform}, Figure~\ref{sup_POT_fig} in Appendix~B gives maps of observed $Y(s,t)$, for August 2001, and the corresponding estimates of $u(s,t)$ (with $\tau = 0.4$), $p_u(s,t)$ and exceedances $Y(s,t)-u(s,t)$.}
\subsection{Partially-interpretable neural networks}
\label{DL_sec}
Similarly to \cite{richards2022}, we represent model parameters using partially-interpretable neural networks (PINNs). Consider a general parameter $\theta(s,t)$, which is a function of space-time locations $(s,t)\in \mathcal{S}\times \mathcal{T}$; for example, we may take one of $\theta:=\sigma$, ${\theta:=u}$, $\theta:=p_0$ or $\theta:=p_u$, for $\sigma, u,$ and $p_u,$ defined in \eqref{MargTransform}, and $p_0$ defined in Subsection~\ref{mod_overview_sec}. For all $(s,t)$, we divide the predictor process $\{\mathbf{X}(s,t)\}$ into two complementary components: ``interpreted" and ``non-interpreted" predictors, denoted $\{\mathbf{X}_\mathcal{I}(s,t)\}$ and $\{\mathbf{X}_\mathcal{N}(s,t)\}$, respectively, where for $I\in\{0,1,\dots,d-1\}$ we have that $\mathbf{X}_\mathcal{I}(s,t)\in\mathbb{R}^I$ and $\mathbf{X}_\mathcal{N}(s,t)\in\mathbb{R}^{d-I}$ are distinct sub-vectors of $\mathbf{X}(s,t)$, with observations of the components denoted $\mathbf{x}_{\mathcal{I}}(s,t)$ and $\mathbf{x}_{\mathcal{N}}(s,t)$, respectively; we also note that the indices mapping $\mathbf{x}(s,t)$ to $\mathbf{x}_{\mathcal{I}}(s,t)$ are consistent across all $(s,t)$, and similarly for $\mathbf{x}_{\mathcal{N}}(s,t)$. We then represent $\theta$ as
\begin{align}
\theta(s,t)=C(s,t)h[m_\mathcal{I}\{\mathbf{x}_\mathcal{I}(s,t)\}+m_\mathcal{N}\{\mathbf{x}_\mathcal{N}(s,t)\}]
\label{PINNmodel}
\end{align}
for functions $m_{\mathcal{I}}:\mathbb{R}^{I}\rightarrow\mathbb{R},m_\mathcal{N}:\mathbb{R}^{d-I}\rightarrow\mathbb{R}$, a link function $h:\mathbb{R}\rightarrow\mathbb{R}$ controlling the range of $\theta$ and a scaling function $C(s,t):(\mathcal{S},\mathcal{T})\rightarrow \mathbb{R}$. In cases where $\theta\in(0,1)$, (e.g., for $\theta:=p_0$ or $\theta:=p_u$), we set $h$ to the logistic function, that is, $h(x)=1/2+\tanh(x/2)/2$. When $\theta>0$, for example, for $\theta:=u$ or $\theta:=\sigma$, we take $h(x)=\exp(x)$. \par
\cite{richards2022} assert that the function $m_{\mathcal{I}}$ must be interpretable, that is, it must be simple and either parametric, for example, linear, or semi-parametric (e.g., additive). We choose to adopt thin-plate splines \citep{wood2003thin} to independently model the effect of each component of $\mathbf{x}_\mathcal{I}(s,t)=(x_{\mathcal{I},1}(s,t),\dots,x_{\mathcal{I},I}(s,t))^T$ on $m_\mathcal{I}$; that is, we let 
\[
m_\mathcal{I}\{\mathbf{x}_{\mathcal{I}}(s,t)\}=\sum^I_{j=1}m_{\mathcal{I},j}\{x_{\mathcal{I},j}(s,t)\},
\] 
where for all $j=1,\dots,I$, each $m_{\mathcal{I},j}$ is a thin-plate spline of $x_{\mathcal{I},j}$.
The function  $m_\mathcal{N}$ is assumed to be of unknown structure and highly non-linear, and is approximated via a neural network with a suitable architecture. As neural networks are not readily-interpretable, we should choose $\mathbf{x}_{\mathcal{I}}(s,t)$ to be the set of predictors for which we wish to infer their influence on the parameter $\theta$; the remaining $d-I$ predictors form $\mathbf{x}_{\mathcal{N}}$. In the case where $m_{\mathcal{I}}(s,t)=0$ for all $(s,t)$ or $I=0$, we describe the model as ``fully-NN". \par
We use representation \eqref{PINNmodel} for each of the parameters $p_0$, $p_u$, $u$, and $\sigma$, albeit with different predictor components $\mathbf{x}_\mathcal{I}$ and $\mathbf{x}_\mathcal{N}$, and functions $h$, $m_\mathcal{I}$ and $m_\mathcal{N}$; see Subsection~\ref{app_overview_sec} for full details. The scaling function $C(s,t)$ also differs between parameters; for $p_0$ and $p_u$, we simply set $C(s,t)=1$ for all $(s,t)$. However, for $u$ and $\sigma$ we found improvements in model fit when setting $C(s,t)={\lambda}(s,t)$, that is, by using burnable-area as an offset term. With this representation, the predictor set $\mathbf{x}(s,t)$ directly influences properties of the distribution of proportion of burnable-area that is burnt by wildfires, rather than the absolute value of burnt area; this accounts for the high variability in $\lambda(s,t)$ observed over the observed domain $\boldsymbol{\omega}\subset \mathcal{S} \times \mathcal{T}$. For models with $C(s,t)=1$, we include $\lambda(s,t)$ as an extra predictor in $\mathbf{x}_\mathcal{N}(s,t)$, raising the number of predictors in these models to $d=39$.\par
\label{pinnrepsec}
\label{architecture_sec}
Different types of neural network exist for estimating $m_\mathcal{N}$ in \eqref{PINNmodel}, which capture different structures in $\mathbf{x}_\mathcal{N}$. For estimating GPD parameters, \cite{rietsch2013network}, \cite{carreau2007hybrid},  \cite{carreau2011stochastic} and \cite{wilson2022deepgpd} employ feed-forward neural networks, whilst \cite{pasche2022neural} use recurrent neural networks built with long-short term memory (LSTM) layers \citep{hochreiter1997long}. In their application, \cite{richards2022} favour the use of feed-forward over recurrent layers for estimating $m_\mathcal{N}$ within the PINN framework, due to the high computational demand and a large number of trainable parameters required for training the latter; we thus adopt their approach and construct a feed-forward PINN. {We use only feed-forward NNs in our model, but employ different types for each parameter (i.e., $p_0$ in \eqref{eq:total_prob} and $\sigma$, $u$, and $p_u$, in \eqref{MargTransform}; see Subsection~\ref{app_overview_sec} for details). We use either a convolutional neural network (CNN) or densely-connected NN; for details of the two types, see Appendix~A.} 
\subsection{Inference}
\label{training_sec}
We use a two-stage procedure to construct a model for the {full distribution of burnt area} $Y(s,t)\;|\;\{\mathbf{X}(s,t)=\mathbf{x}(s,t)\}$. We first simultaneously estimate $u(s,t)$, the $\tau$-quantile of {non-zero spread} $Y(s,t)\;|\;\{Y(s,t)>0,\mathbf{X}(s,t)=\mathbf{x}(s,t)\}$, and {occurrence probability} $p_0(s,t)$; we then use estimates of $u(s,t)$ to estimate $p_u(s,t)$ and the GPD parameters, $\sigma(s,t)$ and $\xi$. Combining the aforementioned functional parameter estimates with the empirical estimator $F^*_{(s,t),0}$, defined in \eqref{emp_eq}, allows us to then estimate the distribution function of {burnt area} $Y(s,t)\;|\;\{\mathbf{X}(s,t)=\mathbf{x}(s,t)\}${; a schematic illustrating this modelling procedure is presented in Figure~\ref{fig:flow}.}

\tikzstyle{decision} = [diamond, draw, fill=blue!20, 
    text width=4em, text badly centered, node distance=3cm and 0.5cm, inner sep=0pt]
\tikzstyle{block} = [rectangle, draw, fill=blue!20, 
    text width=7.4em, text centered, rounded corners, minimum height=4em]
\tikzstyle{line} = [draw, -latex']
\tikzstyle{cloud} = [draw, ellipse,fill=red!20, node distance=3cm,
    minimum height=2em]
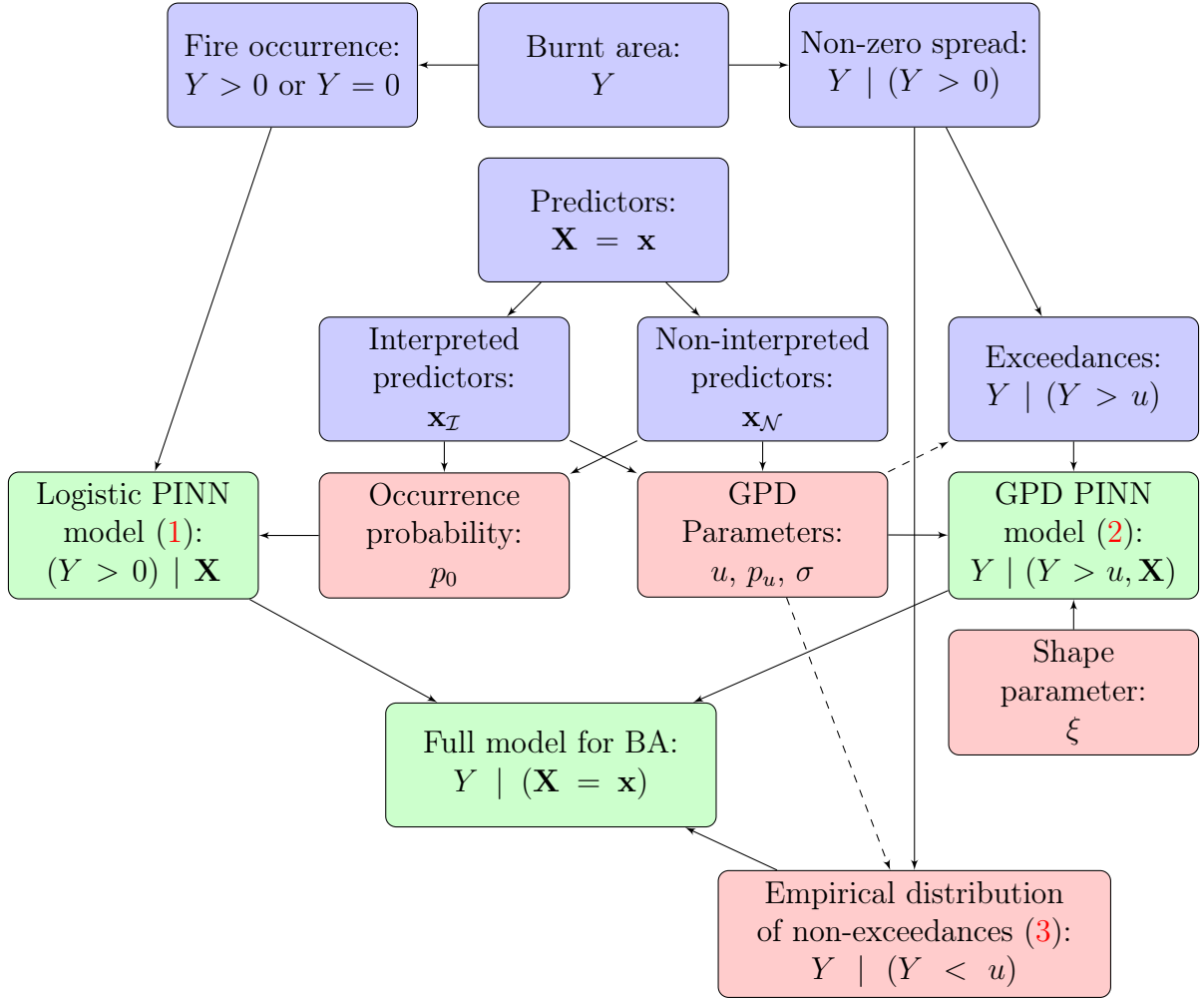
\begin{figure}[t!]
    \centering
\begin{tikzpicture} [node distance=0.4cm and 0.8cm, align=center]
    \node [block] (init) {Burnt area: \\$Y$};
    \node [block, left= of init] (occur) {Fire occurrence: \\$Y>0$ or $Y=0$};
    \node [block, right= of init] (spread) {Non-zero spread: \\$Y\;|\; (Y>0)$};
        \node [block, below=of init] (preds) {Predictors: \\$\mathbf{X}=\mathbf{x}$};
        \begin{scope}[node distance=3cm and 2cm]
        \node [block, below left of=preds] (interp) {Interpreted \\predictors: \\$\mathbf{x}_\mathcal{I}$};
        \node [block, below right of=preds] (non-interp) {Non-interpreted \\predictors: \\$\mathbf{x}_\mathcal{N}$};
        \end{scope}
         \node [block, right=of non-interp] (exceeds) {Exceedances: \\$Y\;|\; (Y>u)$};

        \node [block, below right=of non-interp, fill=green!20] (spread pinn) {GPD PINN \\  model \eqref{MargTransform}: \\$Y\;|\;(Y>u, \mathbf{X})$};
        \node [block, below left=of interp, fill=green!20] (log) {Logistic PINN \\model \eqref{eq:total_prob}: \\$(Y>0)\;|\; \mathbf{X}$};
        \node [block, right=of log, fill=red!20] (occur prob) {Occurrence probability: \\ $p_0$};
        \node [block, below left=of exceeds, fill=red!20] (spread pars) {GPD  \\ Parameters: \\ $u$, $p_u$, $\sigma$};
\node [block, below=of spread pinn, fill=red!20] (xi pars) {Shape\\parameter: \\ $\xi$};
\node [block, below=of spread, yshift = -9.5cm,  text width=12em, fill=red!20] (nonexceeds) {Empirical distribution of non-exceedances \eqref{emp_eq}: \\$Y\;|\; (Y<u)$};
        \node [block, below=of preds, yshift= -5.2cm, xshift= -0.7cm, text width=10em, fill=green!20] (comb) {Full model for BA: \\ $Y\;|\;(\mathbf{X}=\mathbf{x})$};
    \path [line] (init) -- (occur);
    \path [line] (init) -- (spread);
    \path [line] (preds) -- (interp);
    \path [line] (preds) -- (non-interp);
  \path [line] (occur) -- (log);
    \path [line] (spread) -- (exceeds);
        \path [line] (exceeds) -- (spread pinn);
 \path [line] (interp) -- (occur prob);
    \path [line] (interp) -- (spread pars);
     \path [line] (non-interp) -- (occur prob);
    \path [line] (non-interp) -- (spread pars);
     \path [line] (occur prob) -- (log);
    \path [line] (spread pars) -- (spread pinn);
    \path [line, dashed] (spread pars) -- (exceeds);
    \path [line] (xi pars) -- (spread pinn);
         \path [line] (spread) -- (nonexceeds);
         \path [line] (spread pinn) -- (comb);
         \path [line] (log) -- (comb);
                  \path [line] (nonexceeds) -- (comb);
                                    \path [line, dashed] (spread pars) -- (nonexceeds);
\end{tikzpicture}
\caption{{Schematic of methodology to estimate the distribution of burnt area $Y\;|\:(\mathbf{X}=\mathbf{x})$. Dashed lines denote a connection through the exceedance threshold $u$ only. Blue, red and green boxes denote data, parameters, and models, respectively. The space-time index $(s,t)$ has been dropped from notation for brevity.}}
\label{fig:flow}
\end{figure}

The neural networks in each parameter model are trained using the \texttt{pinnEV} package \citep{pinnEV}, which utilises the \texttt{R} interface to \texttt{Keras} \citep{kerasforR}.  We use the Adaptive Moment Estimation (Adam) algorithm \citep{kingma2014adam}\footnote{{with default hyper-parameters, see \url{https://keras.io/api/optimizers/adam/} (accessed 05/23/2023).}}, to minimise the negative log-likelihood associated with each model; for $p_0$ and $p_u$, this is equivalent to the binary cross-entropy loss or the negative log-likelihood for a Bernoulli random variable; for $u(s,t)$, we use the tilted quantile loss \citep{koenker_2005} \[\ell\{u(s,t); y(s,t)\}=\sum_{(s,t)\in\boldsymbol{\omega}}\max[\tau\{y(s,t)-u(s,t)\},(\tau-1)\{y(s,t)-u(s,t)\}].\] For $\sigma(s,t)$ and $\xi$, we minimise the negative log-likelihood for the GPD \citep{coles2001introduction}. {In cases where $y(s,t)$ is treated as missing, we remove its influence from the loss function.} For all model fits, we use a validation and testing scheme to reduce over-fitting and improve out-of-sample prediction, with details provided in Appendix~A; {we do not use any form of dropout during training, as the NNs we use are relatively simple (see Subsection~\ref{app_overview_sec}). To quantify all parameter uncertainty, we utilise a stationary bootstrap scheme \citep{politis1994stationary} with expected block size of two months (see Appendix~A), and with the entire model re-estimated for each bootstrap sample.}

\section{Results}
\label{application_sec}
\subsection{Overview}
\label{app_overview_sec}
We choose to interpret the effect of $I=3$ variables on the occurrence probability of wildfires and the $\sigma$ parameter determining extreme spread; these are VPD, temperature and 3-month SPI, which were chosen as these are important drivers of fire occurrence and are strongly impacted by climate change in the Mediterranean \citep{giorgi2008climate,bevacqua2022precipitation}. As covariate effects on the parameters $p_u$ and $u$ are difficult to interpret, we estimate both using a fully-NN model, that is, we do not interpret the effect of any predictors and set $\mathbf{x}_\mathcal{N}(s,t):=\mathbf{x}(s,t) $ for all $(s,t)$; this improves model fits through added model flexibility. To assess the impact associated with the interpreted predictors, we model $p_0$ and $\sigma$ using the PINN framework described in Subsection~\ref{pinnrepsec}. Before training, predictors are standardised by subtracting and dividing by their marginal means and standard deviations to improve numerical stability.  \par
To perform model selection and determine the optimal hyper-parameters and network architecture, we evaluate scores and diagnostics on test data using the estimated parameters from the fitted models for each bootstrap sample; recall that the test data is not used in model fitting. For $p_0(s,t)$ and $p_u(s,t)$, we compare model fits using the area under the receiver operating characteristic curve, denoted AUC. {For details on comparing fits of the GPD PINN model \eqref{MargTransform}, see Appendix~A.} \par
All neural networks are trained {with mini-batch size equal to the number of observations; a model checkpoint is saved at each of 10,000 epochs and only the estimate that minimises the validation loss is returned}. Under our model selection scheme, we determine that the optimal architecture for estimating $m_\mathcal{N}$ in \eqref{PINNmodel} differs between the four parameters. As inference on $p_0$ uses all observations of {burnt area} $Y(s,t)$, it requires the most complex architecture: a five-layered CNN (see Appendix~A) with 16 filters/nodes per layer. The threshold $u(s,t)$ is estimated with non-exceedance probability $\tau=0.4$, and uses a two-layered CNN with four filters/nodes per layer. We use densely-connected NNs to estimate $p_u$ and $\sigma$; precisely, we use a five-layered network with 16 nodes per layer and a four-layered network with 10 nodes per layer for the former and latter, respectively. \par
We use 250 bootstrap samples to assess model and parameter uncertainty and present results for the quantities of interest (e.g., quantiles and models parameters) as the empirical median and pointwise quantile estimates across all bootstrap samples. The splines used to model $m_\mathcal{I}$ are estimated with knots taken to be marginal quantiles of the predictors $\{\mathbf{x}(s,t):(s,t)\in\boldsymbol{\omega}\}$ with equally-spaced probabilities; splines for $p_0$ and $\sigma$ use ten and six knots, respectively. Inference on $m_\mathcal{I}$ is conducted by considering the individual contribution of each interpreted predictor to $m_\mathcal{I}$, denoted by $\hat{m}_{\mathcal{I},j}(x_j)$ for $j=1,2,3$. Estimates of each spline $\hat{m}_{\mathcal{I},j}(x_j)$ for each bootstrap sample are centred by subtracting the value of the spline evaluated at the median of the predictor values; it is important to centre the curves as we are interested in relative changes in $m_\mathcal{I}$, not its absolute value, and it circumvents issues related to identifiability between $m_\mathcal{I}$ and $m_\mathcal{N}$ \citep{richards2022}. Uncertainty in the centred spline estimates are then presented using functional box-plots \citep{sun2011functional}; these can be considered to be analogues of the classical box-and-whisker plots for functional data. The central region gives an indication of the ``central'' $50\%$ of curves, with the median taken to be the most central. Maximum envelopes (blue curves in Figure~\ref{occur_GAM}) encompass all functions not identified as outliers.  For full details on the definition of ``centrality" and identification of outliers for functional data, see details in  \cite{sun2011functional}.\par
\begin{figure}[t!]
\centering
\begin{minipage}{0.49\linewidth}
\centering
\includegraphics[width=\linewidth]{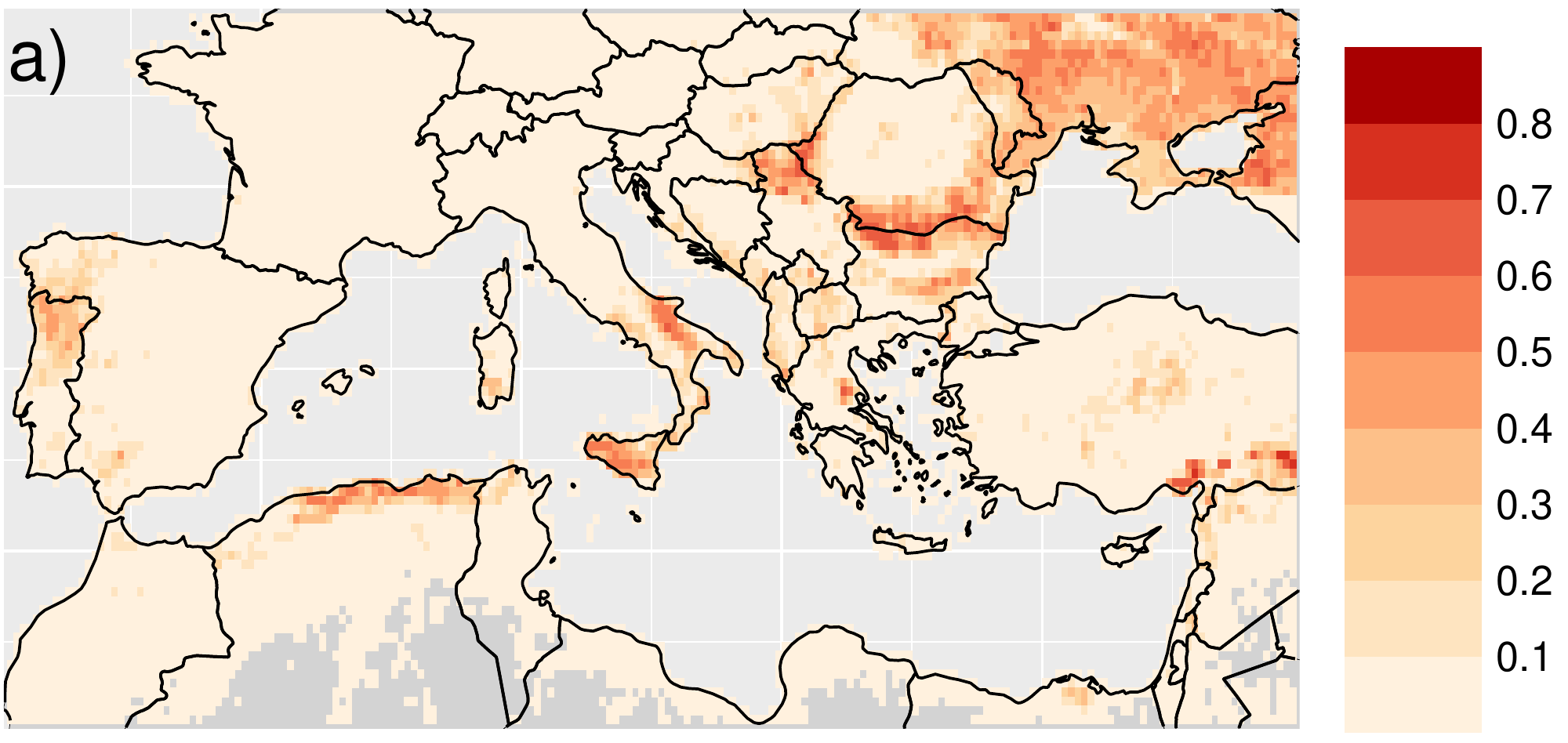} 
\end{minipage}
\begin{minipage}{0.49\linewidth}
\centering
\includegraphics[width=\linewidth]{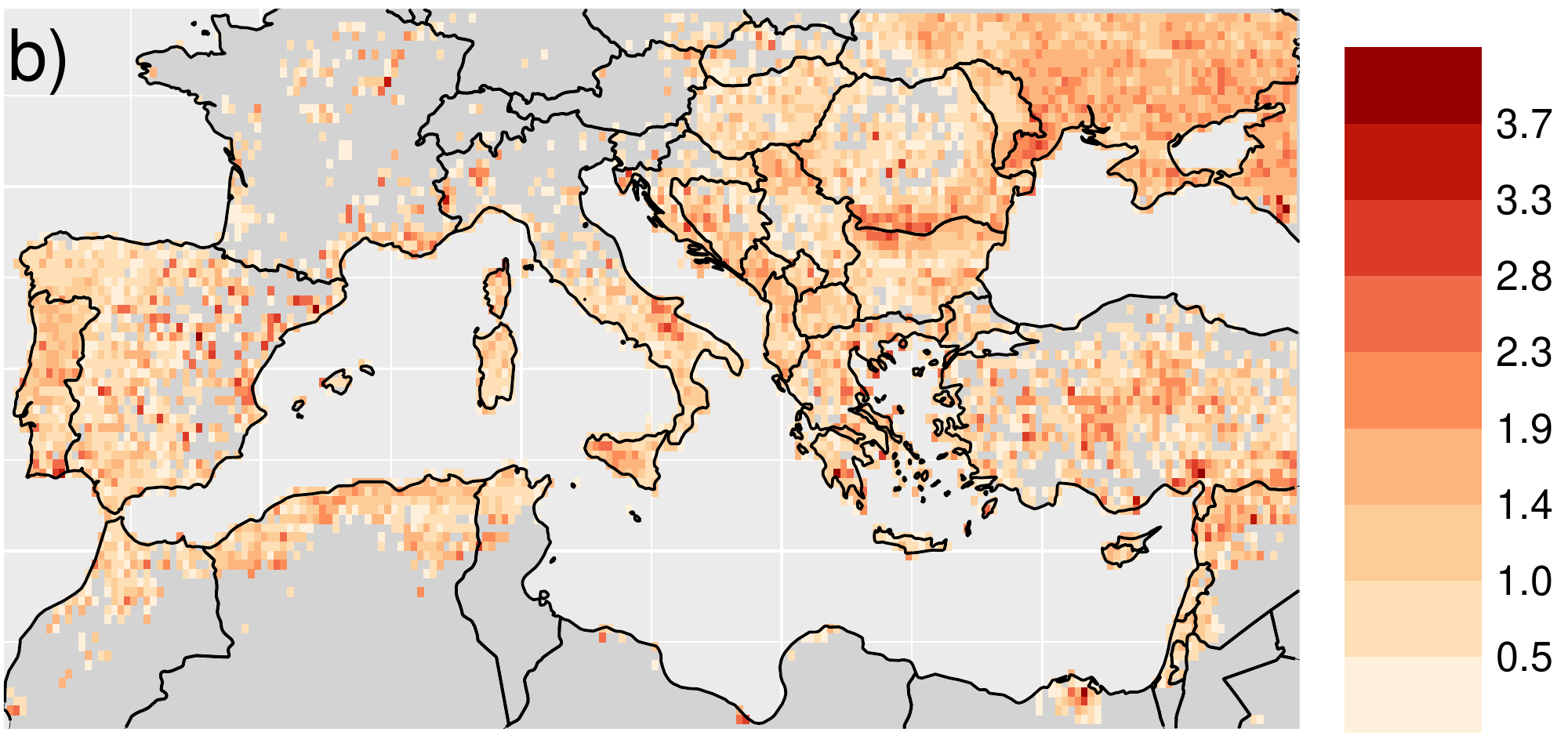} 
\end{minipage}
\begin{minipage}{0.49\linewidth}
\centering
\includegraphics[width=\linewidth]{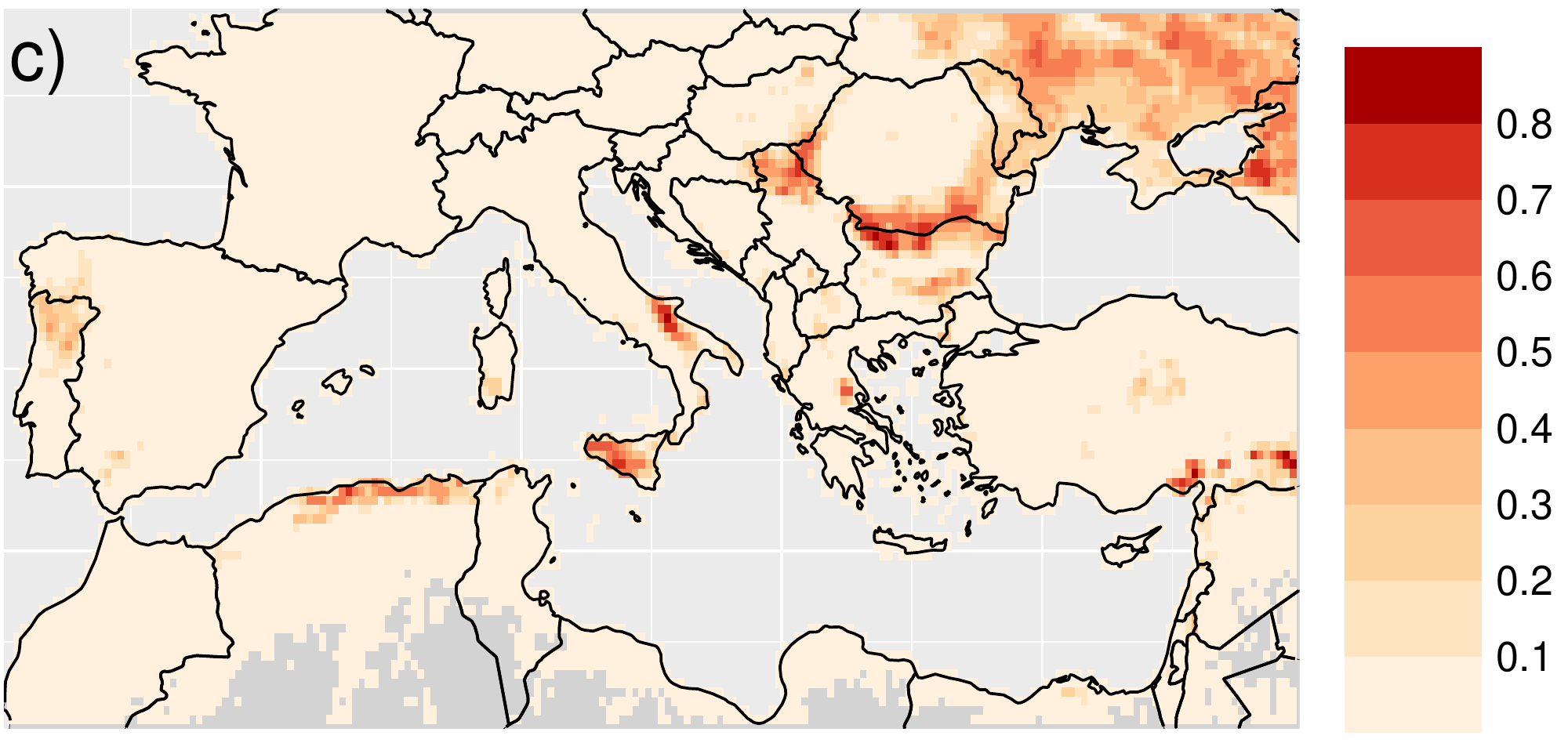} 
\end{minipage}
\begin{minipage}{0.49\linewidth}
\centering
\includegraphics[width=\linewidth]{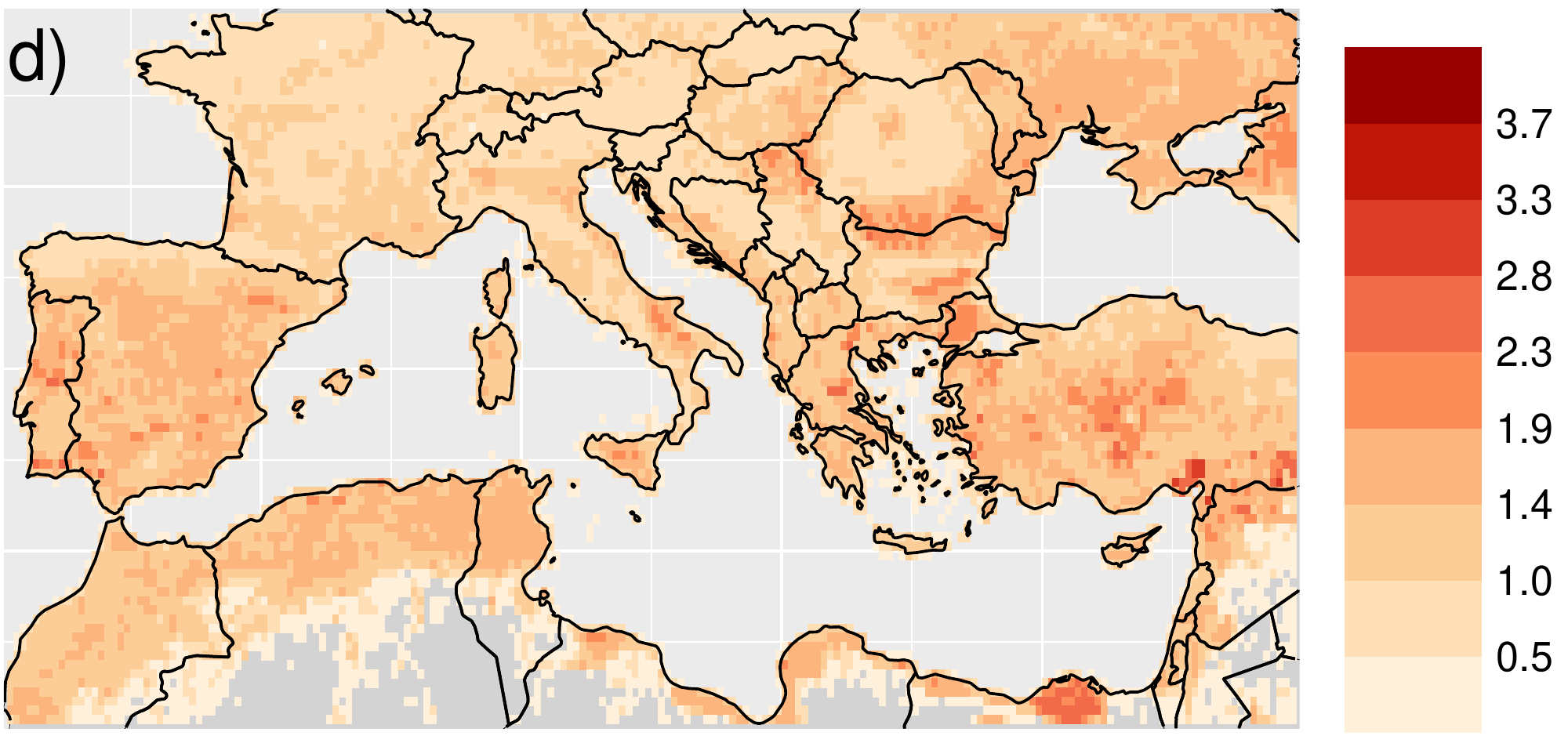} 
\end{minipage}
\caption{Maps of site-wise a) empirical {fire occurrence probability} $p_0(s,t)$ [unitless] and medians of b) observed $\log\{1+Y(s,t)\}\;|\;Y(s,t)>0$ [conditional spread, $\log(\mbox{km}^2)$], c) estimated $p_0(s,t)$ [unitless] and d) estimated $\log\{1+\sigma(s,t)\}$ [conditional spread severity; $\log(\mbox{km}^2)$]. Metrics are averages over all months and years in the observation period. }
\label{clim_maps}
\end{figure}
As an exploratory analysis, we present a climatology of observed, and modelled, wildfire occurrence, and spread, in Figure~\ref{clim_maps}. Here we present site-wise medians of model estimates of wildfire occurrence probability (i.e., $p_0(s,t)$), and {conditional spread}  intensity (i.e., $\sigma(s,t)$), as well as site-wise empirical estimates of $p_0(s,t)$ and median {non-zero spread} $Y(s,t)\:|\;Y(s,t)>0$. Excellent fit is obtained for the occurrence probability, and extreme spread, models, with details of goodness-of-fit discussed in Subsection~\ref{risk_assess_sec}. All metrics are averaged over all months and years in the observation period; model outputs are averaged over all bootstrap samples and empirical estimates of $p_0(s,t)$ are derived by counting the mean number of wildfire occurrences within the observation period. Figure~\ref{clim_maps} highlights areas where fires are likely to occur in any given month and the average site-wise intensity of those wildfires. During 2001--2020, areas of concern include the Nile Delta and Turkey, where the probability of wildfire occurrence is very low (see panels a and c) but extreme wildfires were particularly intense (panels b and d), and Ukraine, which exhibits both high probability of occurrence and intensity; the Alps and Carpathians experience very few, if any, wildfires, and the estimated intensity in these regions is relatively low. We note that, given the short observation period of the data, there is some uncertainty around these estimates and the inferences drawn may not apply to other 20-year periods.
\subsection{Interpretable results}
\label{interp_results_sec}
\begin{figure}[t]
\centering
\begin{minipage}{0.32\linewidth}
\centering
\includegraphics[width=\linewidth]{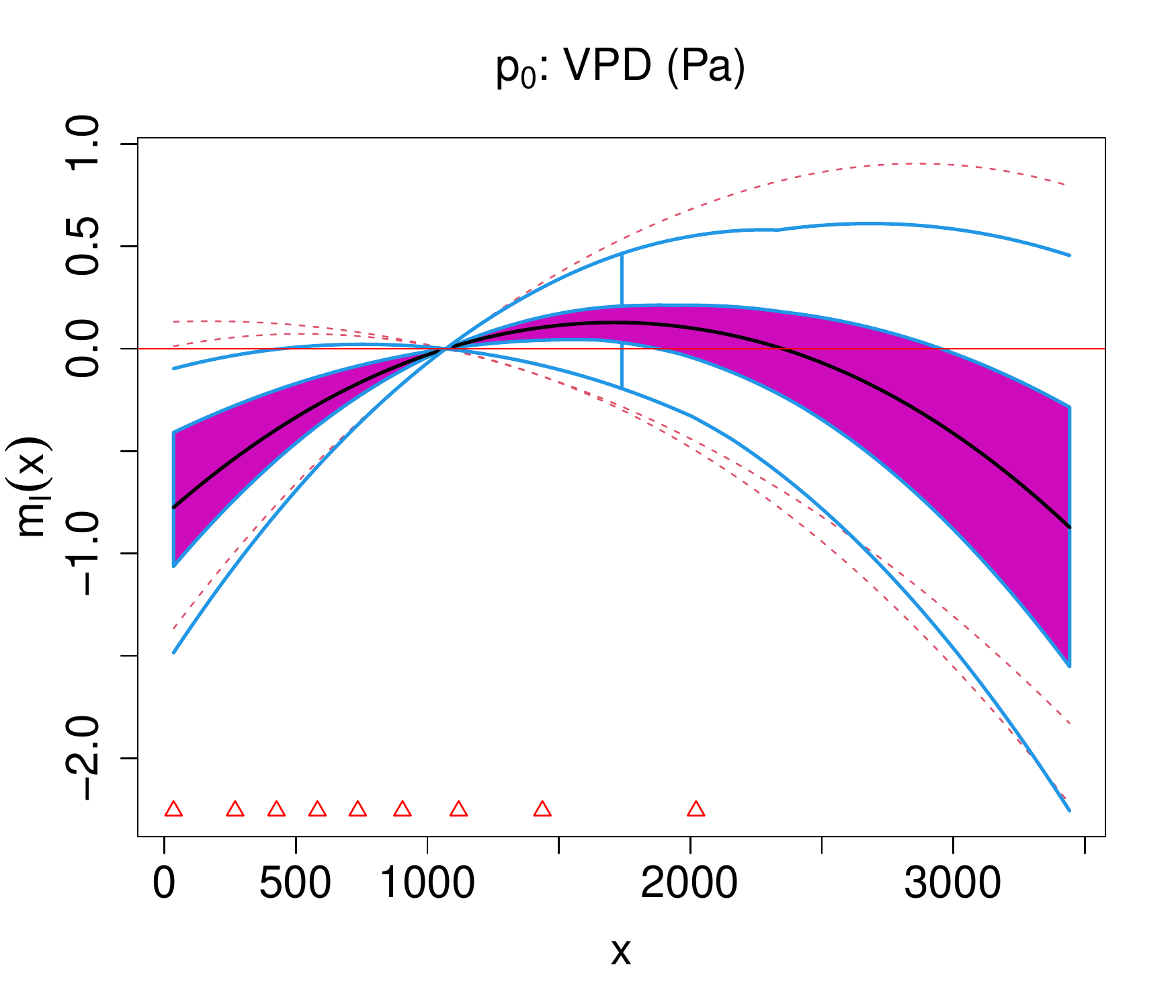} 
\end{minipage}
\begin{minipage}{0.32\linewidth}
\centering
\includegraphics[width=\linewidth]{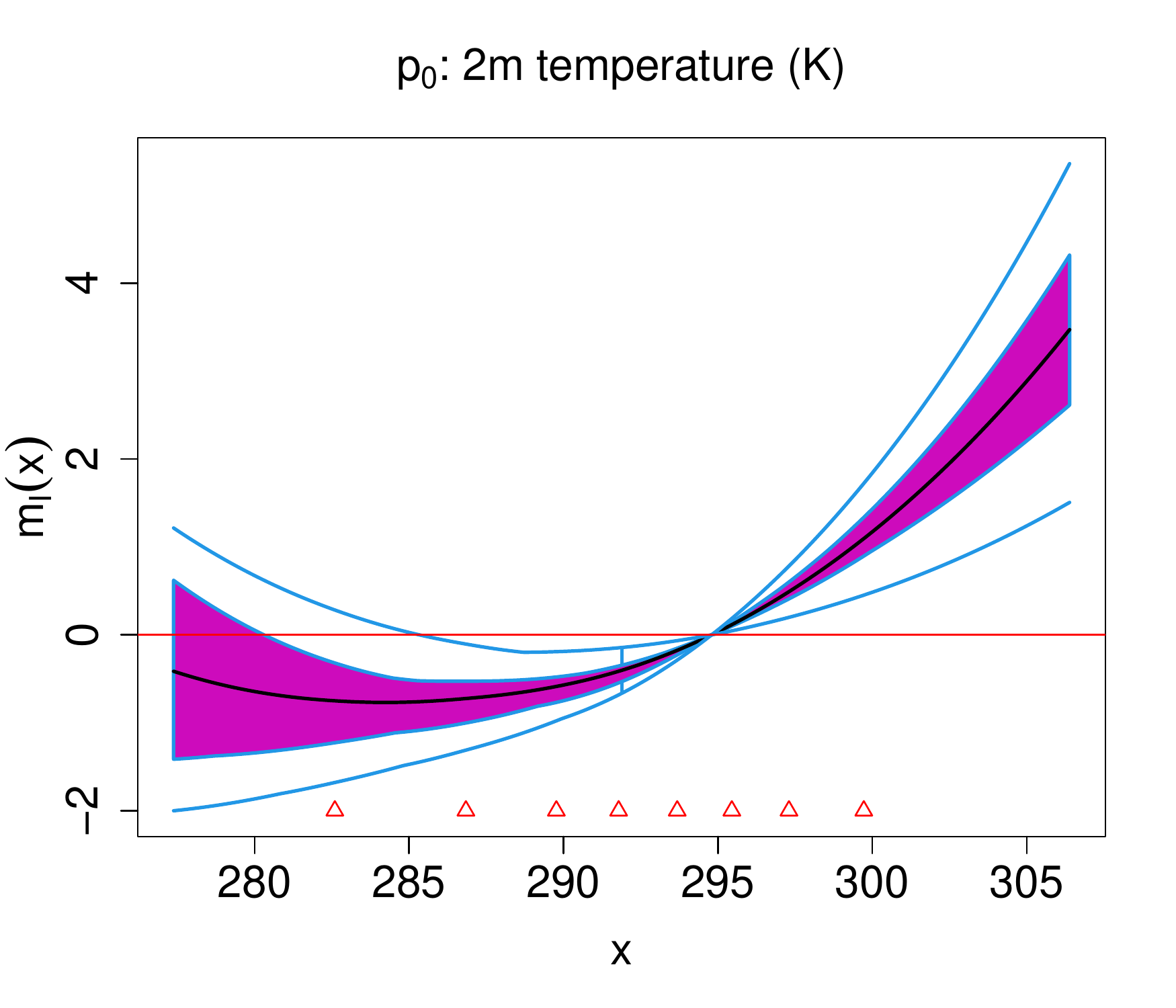} 
\end{minipage}
\begin{minipage}{0.32\linewidth}
\centering
\includegraphics[width=\linewidth]{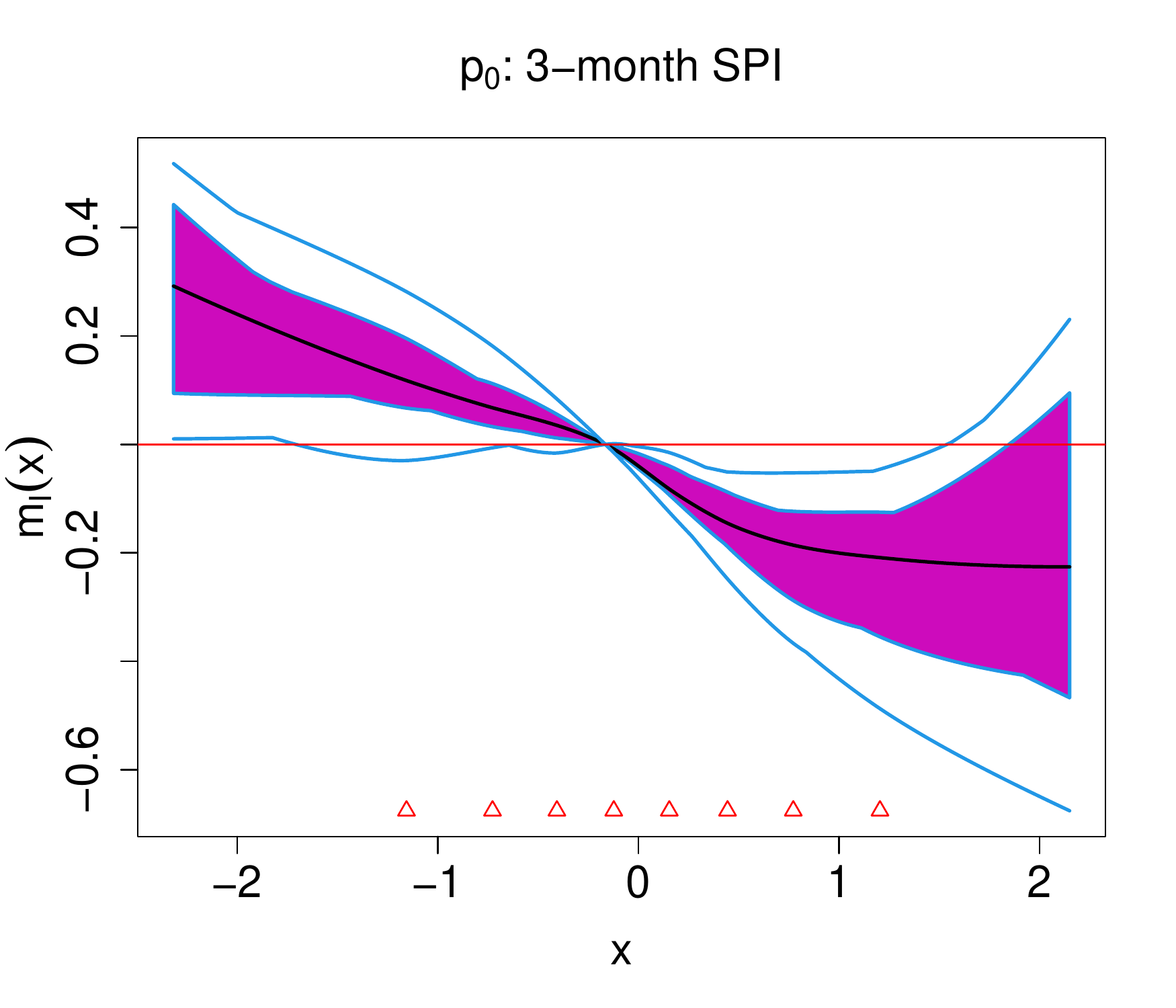} 
\end{minipage}
\begin{minipage}{0.32\linewidth}
\centering
\includegraphics[width=\linewidth]{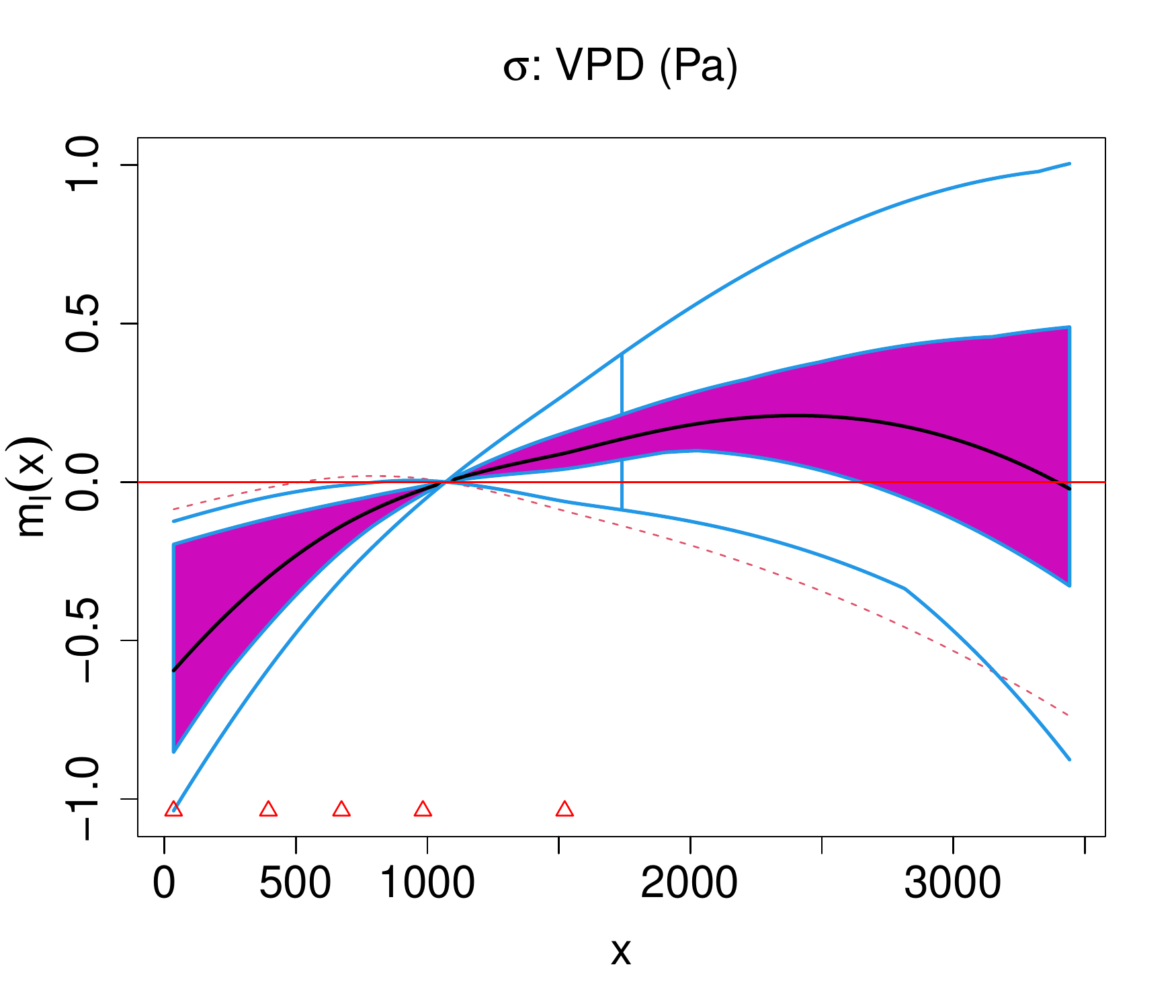} 
\end{minipage}
\begin{minipage}{0.32\linewidth}
\centering
\includegraphics[width=\linewidth]{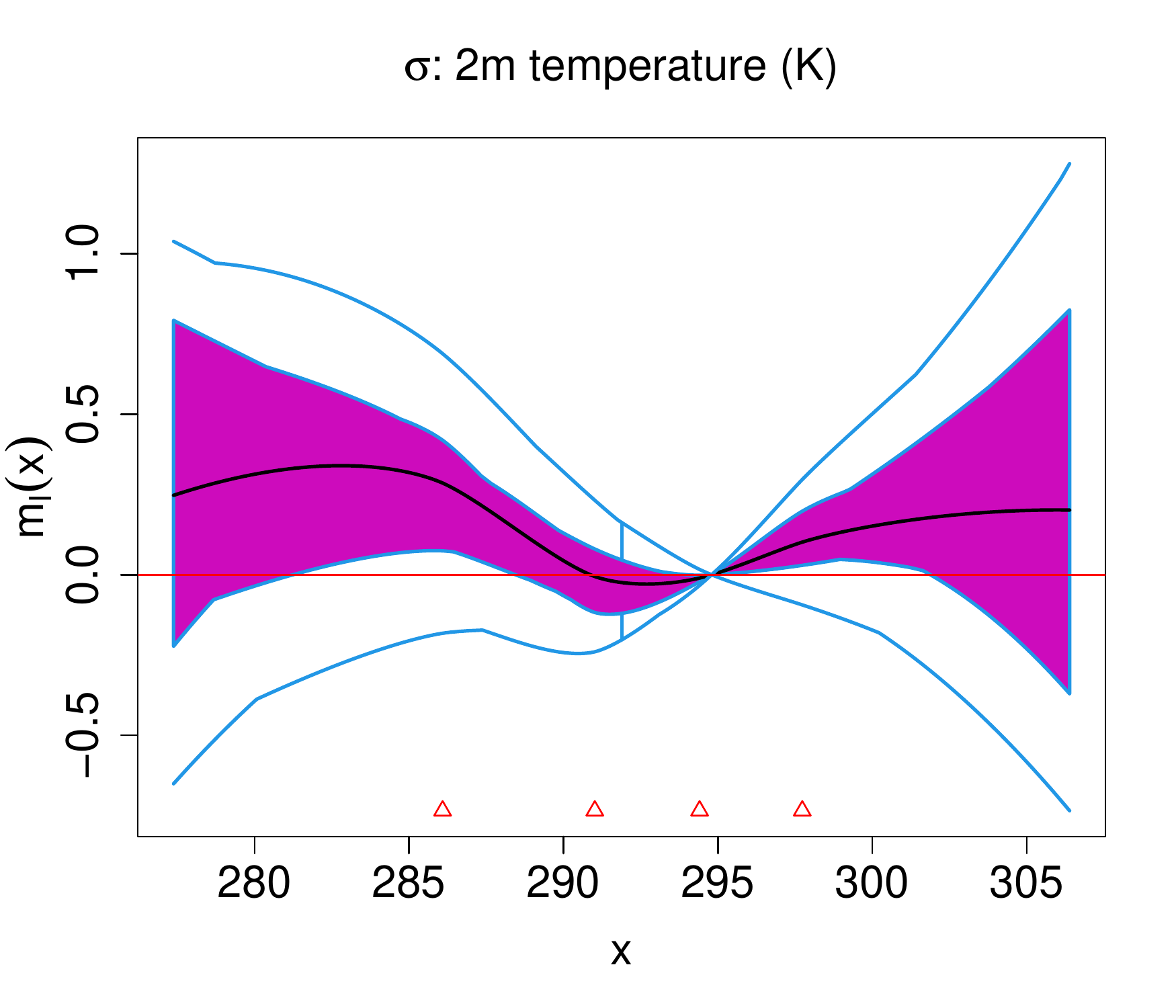} 
\end{minipage}
\begin{minipage}{0.32\linewidth}
\centering
\includegraphics[width=\linewidth]{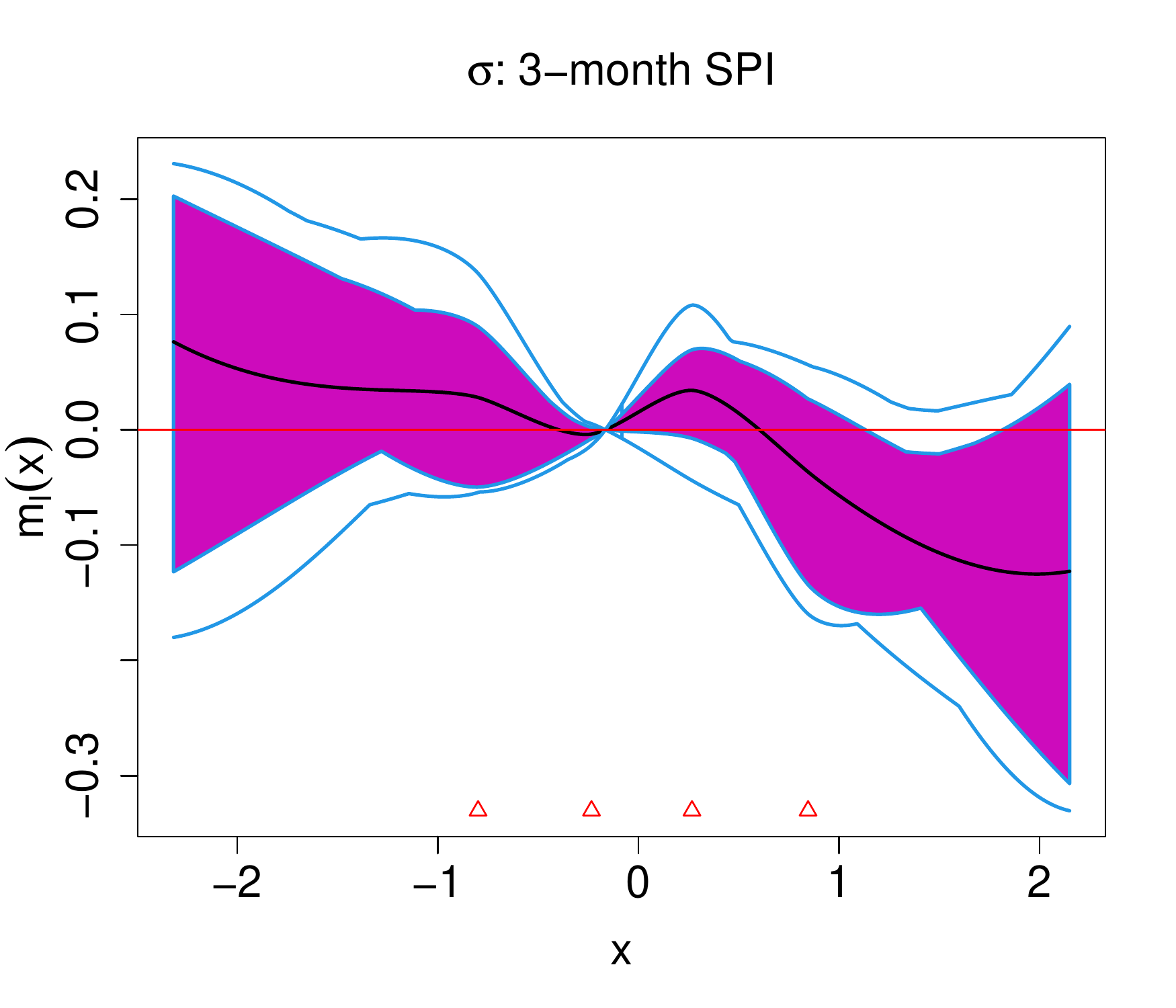} 
\end{minipage}
\caption{Functional box-plots of estimated additive function contributions $m_{\mathcal{I},j}(x), j=1,2,3$ (left to right) to the $\log$-odds of {fire occurrence probability} $p_0$ [top row] and {log conditional spread severity} $\log(\sigma)$ [bottom row]; an increase of one on the $y$-axis corresponds to an increase in one of these parameter functions. Blue curves denote maximum envelopes and the black curve gives the median function over all bootstrap samples. The red dashed curves represent outlier candidates and the $50\%$ central region is given in magenta. Red triangles denote the locations of knots and the red horizontal lines intersect $m_{\mathcal{I},j}(x)=0$.}
\label{occur_GAM}
\end{figure}
Figure~\ref{occur_GAM} gives estimates $\hat{m}_{\mathcal{I},j},j=1,2,3,$ for both the {occurrence probability} $p_0$ and {conditional spread severity} $\sigma$. As the link function $h$ in \eqref{PINNmodel} is taken to be the logistic and exponential functions for $p_0$ and $\sigma$, respectively, the scales on the $y$-axes in the top and bottom rows of Figure~\ref{occur_GAM} indicate the contribution of each predictor to the log-odds of $p_0$ and $\log\sigma$, respectively. We can roughly determine if there is a ``significant'' effect of a predictor on a parameter by observing whether the magenta regions in Figure~\ref{occur_GAM} encompass the horizontal line intersecting the origin. We observe a positive relationship between temperature and $p_0$, and the converse holds for the SPI drought index; this is in line with the fact that hotter and drier climates produce readily-ignitable fuel. The results for vapour-pressure deficit are less easy to interpret. We first note that an initial increase in VPD above zero leads to an increase in $p_0$. As VPD increases, the air gets relatively drier; hence we find agreement with the results for the SPI. However, as VPD increases above approximately 1500Pa, we begin to observe a reduction in $p_0$ with VPD. This may be due to a lack of availability of burnable fuel, as regions with high values of monthly VPD typically have arid climates with little vegetation. Overall, note that temperature has the strongest impact on $p_0$ relative to VPD and SPI (comparing the scales of the $y$-axes in Figure~\ref{occur_GAM}).
\par
We observe major differences in the spline results for $p_0$ and $\sigma$, suggesting that the interpreted predictors do not impact these two parameters in a similar fashion. The scale parameter $\sigma$ can be intuited as a measure of conditional wildfire spread severity; relatively large values of $\sigma$ suggest that, given the occurrence of a wildfire, the magnitude of extreme wildfire spread above the threshold $u$ is likely to be larger. Similar results are found for the effect of VPD on $\sigma$ as for VPD on {occurrence probability} $p_0$; an initial positive relationship between VPD and {conditional spread}  severity is observed as VPD increases above zero, but the reverse holds as VPD increases above approximately 2500Pa. Whilst there appears to be a significant effect of VPD on $\sigma$, the same does not seem to hold for 2m air temperature and 3-month SPI. For temperature, we observe that temperatures larger than the median do not lead to a significant increase in $\sigma$. At the lower end, we observe a small negative relationship between temperatures below the median and $\sigma$, with lower temperatures leading to an increase in conditional wildfire spread severity. This is an interesting result that may be caused by the differences in the distribution of land cover types in regions with lower monthly temperatures {(see Figure~\ref{obs_maps}f)}.\par
Figure~\ref{occur_GAM} suggests that an increase in SPI values greater than zero leads to a small but significant decrease in $\sigma$. Oddly, we find no significant effect for changes in SPI less than zero, which corresponds to space-time locations experiencing less three-monthly rainfall relative to the average conditions, that is, drought conditions; this seems counter-intuitive as we would expect drought conditions to facilitate extreme wildfire spread. The GPD models extreme wildfire spread conditioned on the prerequisite of a wildfire occurrence; this decomposition reveals that drought conditions may facilitate only wildfire occurrence. Once the drought conditions are such that a wildfire ignites, further changes do not impact {extremes of the distribution of the subsequent spread}. Moreover, the lack of a significant effect of air temperature and SPI on $\sigma$ may be evidence to suggest that much of the non-stationarity in the upper-tail of the spread distribution can be accommodated by the non-stationary threshold model $u(s,t)$; with the variability captured in $u(s,t)$, the parameter function $\sigma$, and hence the distribution of extreme spread, is almost stationary with respect to the predictors. {Note that these inferences are made for the upper-tails of the spread distribution only; consideration for non-extreme spread is outside of the scope of our analyses.}
\subsection{Risk assessment}
\label{risk_assess_sec}
\begin{figure}[t!]
\centering
\begin{minipage}{0.49\linewidth}
\centering
\includegraphics[width=\linewidth]{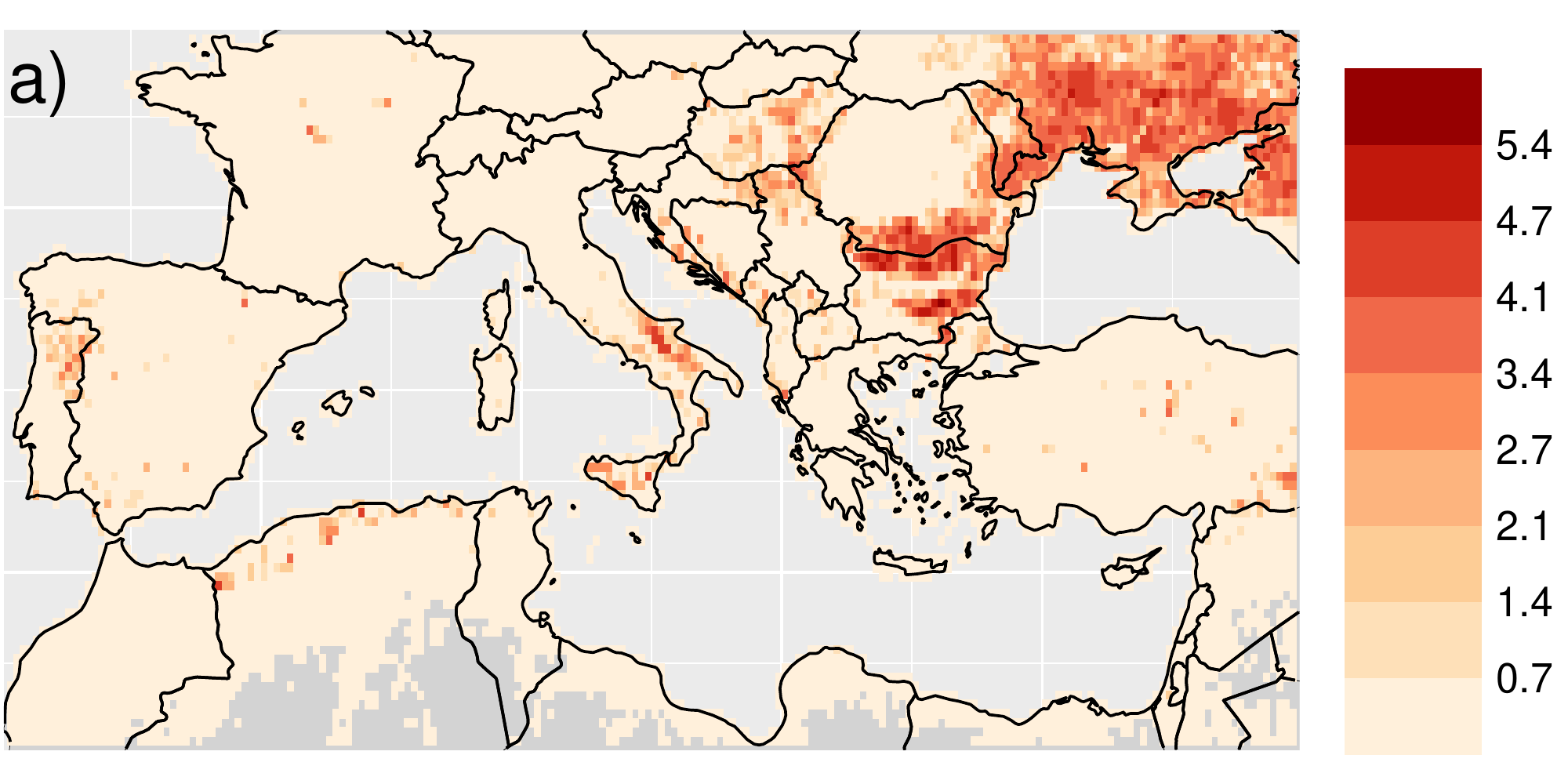} 
\end{minipage}
\begin{minipage}{0.49\linewidth}
\centering
\includegraphics[width=\linewidth]{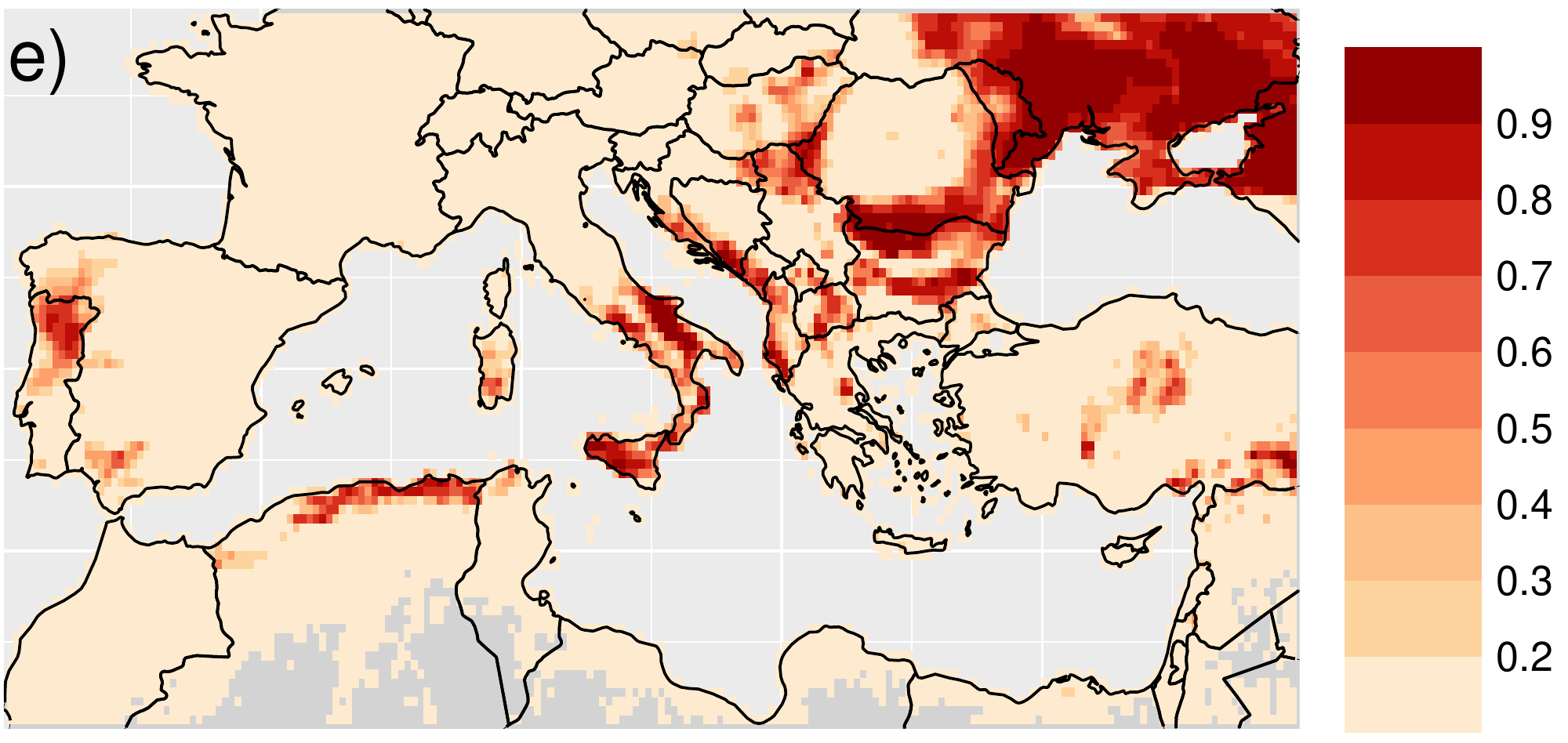} 
\end{minipage}
\begin{minipage}{0.49\linewidth}
\centering
\includegraphics[width=\linewidth]{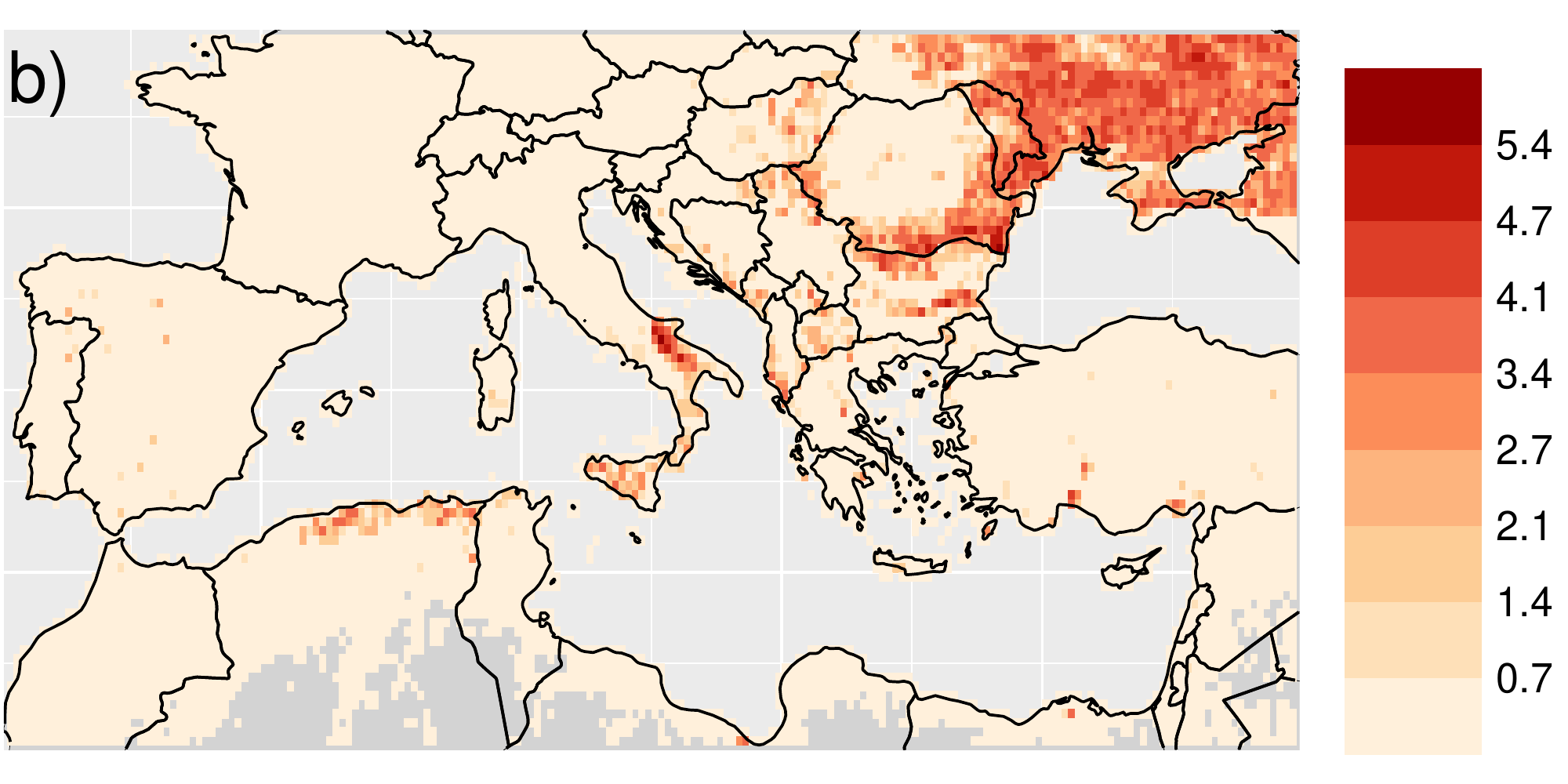} 
\end{minipage}
\begin{minipage}{0.49\linewidth}
\centering
\includegraphics[width=\linewidth]{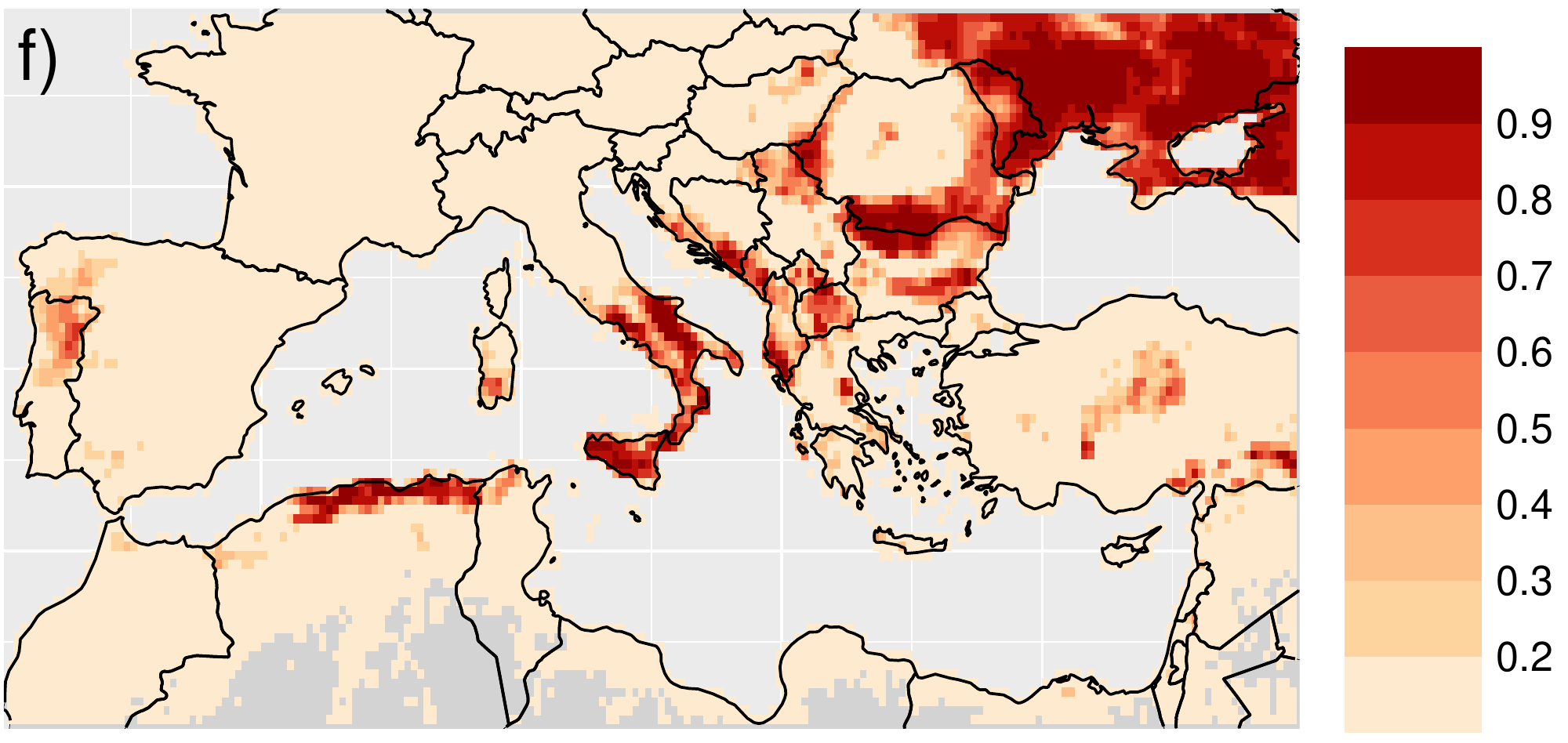} 
\end{minipage}
\begin{minipage}{0.49\linewidth}
\centering
\includegraphics[width=\linewidth]{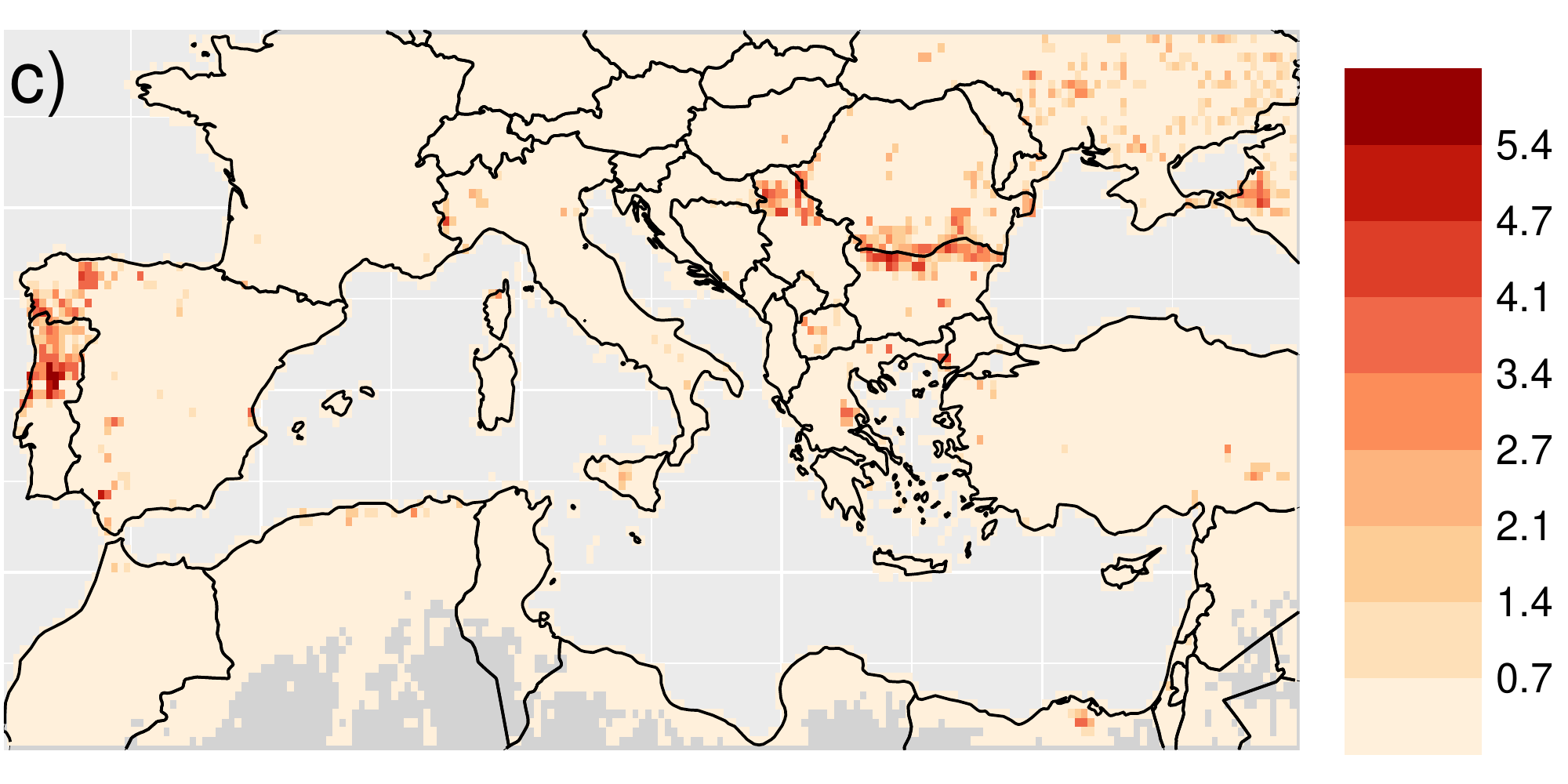} 
\end{minipage}
\begin{minipage}{0.49\linewidth}
\centering
\includegraphics[width=\linewidth]{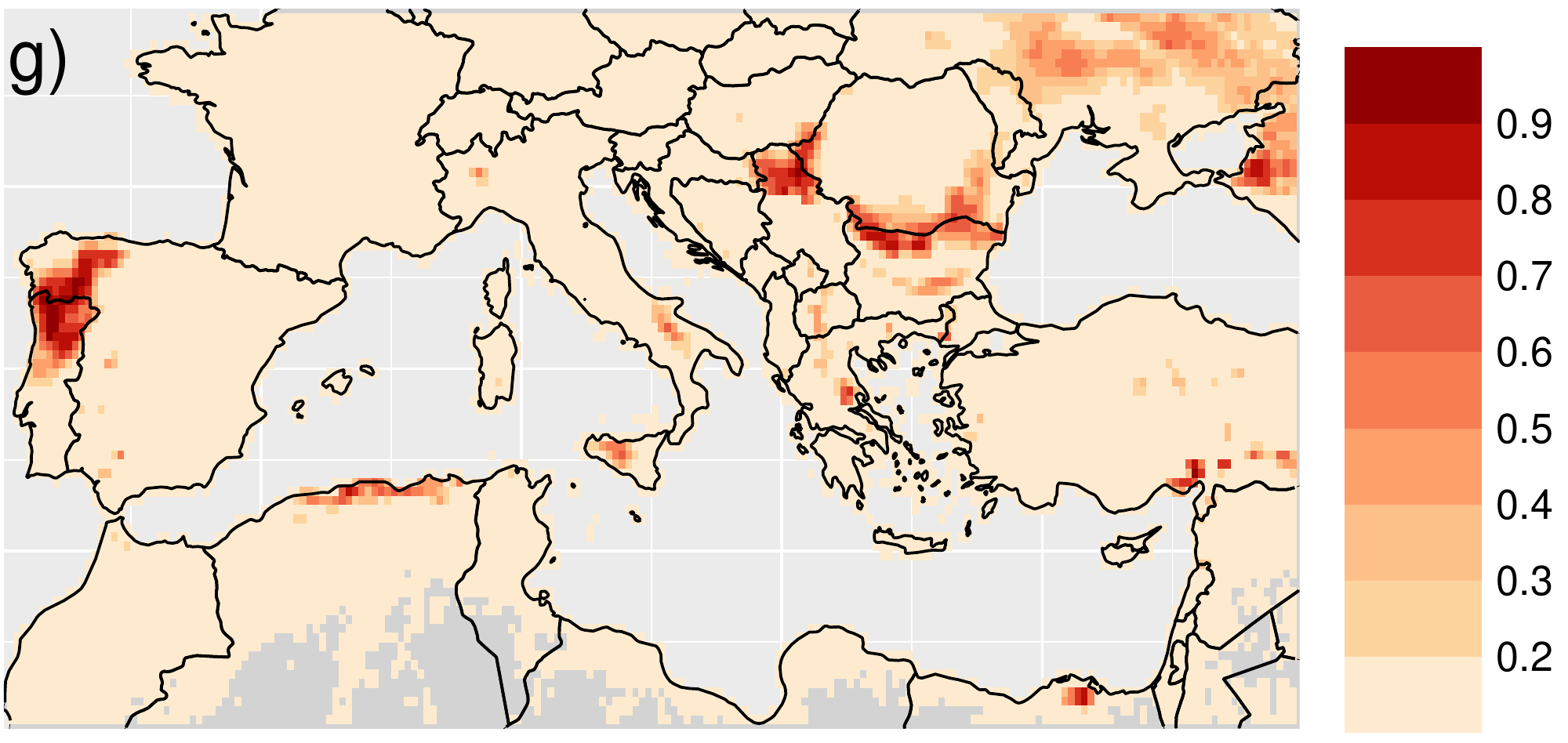} 
\end{minipage}
\begin{minipage}{0.49\linewidth}
\centering
\includegraphics[width=\linewidth]{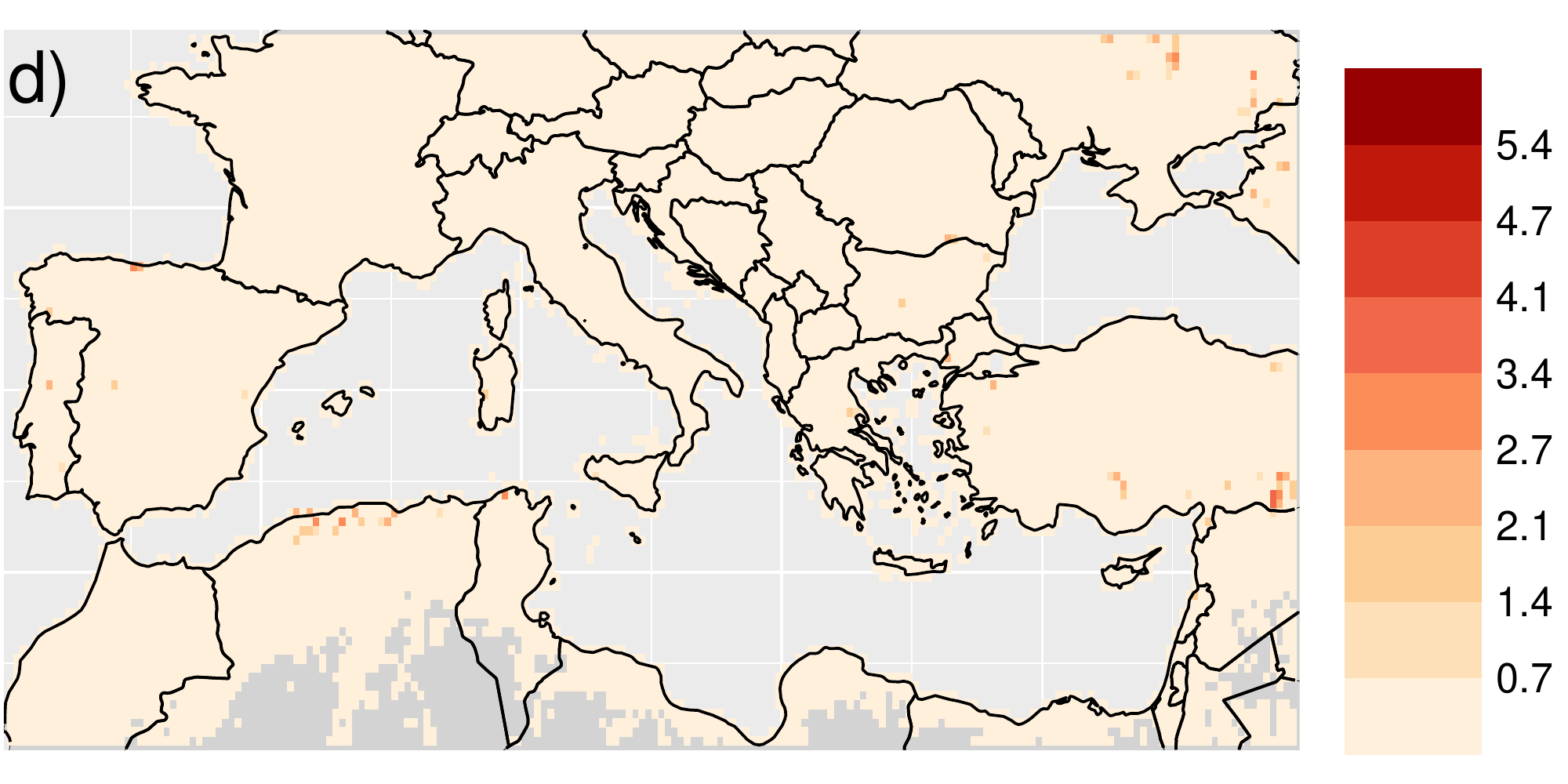} 
\end{minipage}
\begin{minipage}{0.49\linewidth}
\centering
\includegraphics[width=\linewidth]{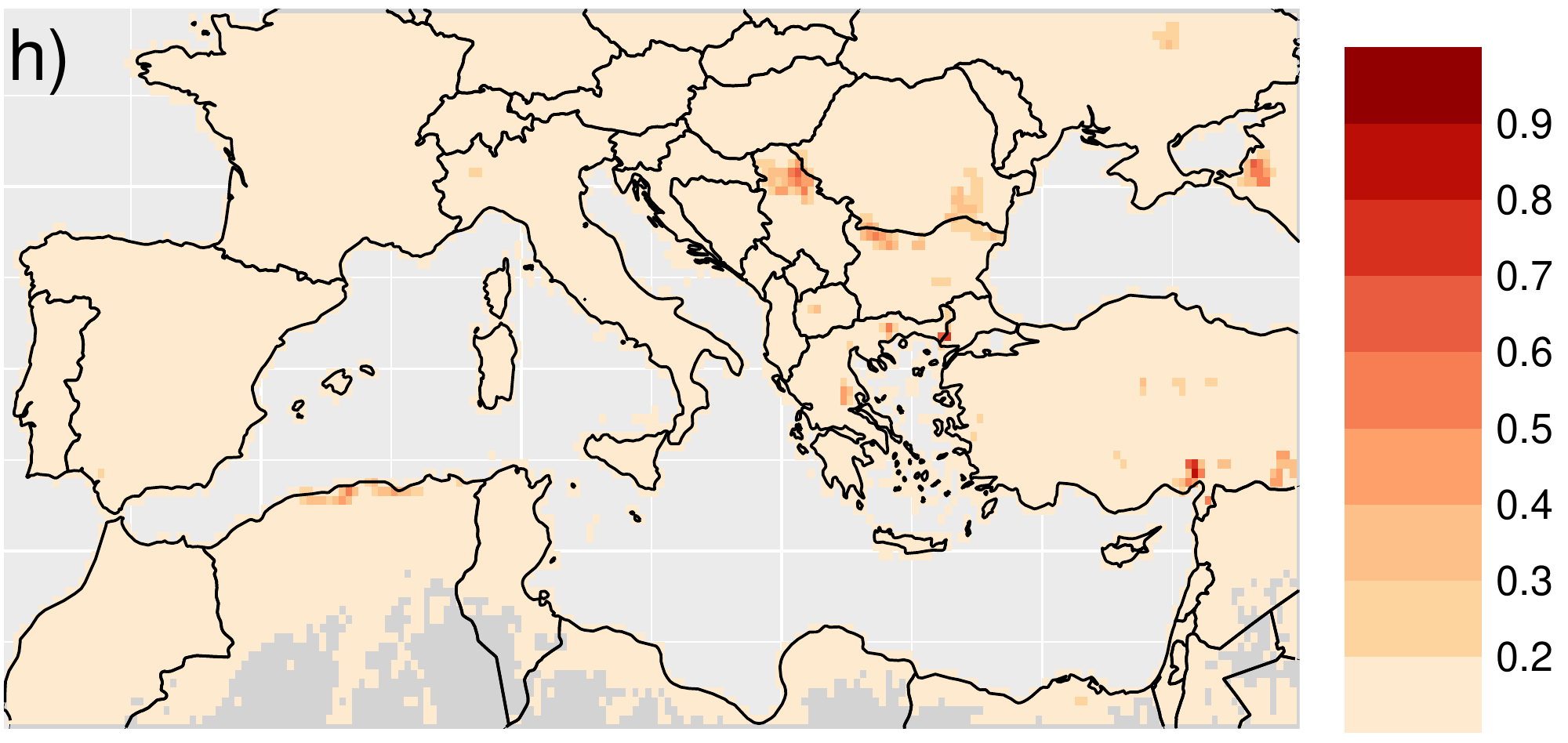} 
\end{minipage}
\caption{First row, a) observed $\log\{1+{Y}(s,t)\}$ [burnt area; $\log(\mbox{km}^2)$] and e) bootstrap median estimated {fire occurrence probability} $p_0(s,t)$ [unitless] for August 2001. The second, third, and fourth, rows are as the first row, but for August 2008, October 2017, and  November 2020, respectively. }
\label{result_maps}
\end{figure}
Figure~\ref{result_maps} provides maps of observations of $\log\{1+Y(s,t)\}$ (burnt area) and the modelled probability of fire occurrence, that is, the median estimated $p_0(s,t)$, for fixed $t$ corresponding to chosen months. Here we consider four months: August 2001, August 2008, October 2016 and November 2020. The first two months are considered as across the observation period these were the most devastating in terms of the total burnt area across the entirety of the spatial domain; we also consider October 2016 as this month observed the largest recorded value of BA\footnote{The largest observed value of BA was $416.9km^2$, occurring within the Lous\~a and Oliveira do Hospital municipalities of Portugal \citep{ribeiro2020extreme}.} and November 2020, which had the lowest total wildfire spread across all months. \par
For each bootstrap model fit, we evaluate the AUC for the fitted {occurrence probability} models on both the entire original data and the test data for the bootstrap sample, that is, an out-of-sample estimate. The median values (2.5$\%$- and 97.5$\%$-quantiles) across all bootstrap samples are 0.947 (0.944, 0.949) and 0.948 (0.940, 0.955), for the original and test data, respectively, suggesting that the chosen architecture provides an excellent predictor for the occurrence of wildfires. These values, alongside the maps of estimated $p_0$ given in Figure~\ref{result_maps}, suggest that the model predicts well the probability of wildfire occurrence; we find agreement in the predicted $p_0$ values and the observations of occurrence, for the period 2001--2020. For the chosen months, notable areas of concern include large portions of eastern Europe, including Ukraine, Romania, Bulgaria and Serbia, as well as the north of Portugal and the Galicia region of Spain, as these regularly experience large estimates of $p_0$; we also observe high probabilities of wildfire occurrence in northern Algeria, Turkey and Italy.      \par
To assess the fit of conditional model \eqref{MargTransform} for {extremes of wildfire spread}, we provide a pooled quantile-quantile (Q-Q) plot \citep{heffernan2001extreme} in Figure~\ref{fig_exp_fit} of Appendix~B; the estimated models are used to transform all original observations of {non-zero spread} $Y(s,t)\;|\;Y(s,t)>0$ onto standard Exponential margins, and we then compare theoretical quantiles against empirical ones derived using the fitted model; this procedure is repeated for each bootstrap sample and we observe good fits, particularly in the upper-tails, as the estimated $95\%$ tolerance bands include the diagonal. Despite the physical constraint {that the burnt area must subcede the total burnable area for a gridbox}, that is, $Y(s,t) \leq \lambda(s,t)$, the distribution of Mediterranean wildfire spread is well-approximated by a heavy-tailed distribution, with the median of the shape parameter estimates ($2.5\%$ and $97.5\%$ quantiles) being estimated as $0.322\; (0.280, 0.353)$. \cite{richards2022} find similar values for the shape parameter estimates of burnt area due to U.S. wildfires; however, they model the square-root of the response, rather than its unadulterated counterpart, which suggests that wildfire spread due to Mediterranean wildfires is considerably lighter-tailed than those occurring in the U.S.
\par
\begin{figure}[t!]
\centering
\begin{minipage}{0.49\linewidth}
\centering
\includegraphics[width=\linewidth]{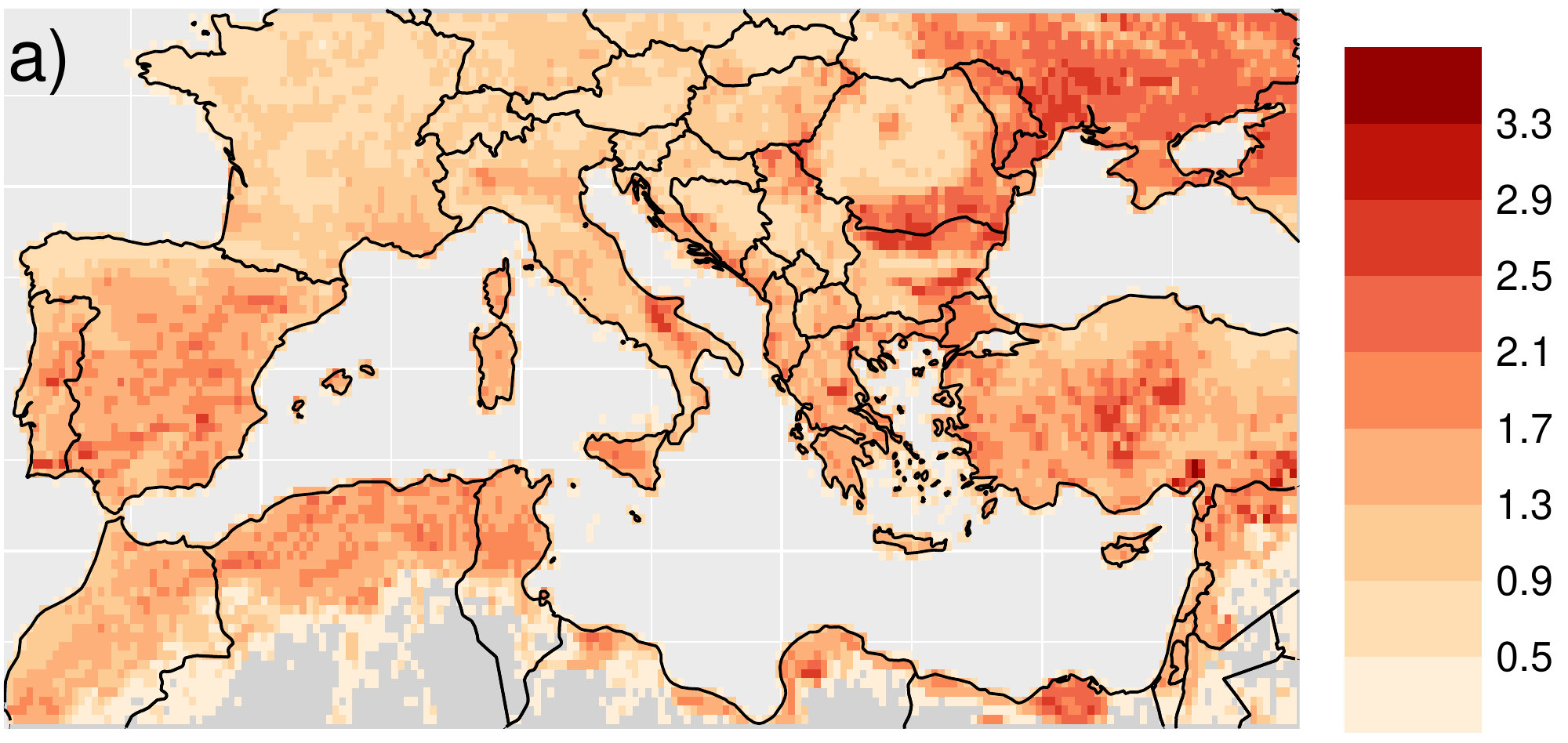} 
\end{minipage}
\begin{minipage}{0.49\linewidth}
\centering
\includegraphics[width=\linewidth]{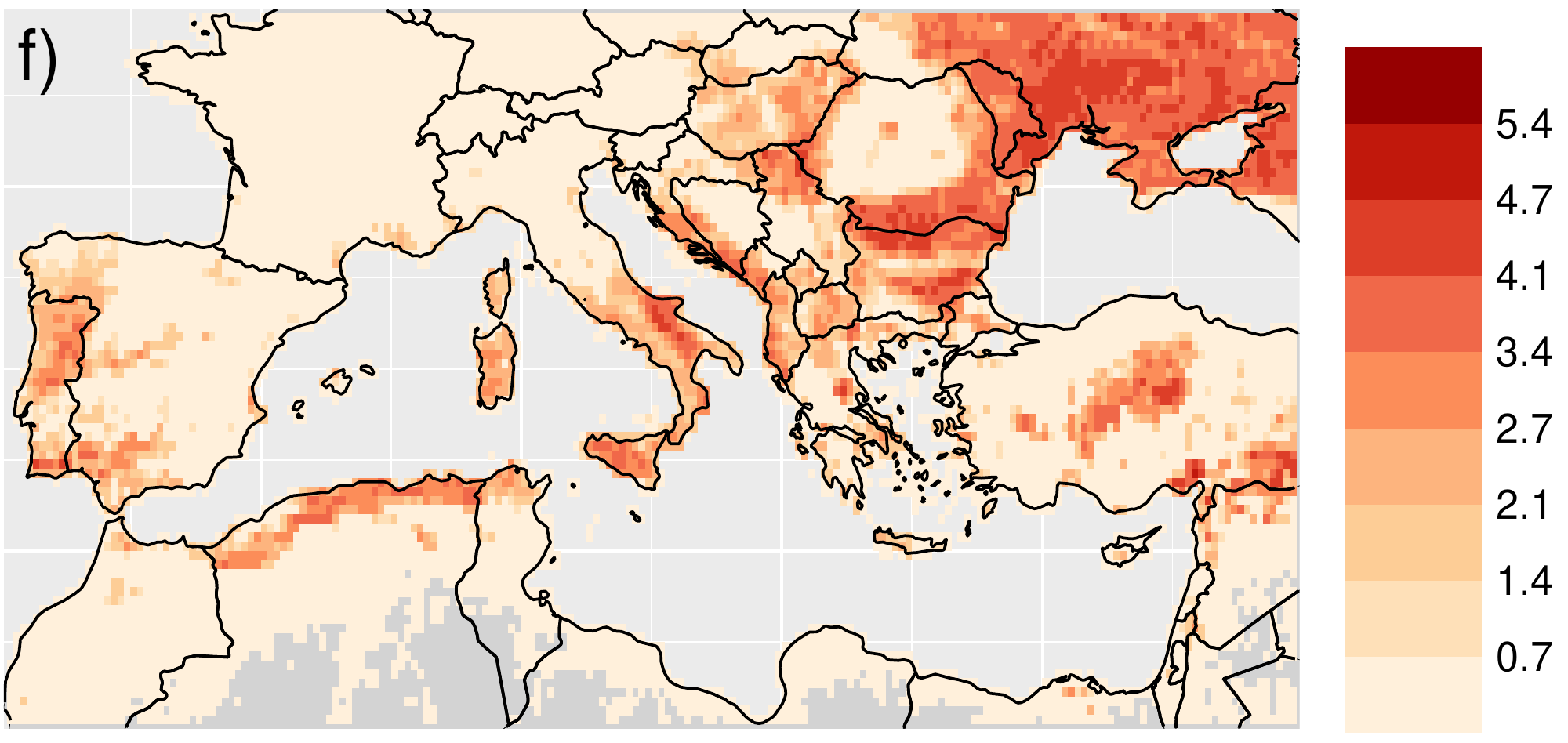} 
\end{minipage}
\begin{minipage}{0.49\linewidth}
\centering
\includegraphics[width=\linewidth]{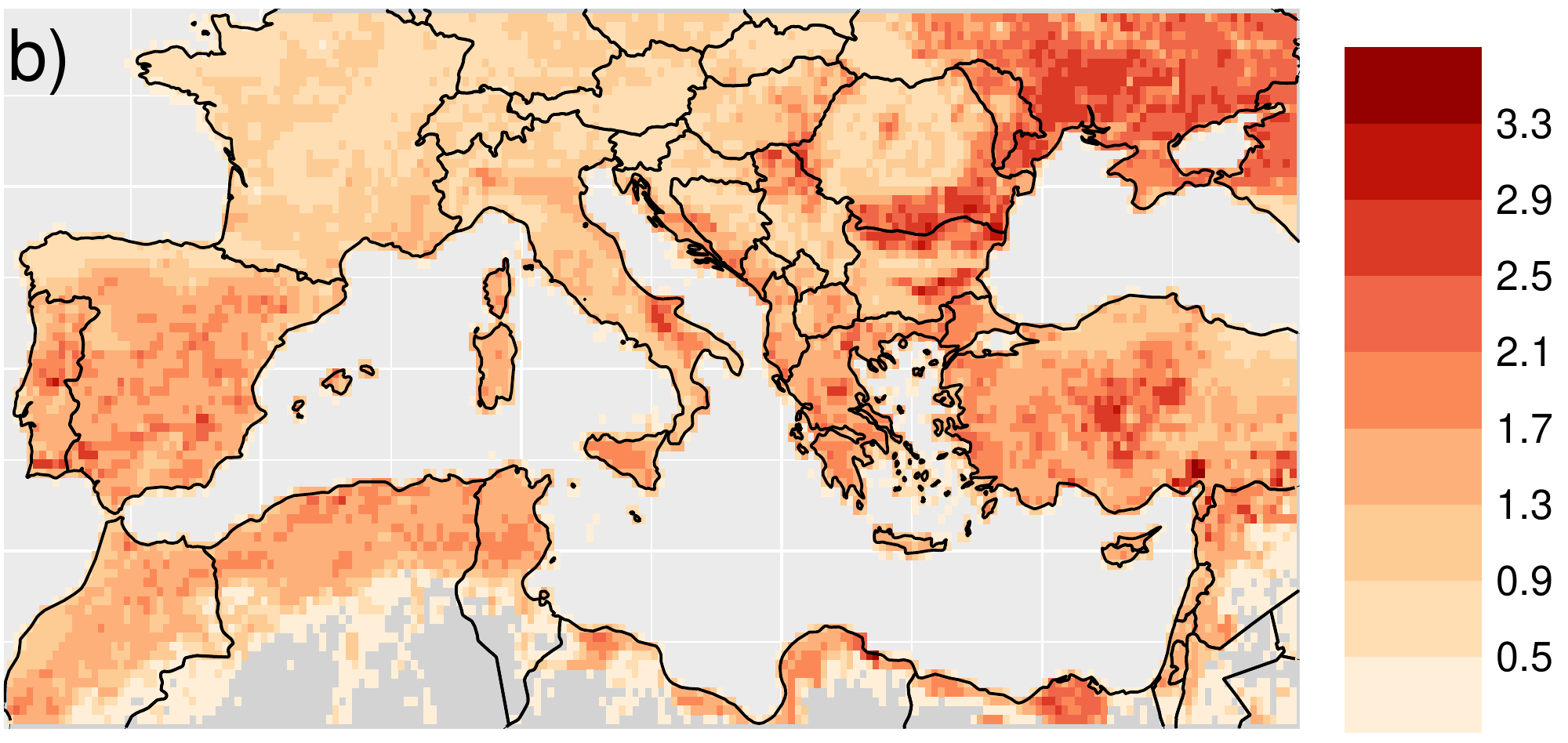} 
\end{minipage}
\begin{minipage}{0.49\linewidth}
\centering
\includegraphics[width=\linewidth]{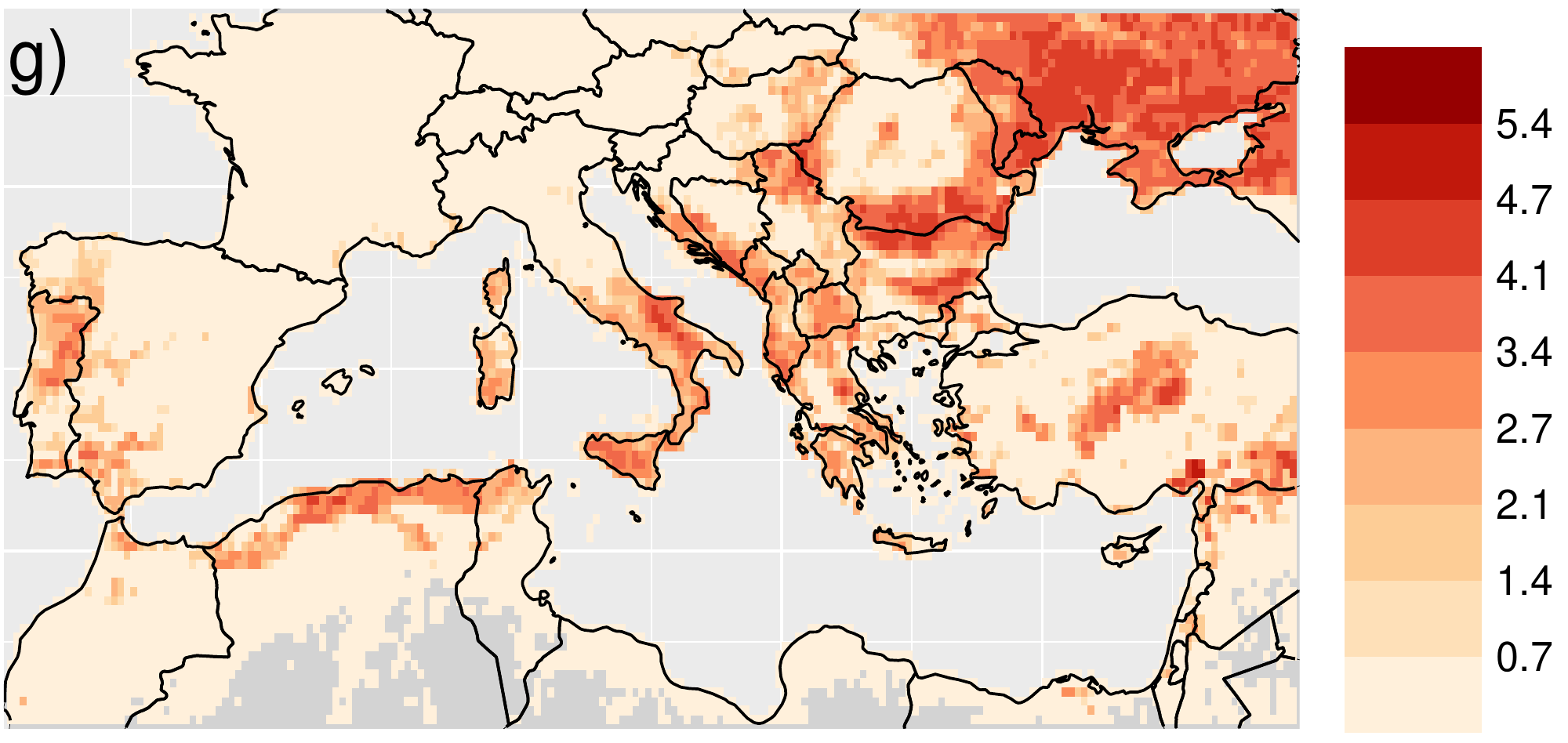} 
\end{minipage}
\begin{minipage}{0.49\linewidth}
\centering
\includegraphics[width=\linewidth]{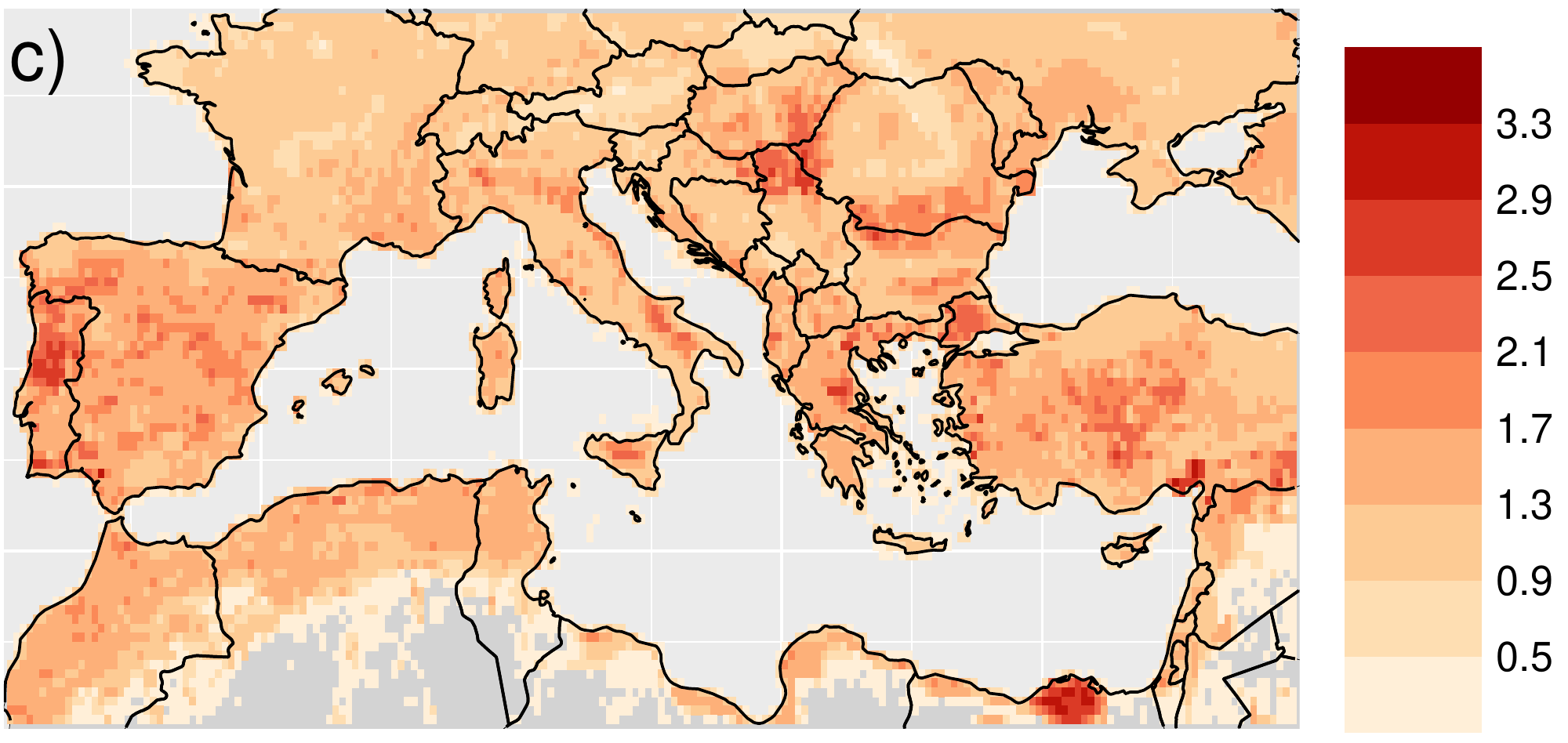} 
\end{minipage}
\begin{minipage}{0.49\linewidth}
\centering
\includegraphics[width=\linewidth]{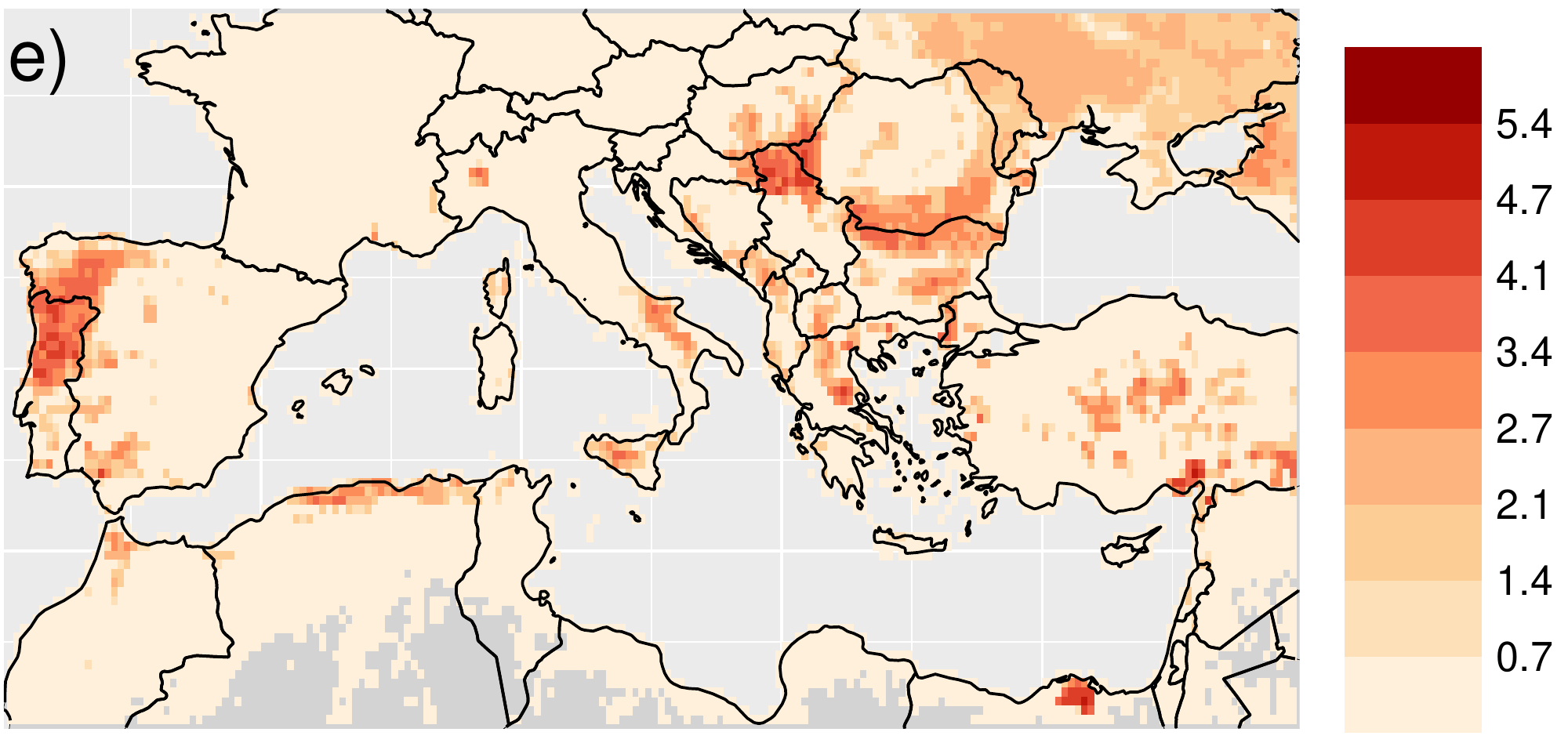} 
\end{minipage}
\begin{minipage}{0.49\linewidth}
\centering
\includegraphics[width=\linewidth]{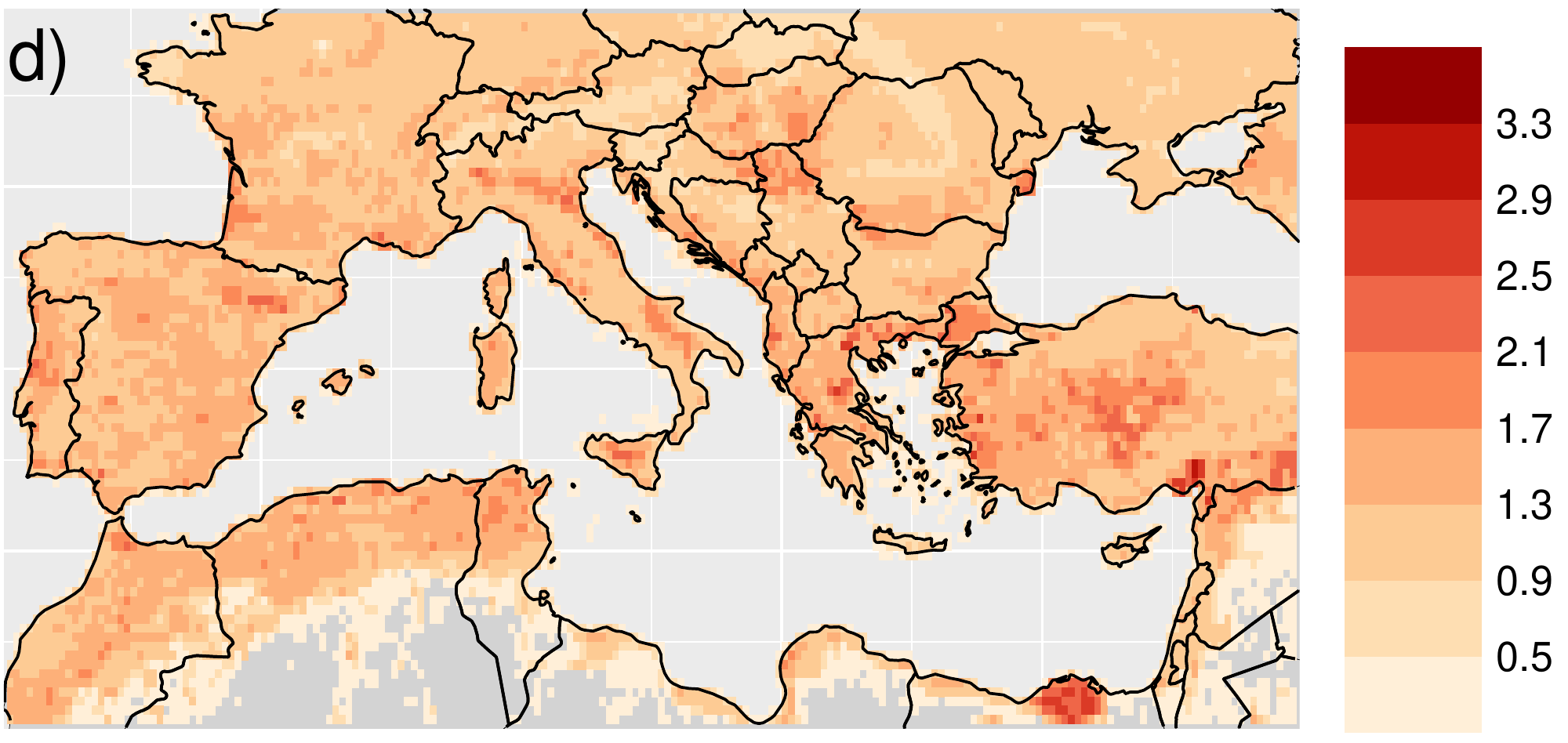} 
\end{minipage}
\begin{minipage}{0.49\linewidth}
\centering
\includegraphics[width=\linewidth]{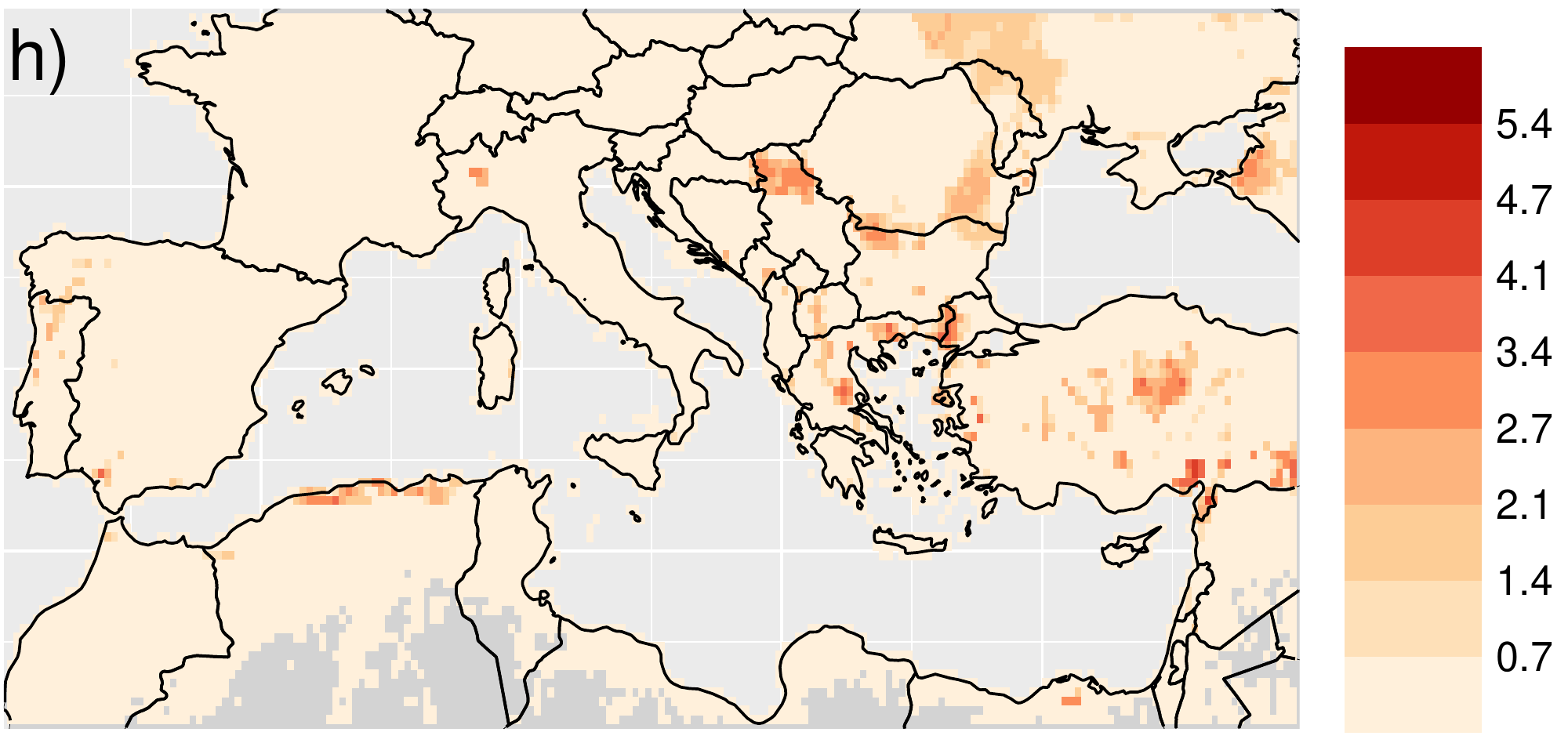} 
\end{minipage}
\caption{First row, bootstrap median estimated a) $\log\{1+\sigma(s,t)\}$ [{conditional spread}  severity; $\log(\mbox{km}^2)$] and e) $95\%$-quantile of $\log\{1+Y(s,t)\}\;|\;\mathbf{X}(s,t)$ [burnt area; $\log(\mbox{km}^2)$] for August 2001. The second, third, and fourth, rows are as the first row, but for August 2008, October 2017, and  November 2020, respectively. }
\label{result_maps2}
\end{figure}
Figure~\ref{result_maps2} provides maps of the bootstrap median estimated $\log\{1+\sigma(s,t)\}$ ({conditional spread} severity) and the bootstrap median of estimated $95\%$-quantiles of $\log\{1+Y(s,t)\}\;|\;\mathbf{X}(s,t)$ (burnt area), for the four considered months; we note that the latter metric concerns quantiles of all values of burnt area, not just strictly positive values (i.e., spread), and so can be considered as a measure of compound risk which combines both the probability of wildfire occurrence and the spread distribution. Through Figures~\ref{result_maps} and \ref{result_maps2}, we observe that there is not necessarily a one-to-one correspondence between wildfire occurrence probability, 
 {conditional spread}  severity and compound risk, that is, locations with high $p_0$ also have high $\sigma$ and high $95\%$-quantile estimates. Notable regions include the northern parts of Africa, particularly the Nile Delta in Egypt, as well as parts of Spain; here we see that the climate conditions and fuel type suggest a particularly high wildfire spread severity across the months, but as $p_0$ remains low in these locations, they exhibit relatively low compound risk.  We quantify uncertainty for the maps in Figures~\ref{result_maps} and \ref{result_maps2} via the $2.5\%$ and $97.5\%$  bootstrap quantiles of occurence probability $p_0(s,t)$, {conditional spread}  severity $\sigma(s,t)$ and the estimated $95\%$-quantile of $\log\{1+Y(s,t)\}\;|\;\mathbf{X}(s,t)$ (burnt area); these are provided in Figures~\ref{p0_uncertain}, \ref{sig_uncertain} and \ref{cr_uncertain}, respectively, of Appendix~B. 

\subsection{Impacts of long-term climate trends}
\label{climate_change_sec}
To gain insights into the impact of climate trends on extreme wildfire events, we focus on the month of August 2001, which was chosen as it exhibits the highest total burnt area throughout the observation period. We estimate how the distribution of wildfires in August 2001 may have looked like under observed changes in the interpreted predictors during 2001--2020 \citep{maraun2022severe}. Given that we ultimately aim at gaining insights into how climate-change-driven trends in interpretable predictors will affect wildfires, here we considered observed trends in temperature and VPD only. This is because observed temperature and VPD trends are already generally in line with trends expected from anthropogenic climate change \citep{hawkins2012time}, while observed SPI trends are yet largely dominated by internal climate variability. In other words, climate change trends in SPI are expected to emerge from the noise of internal climate variability in the future \citep{maraun2013will,Zappa2021}. Overall, considering trends in VPD and temperature in August, which are critical drivers of wildfire changes, allows for gaining insights into the ongoing impact of climate-change-driven trends on fire activity. {However, we note that changes in wintertime and spring conditions, as well as snowpack changes, will further shape wildfire activity changes.} \par
In practice, we investigate the impact of changing temperature and VPD on the wildfire distribution estimates \citep{bevacqua2020more}.  {For each bootstrap sample, we estimate the model using input predictors $\mathbf{x}$ determined by one of two scenarios: scenario i), we use the observed conditions in August 2001 (as in Figures~\ref{result_maps} and \ref{result_maps2}); and scenario ii), the value of VPD at each site is perturbed by adding, to the observed values in August 2001, the long-term trends in August values of VPD calculated over the period 2001 to 2020. We then derive the site-wise differences in estimates of two distribution-related metrics that arise under these two scenarios; we investigate changes in $p_0(s,t)$, which describes the occurrence probability of wildfires, and $q^+_{0.9}(s,t)$, the $90\%$ quantile of non-zero {spread} $Y(s,t)\;|\;\{\mathbf{X}(s,t),Y(s,t)>0\}$, which is a measure of intensity, conditional on fire occurrence, derived using parameters\footnote{To derive burnt area quantiles, we also use values of the parameter $p_u$; we keep this fixed with respect to changes in VPD and 2m air temperature.} $u$ and $\sigma$. We consider $q^+_{0.9}$, rather than quantiles of unconditional burnt area (i.e., only $Y$), as changes in $q^+_{0.9}$ are driven by changes in $u$ and $\sigma$ only. Changes in $p_0$ have no impact on $q^+_{0.9}$ and so we are able to disentangle the effects of long-term climate trends on the frequency, and severity, of Mediterranean wildfires.  Note that in scenario ii) above, only values of VPD are changed, whilst the other $d-1$ predictor variables remain the same; in this way, estimates from scenario ii) give an indication as to how the distribution of wildfires in August 2001 may have looked like with VPD conditions having experienced the 20-year observed trend. This procedure is repeated with 2m air temperature replacing VPD as the chosen perturbed predictor. }
\par
\begin{figure}[h!]
\begin{minipage}{0.49\linewidth}
\centering
\includegraphics[width=\linewidth]{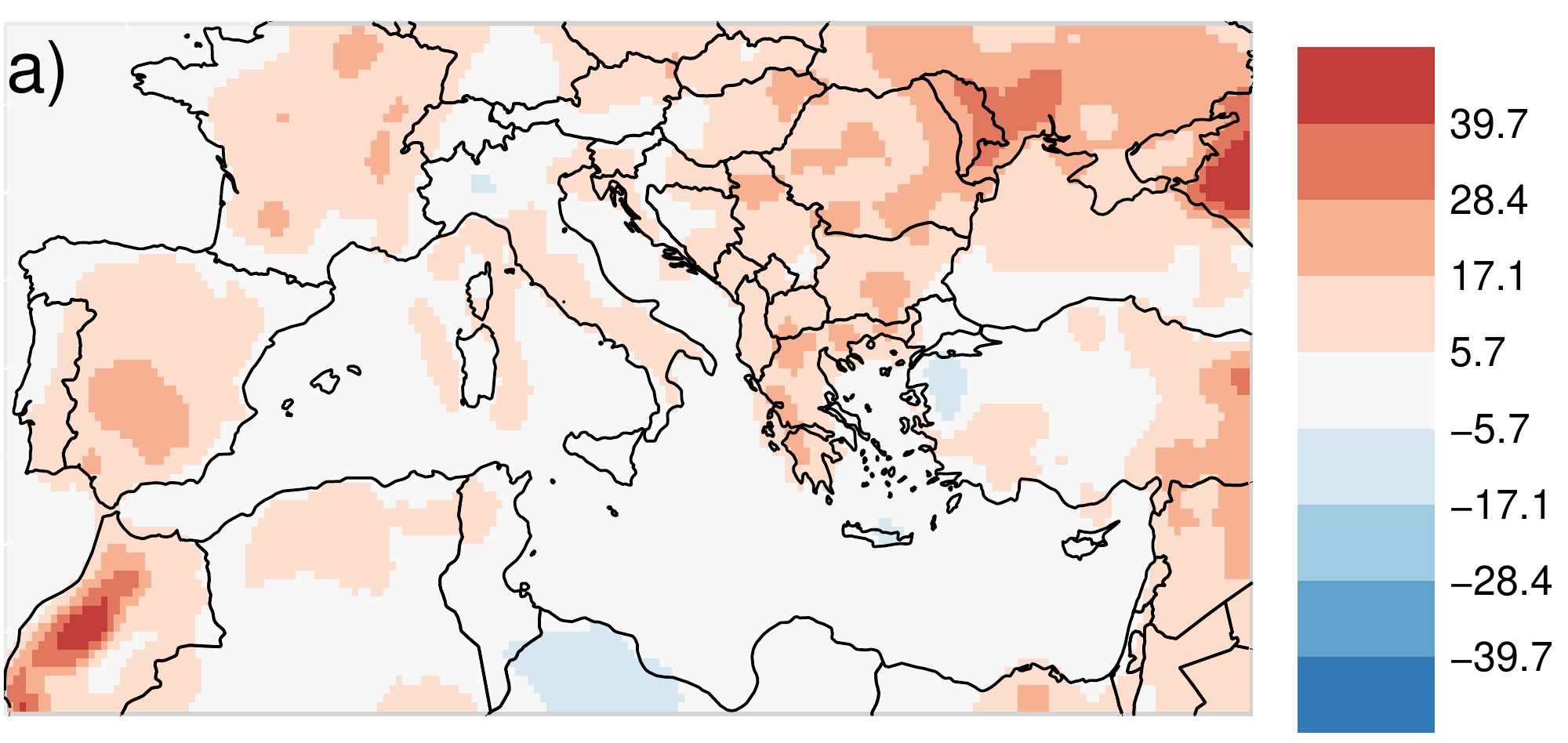} 
\end{minipage}
\begin{minipage}{0.49\linewidth}
\centering
\includegraphics[width=\linewidth]{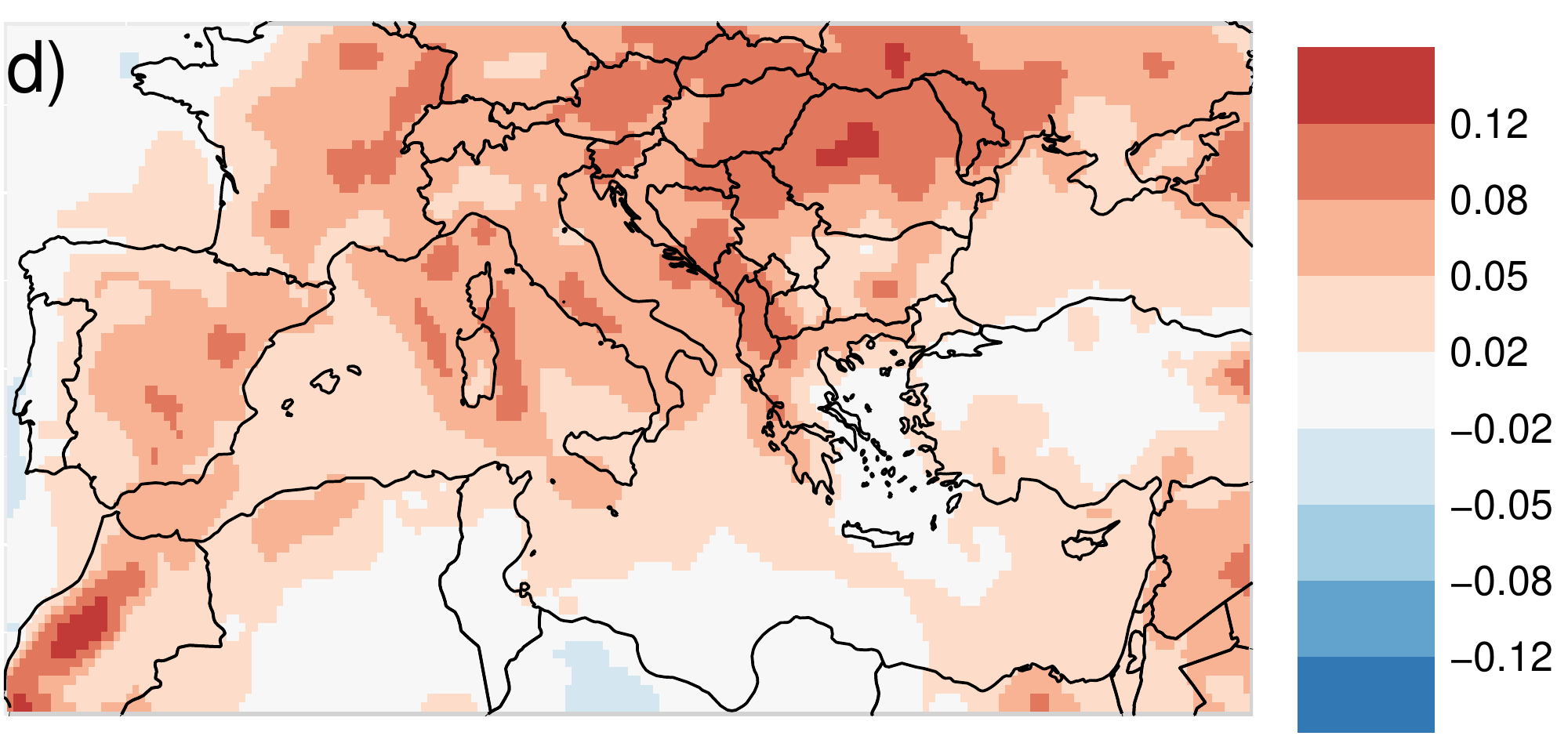} 
\end{minipage}
\begin{minipage}{0.49\linewidth}
\centering
\includegraphics[width=\linewidth]{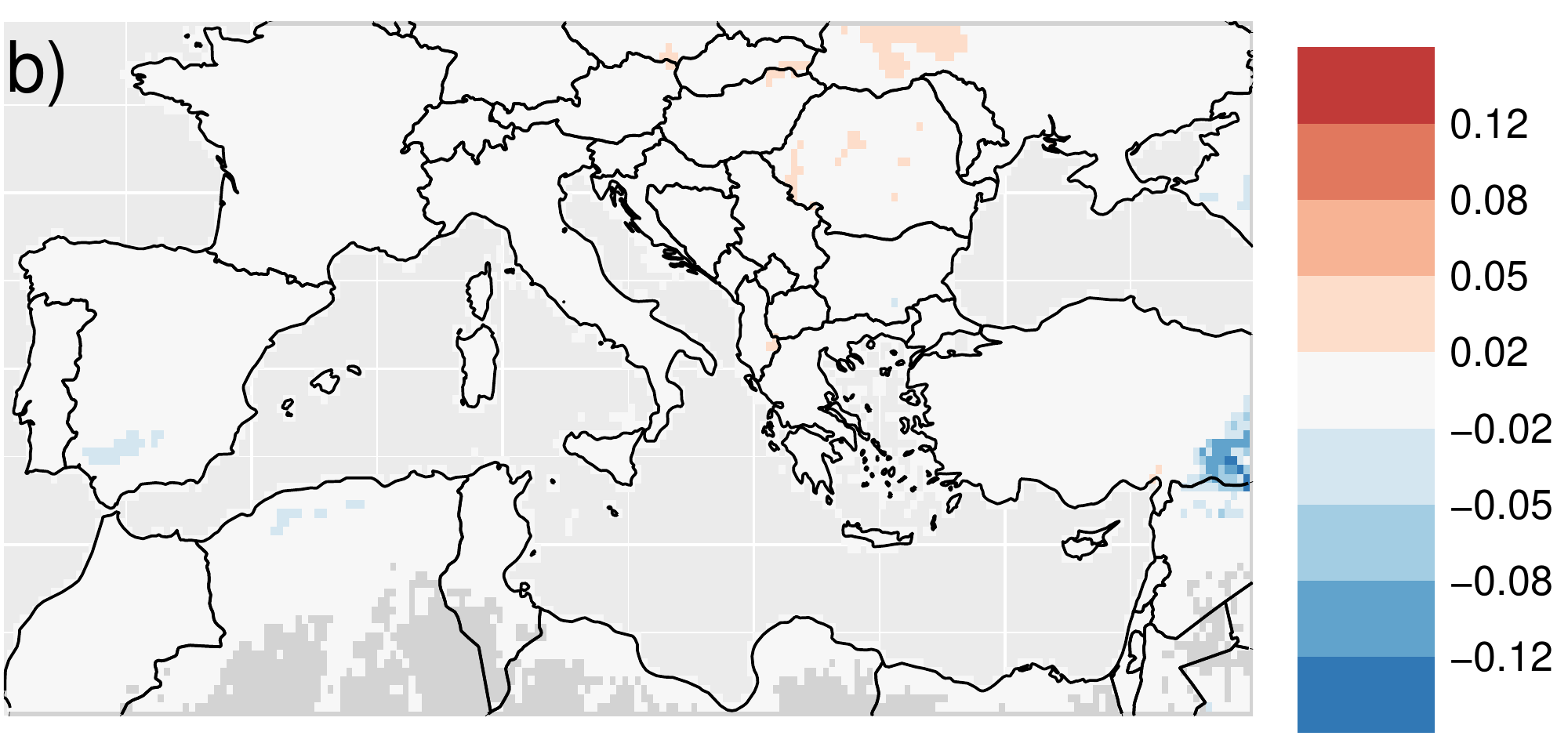} 
\end{minipage}
\centering
\begin{minipage}{0.49\linewidth}
\centering
\includegraphics[width=\linewidth]{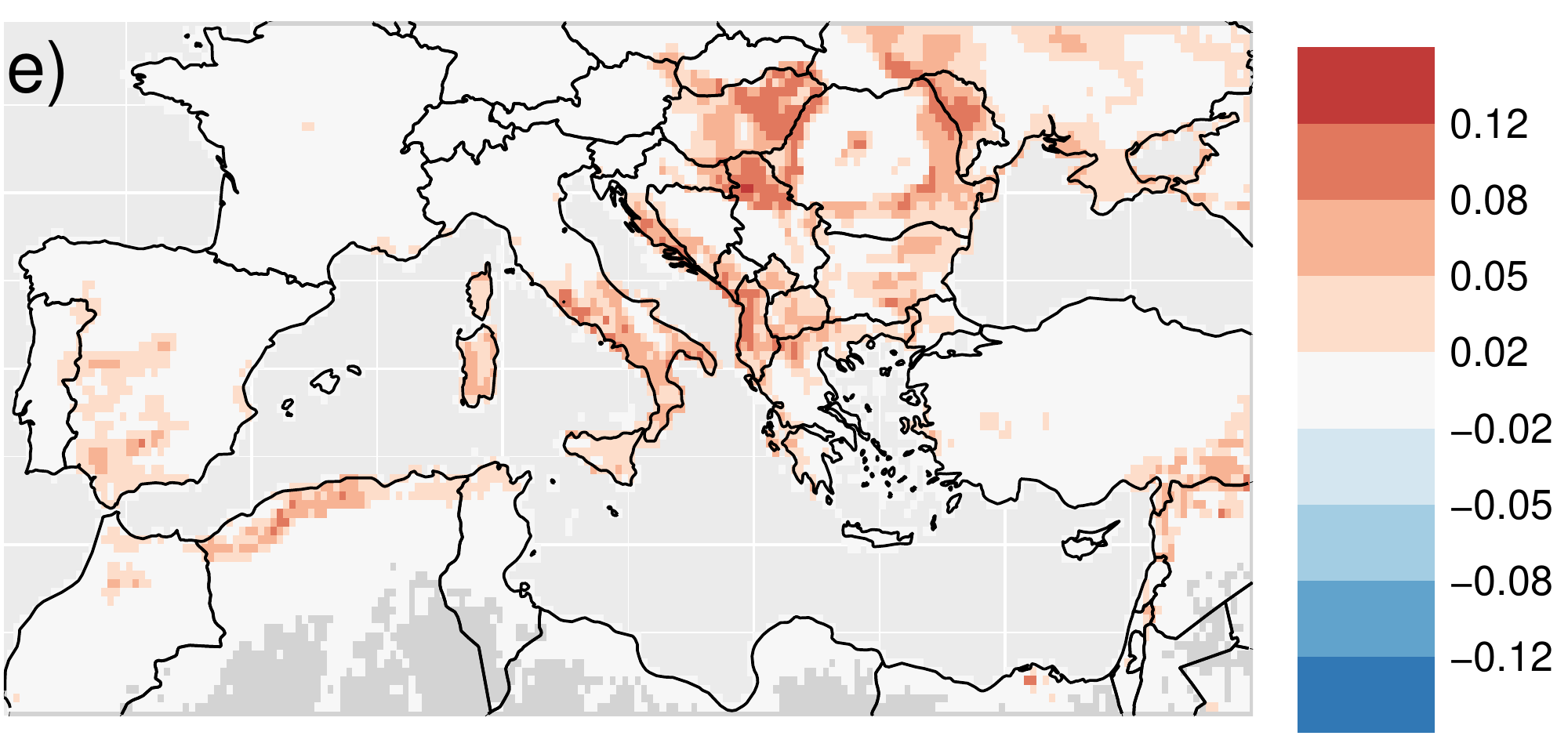} 
\end{minipage}
\begin{minipage}{0.49\linewidth}
\centering
\includegraphics[width=\linewidth]{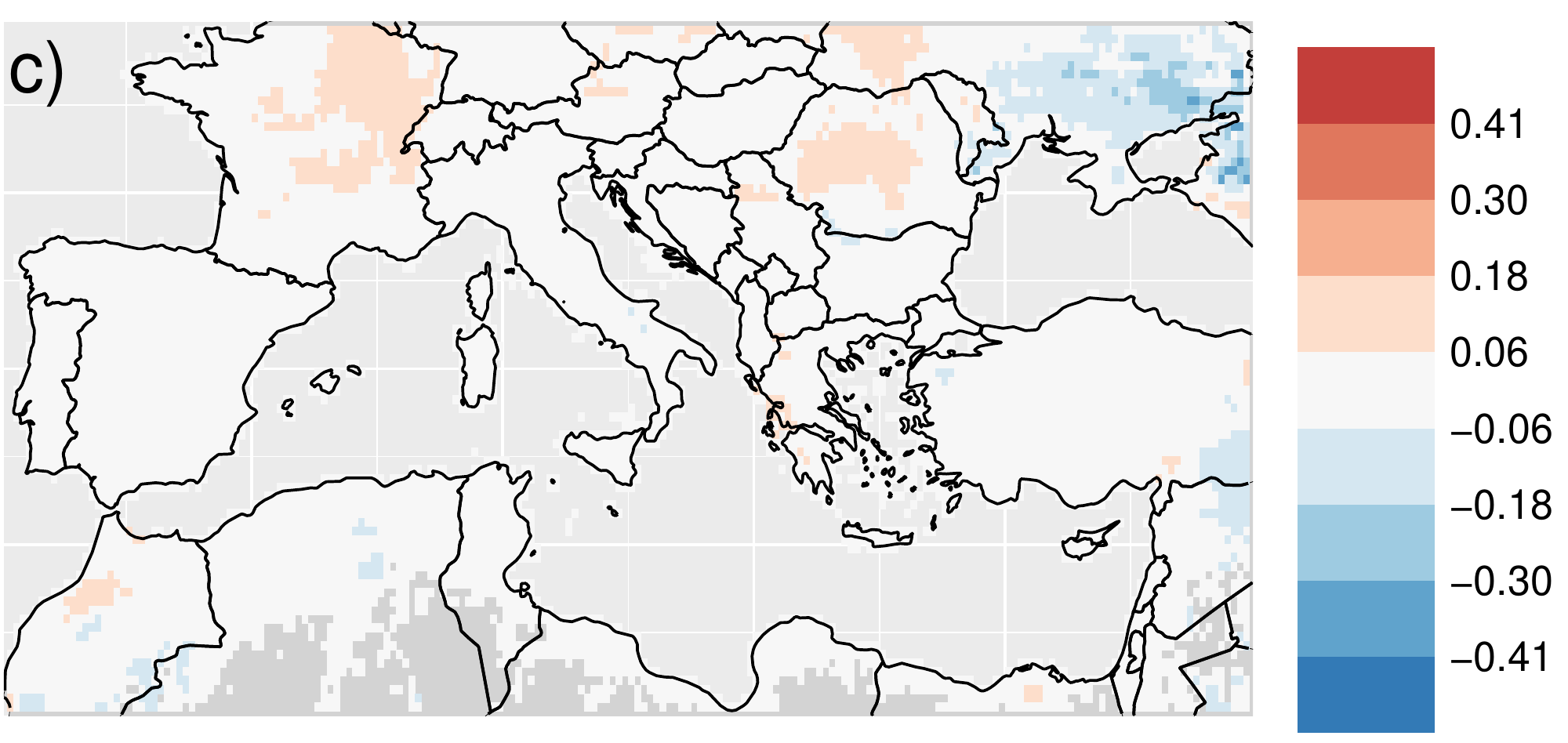} 
\end{minipage}
\begin{minipage}{0.49\linewidth}
\centering
\includegraphics[width=\linewidth]{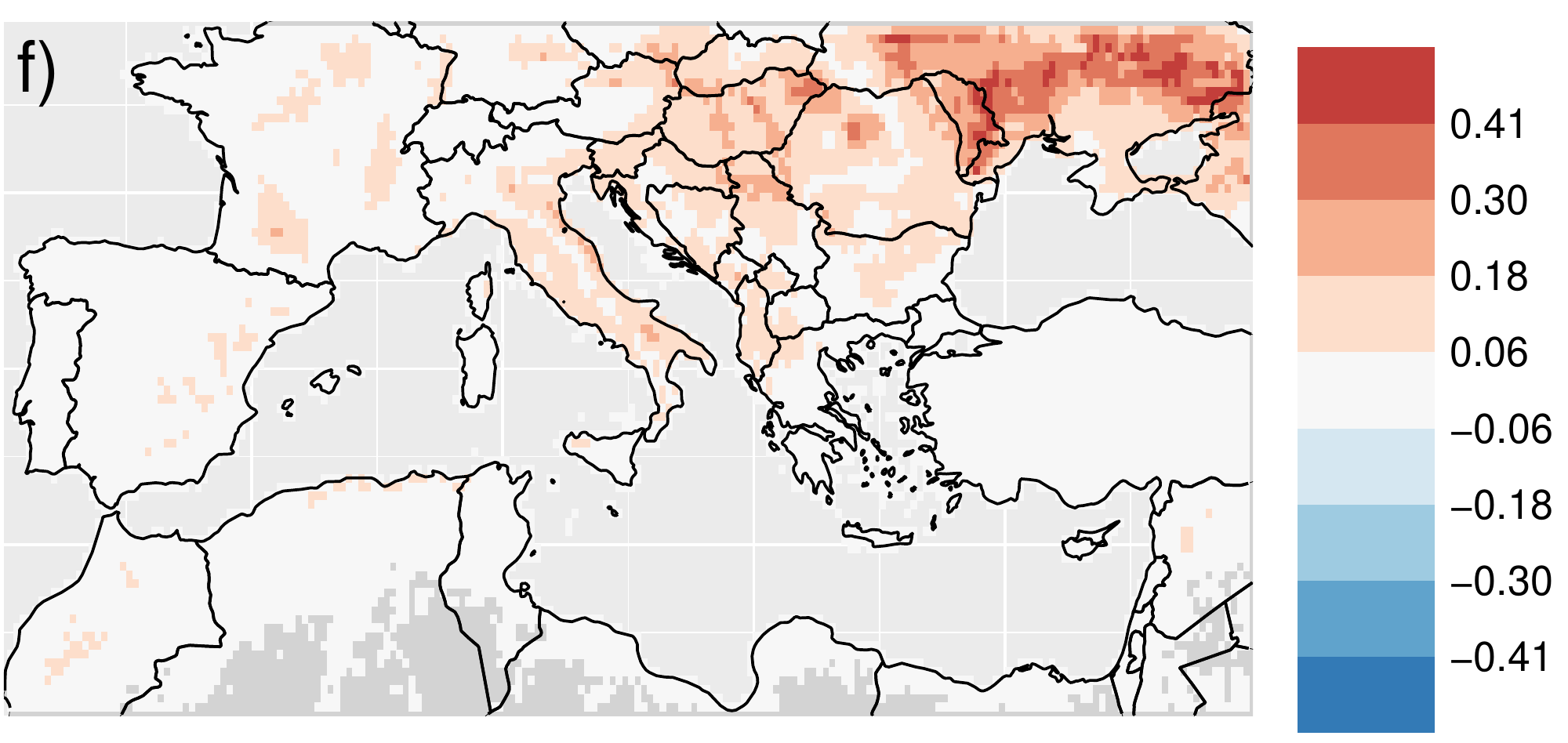} 
\end{minipage}
\caption{{Left column: a) site-wise estimated trends (change per year) in August VPD [Pa] observed over the period 2001--2020. Under these trends, maps of the site-wise (bootstrap median) changes in estimates of the occurrence probability $p_0(s,t)$ [unitless; b] and $\log\{1+q^+_{0.9}(s,t)\}$ [$\log(\mbox{km}^2)$; c] (with $q^+_{0.9}(s,t)$, the $90\%$ quantile of the spread $Y(s,t)\;|\;\{Y(s,t)>0,\mathbf{X}(s,t)\}$, a measure of wildfire intensity) for the month of August 2001. Right column is as the left column, but using trends in August 2m air temperature [K].} }
\label{CCmaps}
\end{figure}

{We predict values of 2m air temperature and VPD for August 2020 at each site $s\in \mathcal{S}$ using expectation regression, under the assumption that the  variable follows a linear temporal trend (i.e., a simple linear regression model with only year as a covariate). Trends are estimated using the ERA5 data for August only, across 2001 to 2020. Although the trends are assumed to be heterogeneous over $\mathcal{S}$ and are estimated at each site independently, we pool data from adjacent grid-cells to improve estimation. The median ($2.5\%$ and $97.5\%$ quantiles) of the estimated yearly site-wise changes, across the entire domain, are $+4.82$Pa ($-4.29$, $26.4$) for VPD and $0.042$K ($-0.012$, $0.103$) for air temperature. Maps of the estimated trends (i.e., the regression coefficients from the fitted linear models) are provided in Figure~\ref{CCmaps} for both air temperature and VPD. As discussed above, in line with climate change trends \citep{Zappa2021}, for both variables we observe generally positive trends across the entirety of Mediterranean Europe. Negative trend estimates are rare; Figure~\ref{CCmaps} does illustrate a slight predicted reduction in both VPD and 2m air temperature in northern Libya, and in VPD only in northern Italy and west Turkey. }\par
Figure~\ref{CCmaps} present maps of the median site-wise differences (for separate perturbations in VPD and air temperature) in $p_0$ and $q^+_{0.9}$ across all bootstrap model fits. Positive increases in $p_0$ and $q^+_{0.9}$ correspond to an increased expected frequency and intensity, respectively, of wildfires under observed trends for VPD and air temperature. As noted in Subsection~\ref{risk_assess_sec}, there is no clear one-to-one correspondence between wildfire occurrence and spread; areas that exhibit positive trends in wildfire frequency do not necessarily exhibit positive trends in intensity, and vice versa. Under perturbations in VPD, we observe a small increase in the intensity of extreme wildfires across some of Europe (Figure~\ref{CCmaps}c), but observe little increase in the expected frequency of wildfires anywhere in the domain (Figure~\ref{CCmaps}b); $p_0$ increases only in Ukraine and Romania. Our model predicts reductions in $p_0$ in regions such as northern Algeria and southern Spain, with relatively large reductions observed in eastern Turkey and Syria (Figure~\ref{CCmaps}b); for the latter two areas, a concomitant reduction in wildfire intensity is also predicted (see decreasing $q^+_{0.9}$ in Figure~\ref{CCmaps}c). Figure~\ref{CCmaps}c illustrates surprisingly large reductions in wildfire intensity in eastern Ukraine and western Russia, as well as small decreases in parts of Algeria, Morocco and northern Saudi Arabia. {Concerning relative, rather than absolute, differences, the median ($2.5\%$ and $97.5\%$ quantiles) of the site-wise estimates in frequency and severity were $1.2\%$ ($-29.4\%$, $16.4\%$) and $3.6\%$ ($-9.5\%$, $12.3\%)$, respectively.}

 \par
Figures~\ref{CCmaps}e,f show a much stronger signal in both {occurrence probability} $p_0$ and {conditional spread intensity} $q^+_{0.9}$ when air temperature is perturbed. Under changes in 2m air temperature, the frequency of August wildfires appears to be increasing at a much faster rate across Europe and north Africa, with no areas exhibiting an obvious reduction in their respective $p_0$ estimates. Unlike the estimated intensity trends observed under changes in VPD, Figure~\ref{CCmaps}f illustrates that changes in temperature appear to adversely effect Ukraine and western Russia; we observe strong positive changes in $q^+_{0.9}$ these regions, compared to Figure~\ref{CCmaps}c. Figure~\ref{CCmaps}f also illustrates positive trends in $q^+_{0.9}$ across the vast majority of Europe, with no regions in the domain exhibiting a reduction in wildfire intensity. {The median ($2.5\%$ and $97.5\%$ quantiles) of relative site-wise changes in frequency and severity were $17.1\%$ ($-3.1\%$, $60.6\%$) and $1.6\%$ ($-5.0\%$, $6.9\%)$, respectively. This suggests that, under our observed warming trends, fires are becoming more frequent and serve, but not at a proportional rate.} Uncertainty in Figures~\ref{CCmaps}b,c,e,f is quantified via the $2.5\%$ and $97.5\%$ bootstrap quantiles of changes in $p_0$ and $q^+_{0.9}$ for the two scenarios; these are presented in Figure~\ref{clim_map_uncertain} of Appendix~B. Figures~\ref{clim_map_uncertain}a,c,i,k cast some doubt on the significance of the results observed in Figures~\ref{CCmaps}b,c, as most of the site-wise $95\%$ confidence intervals include zero; the converse holds for Figures~\ref{CCmaps}e,f, suggesting that the worrying positive trends observed for $p_0$ and $q^+_{0.9}$, under perturbations in air temperature, are significant.  
\section{Conclusion}
\label{conclusion_sec}
We develop a hybrid statistical deep-learning framework that combines asymptotically-justified extreme value models, generalised additive regression, and neural networks, to investigate the drivers of both wildfire occurrence and extreme spread in Mediterranean Europe. The impact of vapour-pressure deficit (VPD), 2m air temperature, and a 3-month standardised precipitation index (SPI), on model parameters determining the occurrence probability, and the scale of extreme spread, of wildfires is accommodated through the use of thin-plate splines, whilst the effect of a large number of other covariates determining meteorological and land-surface conditions is modelled using neural networks. We investigate spatio-temporal trends in wildfire frequency and intensity relative to perturbations in VPD and air temperature by estimating their spatially-localised linear trends; these are then fed into the model to determine how these trends would have affected wildfire occurrence probabilities and the spread distribution parameter for the month of August 2001, which is the month with the highest observed total burnt area.\par
Our analysis reveals that different drivers impact Mediterranean Europe wildfire occurrence and extreme spread. Whilst VPD, air temperature, and SPI, all affect the former, only VPD appears to have a significant effect on extreme wildfire spread; although similar conclusions about wildfire occurrence have been drawn by many studies, for example, by \cite{turco2014climate, turco2017key},  \cite{ruffault2018extreme}, \cite{parente2019drought},  and \cite{de2021climate,de2022convergence}, the conclusions regarding extreme spread are less consistent with the literature, which further supports our claim, in Subsection~\ref{interp_results_sec}, that non-stationarity in the upper-tails of the spread distribution are captured in the model for $u(s,t)$. As the threshold $u(s,t)$ is not easily-interpretable, we chose not to represent this aspect of the model using the PINN framework described in Subsection~\ref{pinnrepsec}; in order to better understand the impact of the interpretable predictors on the extreme values of  wildfire spread, we may be better suited replacing the GPD tail model with an entire bulk-tail model that foregoes pre-specification of an exceedance threshold $u(s,t)$, such as those proposed by \cite{papastathopoulos2013extended},  \cite{naveau2016modeling} and \cite{stein2021parametric}.  We further find that, when comparing month-specific maps of estimated wildfire occurrence probability and extreme quantiles of wildfire spread across Mediterranean Europe, the meteorological and land-surface conditions that lead to high estimates of occurrence probabilities do not necessarily lead to high estimates of spread quantiles. Maps of estimated extreme quantiles of wildfire spread can facilitate risk assessment as they provide a means of identifying high-risk areas of Europe; we highlight areas of major concern in eastern Europe and the Iberian Peninsula. {By focusing on the extreme wildfires in August 2001, we find that the impact on wildfire activity of ongoing trends in VPD and temperature, which are critical in view of climate change, may not be homogeneous across Mediterranean Europe; whilst many regions exhibit worrying positive trends in wildfire frequency and intensity, some locations do exhibit a reduction in extreme wildfire risk. We further find that ongoing trends in VPD and temperature may lead to substantially different changes in the expected frequency and severity of wildfires. For wildfires in August 2001 and on average over Europe, observed trends in air temperature lead to a relative increase of $17.1\%$ and $1.6\%$ in the expected frequency and severity, respectively, whilst changes in VPD suggest respective increases of $1.2\%$ and $3.6\%$.}
\section*{Acknowledgments} The research reported in this publication was supported by funding from King Abdullah University of Science and Technology (KAUST) Office of Sponsored Research (OSR) under Award No. OSR-CRG2020-4394. Support from the KAUST Supercomputing Laboratory is gratefully acknowledged. This project has received funding from the European Union’s Horizon 2020 research and innovation programme under grant agreement no. 101003469. J.Z. acknowledges funding from the Helmholtz Initiative and Networking Fund (Young Investigator Group COMPOUNDX, Grant Agreement VH-NG-1537).
\section*{Data accessibility}
The data that supports our findings and the code for fitting the PINN models are both available in the \texttt{R} package \texttt{pinnEV} \citep{pinnEV}.
\baselineskip=14pt
\begingroup
\setstretch{0.75}
\bibliographystyle{apalike}
\bibliography{ref}
\endgroup
\baselineskip 10pt

\pagebreak
\begin{appendix}
\renewcommand{\theequation}{A.\arabic{equation}}
\renewcommand{\thefigure}{A\arabic{figure}}
\renewcommand{\thetable}{A\arabic{table}}
\renewcommand{\thesection}{A\arabic{section}}

\renewcommand{\theHfigure}{A\arabic{figure}}
\setcounter{figure}{0}
\setcounter{table}{0}
\setcounter{equation}{0}
\section{Appendix A}
\subsection{Neural network architecture}
\label{sup_NN_sec}
We now describe a feed-forward neural network with $J$ hidden layers, each with $n_j$ neurons for $j=1,\dots,J$; first, note that $\mathbf{x}_\mathcal{N}(s,t)=(x_1(s,t),\dots,x_{I-d}(s,t))^T$ for all $(s,t)\in\mathcal{S}\times\mathcal{T}$. We describe the output of layer $j$, neuron $i$, through the function $m_{j,i}(s,t)$. Define the vector of outputs from layer $j$ by $\mathbf{m}_j(s,t)=(m_{j,1}(s,t),\dots,m_{j,n_j}(s,t))^T$, with the convention that $n_0=I-d$ and $m_{0,i}(s,t):=x_i(s,t)$ for $i=1,\dots,I-d$. The output from the final layer is
\[
m_{\mathcal{N}}\{\mathbf{x}(s,t)\}=(\mathbf{w}^{(J+1)})^T\mathbf{m}_{J}(s,t)+b^{(J+1)},
\]
for a vector of weights $\mathbf{w}^{(J+1)}\in \mathbb{R}^{n_J}$ and a bias $b^{(J+1)}\in\mathbb{R}$. The output of the $j$-th layer at the $i$-th node, for $j=1,\dots,J$ and $i=1,\dots,n_j$, can be written as
\begin{equation}
\label{layer_eq}
m_{j,i}(s,t)=a_j\left\{g_{j,i}(\{\mathbf{m}_{j-1}(s,t):s \in \mathcal{S}\})+b^{(j,i)}\right\},
\end{equation}
for a bias $b^{(j,i)}\in\mathbb{R}$, a fixed activation function $a_j$ and a function $g_{j,i}$ that is dependent on some real weights. As advocated by \cite{glorot2011deep}, we use only the rectified linear unit (ReLu) activation function and set $a_j(x)=\max\{0,x\}$ for all $j=1,\dots,J$. If \eqref{layer_eq} is a densely-connected layer, then $g_{j,i}(\{\mathbf{m}_{j-1}(s,t):s \in \mathcal{S}\})=(\mathbf{w}^{(j,i)})^T\mathbf{m}_{j-1}(s,t)$ for weights $\mathbf{w}^{(j,i)}\in \mathbb{R}^{n_{j-1}}$ and for all $i=1,\dots,n_j$.\par
Through specification of $g_{j,i}$, we can define \eqref{layer_eq} as a convolutional layer; however, these can only be applied in the case where $\mathcal{S}$ can be represented as a $D_1 \times D_2$ grid of locations. In such a case, a convolution filter with pre-specified dimension $d_{1,j}\times d_{2,j}$ for $d_{1,j},d_{2,j}$ odd integers, $d_{1,j}\leq D_1, d_{2,j}\leq D_2$, can be created for the $j$-th layer; we keep the filter dimension consistent over all layers and subsequently drop the subscript $j$ from the notation. Define a neighbourhood $\mathcal{L}(s)$ for $s\in\mathcal{S}$ as the set of locations that form a $d_{1}\times d_{2}$ grid with site $s$ at the centre and let $M^{(j-1,i)}(s,t)$ be a $d_{1}\times d_{2}$ matrix created by evaluating $m_{j-1,i}(s,t)$ on the grid $\mathcal{L}(s)$. For a $d_{1}\times d_{2}$ weight matrix $W^{(j,i)}$ with real entries, the $i$-th convolutional filter for layer $j$ is of the form $\mathcal{C}_{j,i}\{m_{j-1,i}(s,t)\}=\sum_{k=1}^{d_{1}}\sum_{l=1}^{d_{2}}W^{(j,i)}[k,l]\{M^{(j-1,i)}(s,t)\}[k,l]$ with operations taken entry-wise; here $A[k,l]$ denotes the $(k,l)$-th entry of matrix $A$. The layer \eqref{layer_eq} is then considered convolutional if 
\[
g_{j,i}(\{\mathbf{m}_{j-1}(s,t):s \in \mathcal{S}\})=\sum^{n_{j-1}}_{k=1}\mathcal{C}_{j,k}\{m_{j-1,k}(s,t)\}
\]
for all $i=1,\dots,n_j$. To apply a convolutional filter at the boundaries of $\mathcal{S}$, we pad each input map, that is, $\{\mathbf{x}_i(s,t):s\in\mathcal{S}\}$ for $i=1,\dots,d$, by mapping $\mathcal{S}$ to a new domain $\mathcal{S}^*$, taken to be a $\{D_1+d_{1}-1\} \times \{D_2+d_{2}-1\}$ grid of sites with $\mathcal{S}$ at the centre. Then, for all $i=1,\dots,d$, we set $\mathbf{x}_i(s,t)=0$ for all $s$ within distance $(d_{1}-1)/2$ and $(d_2-1)/2$ of the first and second dimension, respectively, of the boundaries of $\mathcal{S}^*$; a similar padding is used for all $\{\mathbf{m}_{j,k}(s,t):s\in\mathcal{S}\}$ for $k=1,\dots,n_j$ (i.e., the output from the $j$-th layer). Throughout the main text we follow \cite{richards2022} and use only $3\times 3$ filters, with $d_{1}=d_{2}=3$. We denote neural networks that use only convolutional or densely-connected layers as convolutional neural networks (CNNs) and densely-connected NNs, respectively.
\subsection{Uncertainty assessment}
To quantify parameter uncertainty, we utilise a stationary bootstrap scheme \citep{politis1994stationary} with expected block size $k$. To create a single bootstrap sample, we repeat the following until obtaining a sample of length greater than or equal to $|\mathcal{T}|$; draw a starting time $t^*\in\mathcal{T}$ uniformly at random and a block size $K$ from a geometric distribution with expectation $k$, then add the block of observations $\{y(s,t):s \in \mathcal{S}, t\in \{t^*,\dots,t^*+K-1\}\}$ to the bootstrap sample. In cases where $t^*$ is generated with $t^*+K-1 > |\mathcal{T}|$, we instead add $\{y(s,t):s\in\mathcal{S}, t\in \{1,\dots,t^*+K-|\mathcal{T}|-1\}\cup\{t^*,\dots,|\mathcal{T}|\}\}$; the sample is then truncated to have length $|\mathcal{T}|$. The full model is estimated for each bootstrap sample.
\subsection{Validation and testing}
\label{partionalg}
For all model fits, we use a validation and testing scheme to reduce over-fitting and improve out-of-sample prediction. Before estimating any parameters, each bootstrap sample is partitioned into a 80-10-10 split for training, validation and testing data.  Partitioning the data is performed at random;  we assign data for testing and validation by removing space-time clusters of observations. For each distinct three month block\footnote{A three month block was chosen to reduce computational expense.} of observed space-time locations (i.e., $\boldsymbol{\omega}$), we simulate a standard space-time Gaussian process $\{Z(s,t)\}$ with separable correlation function $\rho\{(s_i,t_i),(s_j,t_j)\}={\exp(-\|s_i-s_j\|^*/\lambda_s)\exp(-| t_i-t_j|^*/\lambda_t)}$, where $\|\cdot\|^*$ denotes the (geodesic) great-Earth distance in km. The range parameters $\lambda_s,\lambda_t >0$ are tunable hyper-parameters that control the average size of a space-time cluster. In the application described in Section~\ref{application_sec}, we set $\lambda_s=240$ and $\lambda_t=5$, as this gave reasonably large clusters. After simulating a single realisation of $Z(s,t)$ for each three-month block, we assign observations $y(s,t)$ to validation if the realisation $z(s,t)$ falls below the $10\%$ quantile of $\{z(s,t):(s,t)\in\boldsymbol{\omega}\}$: similarly, we assign to testing observations $y(s,t)$ for which $z(s,t)$ falls above the $90\%$ quantile of $\{z(s,t):(s,t)\in\boldsymbol{\omega}\}$. Models are then fitted using only the training and validation data; each neural network is trained for a finite number of epochs, by minimising the associated loss evaluated on the training dataset. However, the model state after each epoch is recorded, and then the final model fit is taken to be that which minimises the validation loss, that is, the loss evaluated on the validation data, across all epochs. The testing data is not used in model fitting whatsoever; instead, we evaluate the associated model loss on these data to perform model selection, (i.e., choosing the optimal architecture, see Subsection~\ref{app_overview_sec}). Note that whilst we use different training/validation/testing partitions for each bootstrap sample, the same partition is used to fit the four models $p_0$, $u$, $p_u$ and $\sigma$ with a single bootstrap sample.\par
To compare fits of model \eqref{MargTransform} for {conditional spread} $Y(s,t)\;|\;Y(s,t)>0$, we utilise the threshold-weighted continuous ranked probability score (twCRPS) \citep{Gneiting2011}, which is a proper scoring rule. Define $\mathcal{P}\in \mathcal{S} \times \mathcal{T}$ as the set of out-of-sample space-time locations used for testing and let $\hat{F}_{(s,t),0}$ denote estimates of $F_{(s,t),0}$. Then for a sequence of increasing thresholds $\{v_1,\dots,v_{n_v}\}$ and a weight function $r(x)=\tilde{r}(x)/\tilde{r}(v_{n_v}), \tilde{r}(x)=1-(1+(x+1)^2/10000)^{-1/4}$, the empirical estimator of the twCRPS, provided by \cite{opitz2022editorial}, is
\[
\mbox{twCRPS}=\sum_{(s,t)\in\mathcal{P}}\sum^{n_v}_{i=1}r(v_i)[\mathbbm{1}\{y(s,t)\leq v_i\}-\hat{F}_{(s,t),0}(v_i)]^2,
\]
with $r(x)$ putting more weight on higher-valued predictions. We use 17 irregularly spaced thresholds ranging from $v_1=1$ to $v_{17}=750 > \max_{(s,t)\in\boldsymbol{\omega}}\{y(s,t)\}$.
\newpage
\section{Appendix B - Supplementary figures}
\renewcommand{\theequation}{B.\arabic{equation}}
\renewcommand{\thefigure}{B\arabic{figure}}
\renewcommand{\thetable}{B\arabic{table}}
\renewcommand{\thesection}{B\arabic{section}}

\renewcommand{\theHfigure}{B\arabic{figure}}
\setcounter{figure}{0}
\setcounter{table}{0}
\setcounter{equation}{0}
\label{sup_fig_sec}

\begin{figure}[h!]
\centering

\begin{minipage}{0.45\linewidth}
\centering
\includegraphics[width=\linewidth]{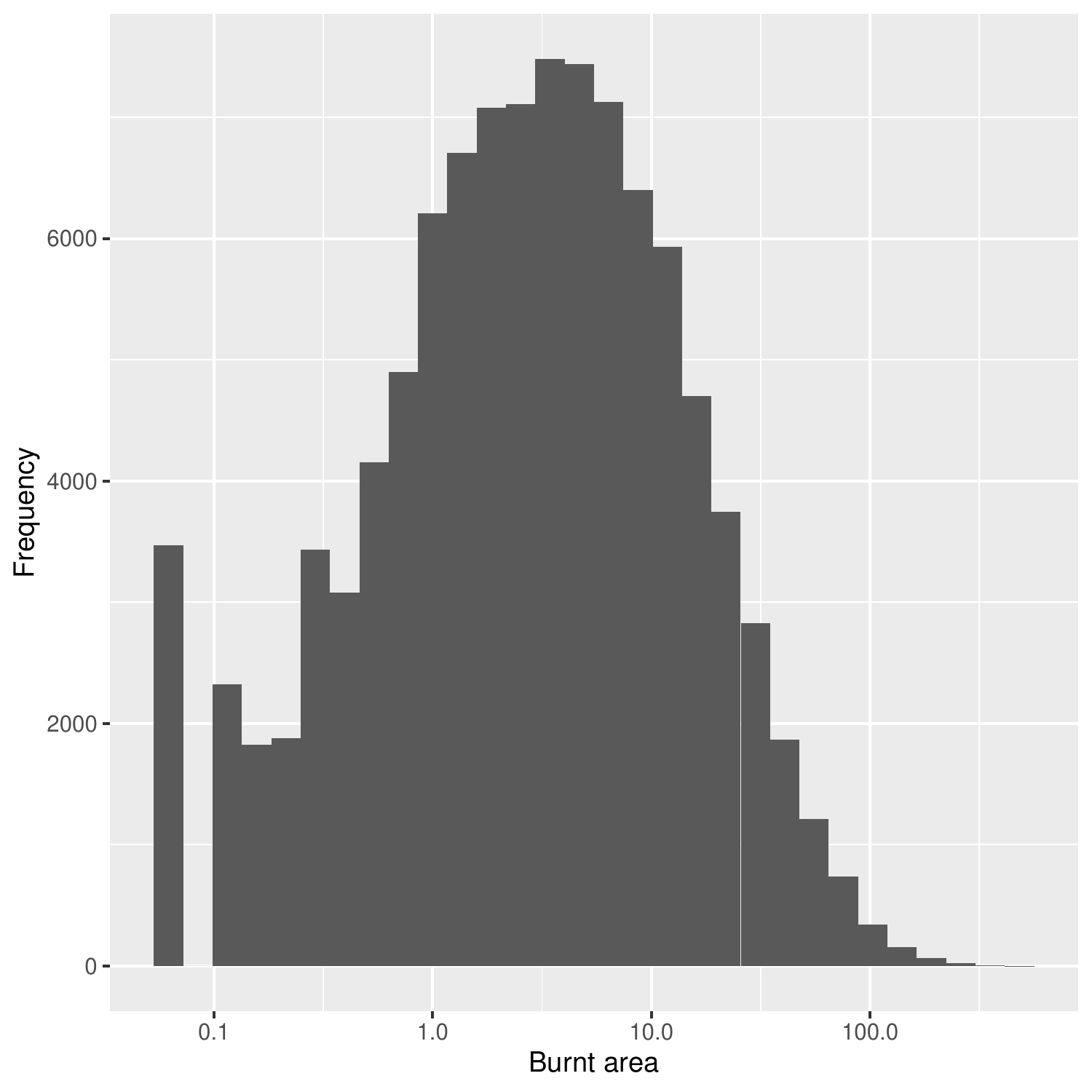} 
\end{minipage}
\begin{minipage}{0.45\linewidth}
\centering
\includegraphics[width=\linewidth]{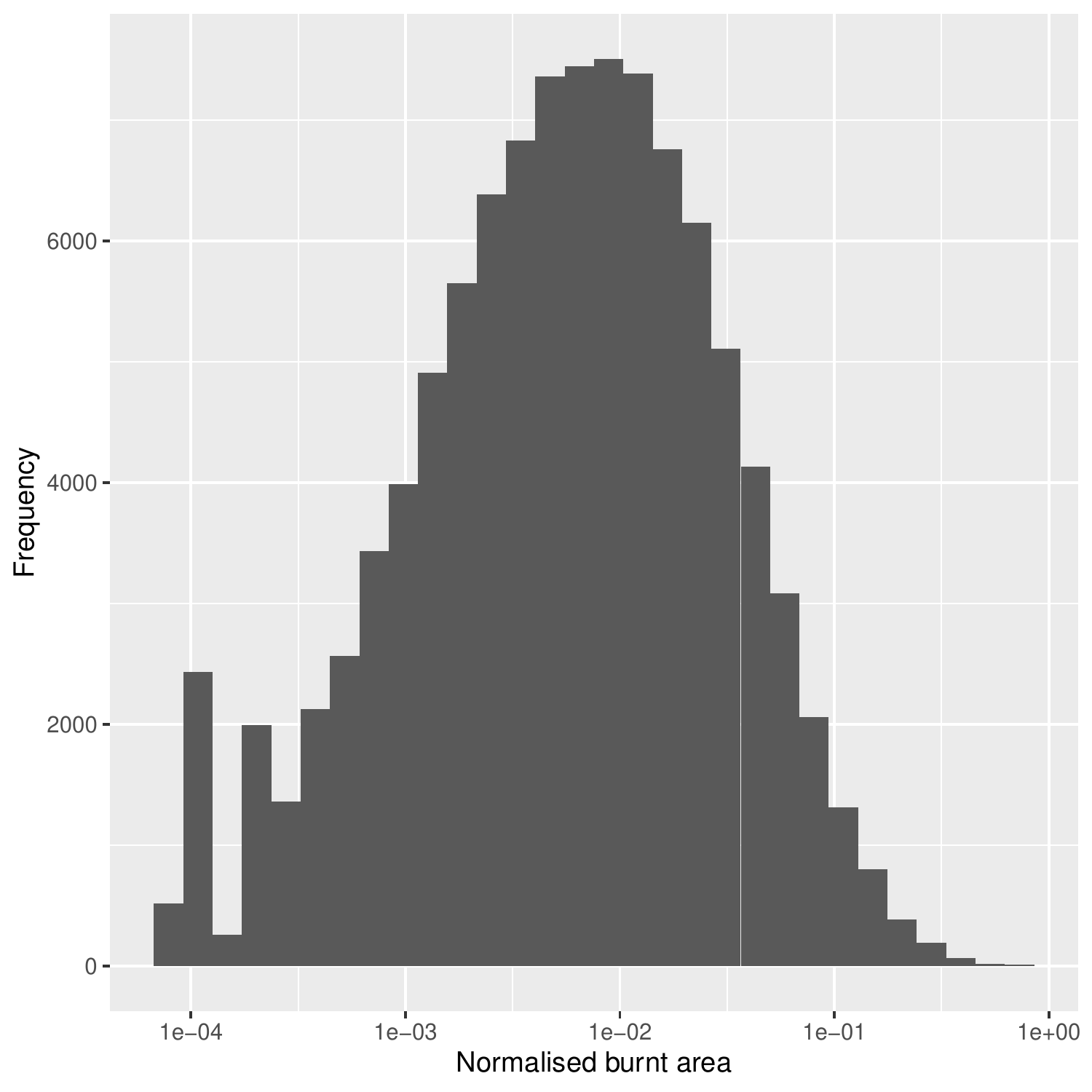} 
\end{minipage}
\caption{
Histogram of all observations of left) burnt area [BA; km$^2$] and right) BA normalised by the burnable area [unitless]. Note that the $x$-axis is on the log scale.}
\label{hists}
\end{figure}
\begin{figure}[h!]
\centering
\includegraphics[width=0.4\linewidth]{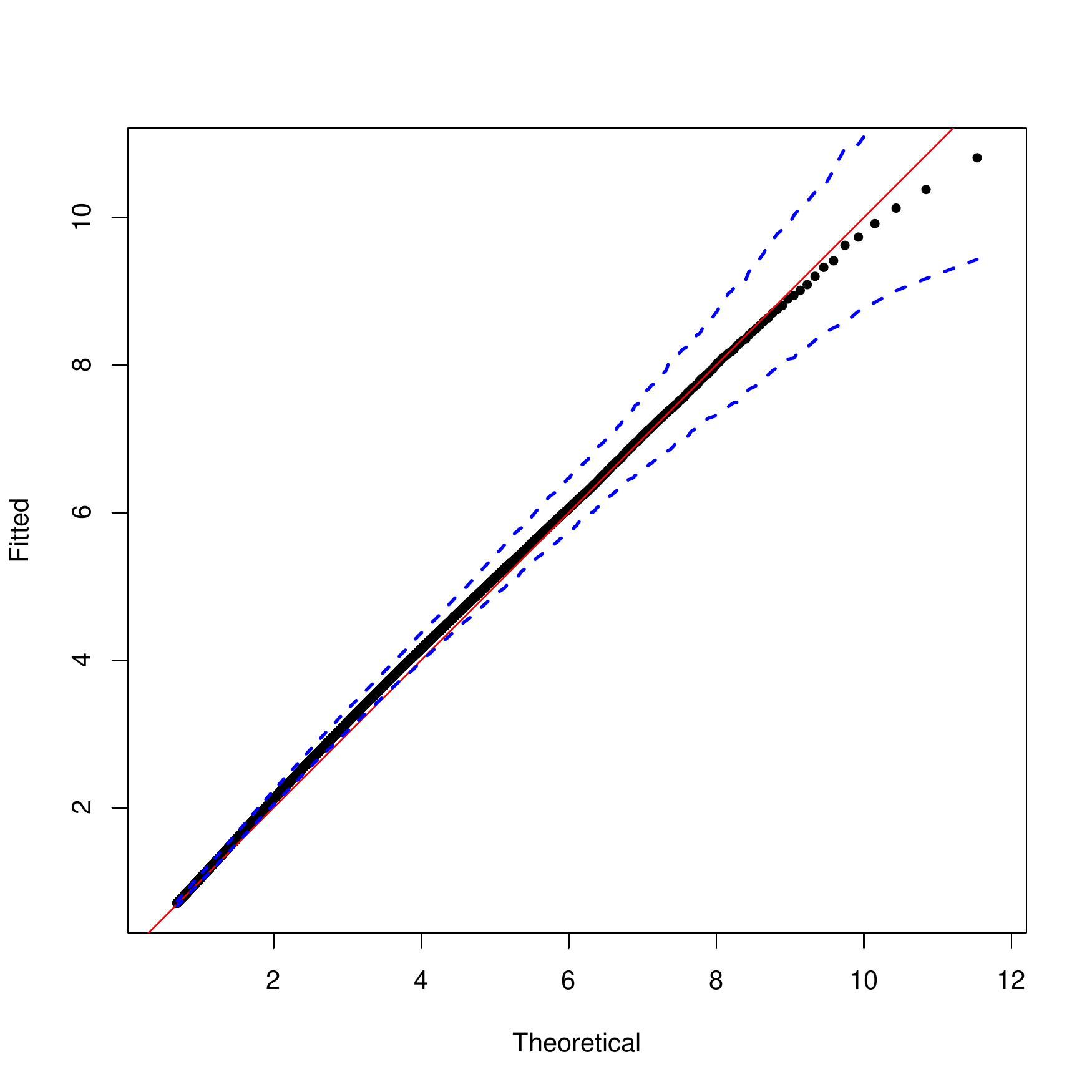} 
\caption{Q-Q plot for the pooled marginal fit of fire spread on standard exponential margins, averaged across all bootstrap samples. $95\%$ tolerance bounds are given by the dashed lines. Black points give the median quantiles across all samples, with the quantile levels ranging from $0.5$ to a value corresponding to the maximum observed value.}
\label{fig_exp_fit}
\end{figure}

\begin{figure}[h!]
\centering
\begin{minipage}{0.49\linewidth}
\centering
\includegraphics[width=\linewidth]{Images/obs_map.pdf} 
\end{minipage}
\hfill
\begin{minipage}{0.49\linewidth}
\centering
\includegraphics[width=\linewidth]{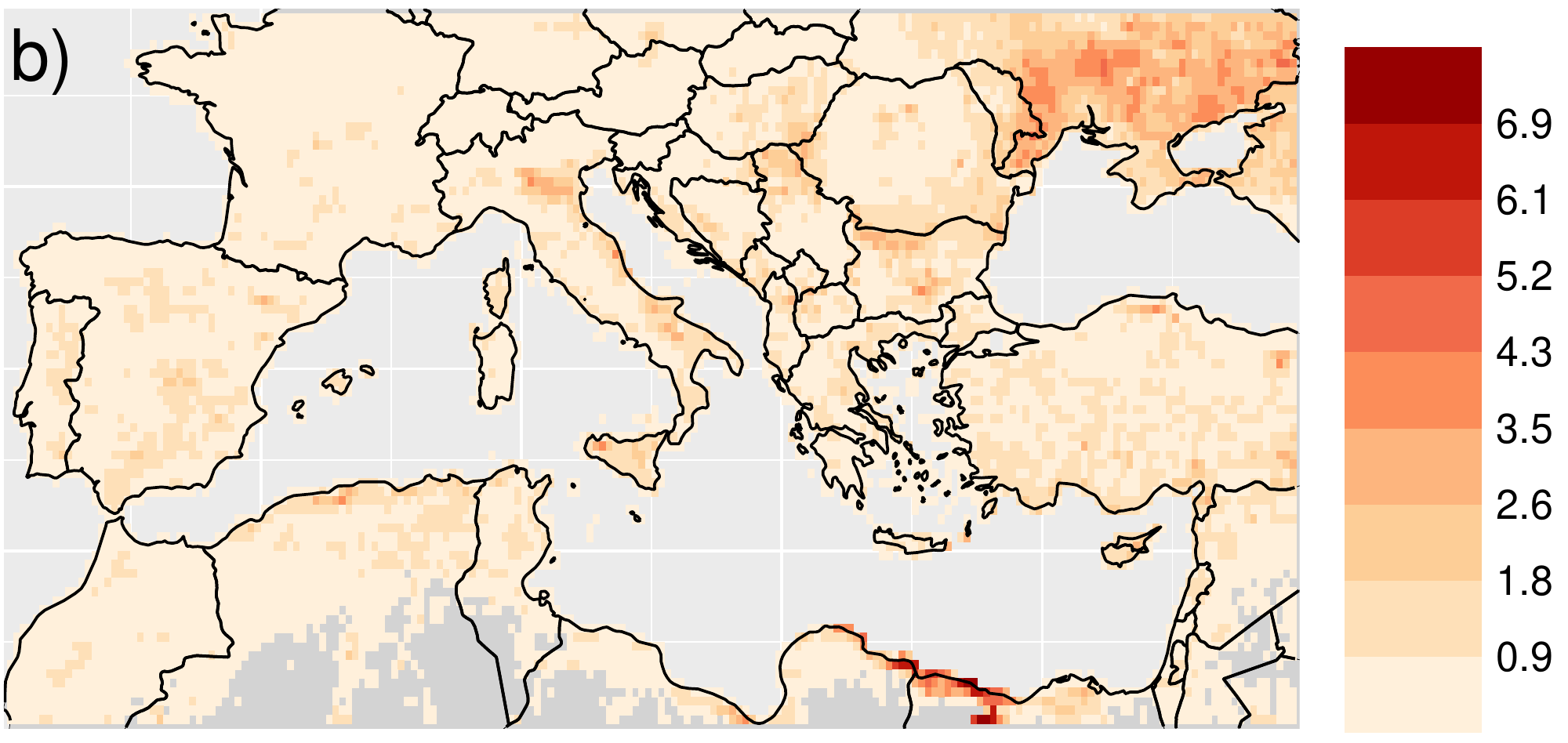} 
\end{minipage}
\centering
\begin{minipage}{0.49\linewidth}
\centering
\includegraphics[width=\linewidth]{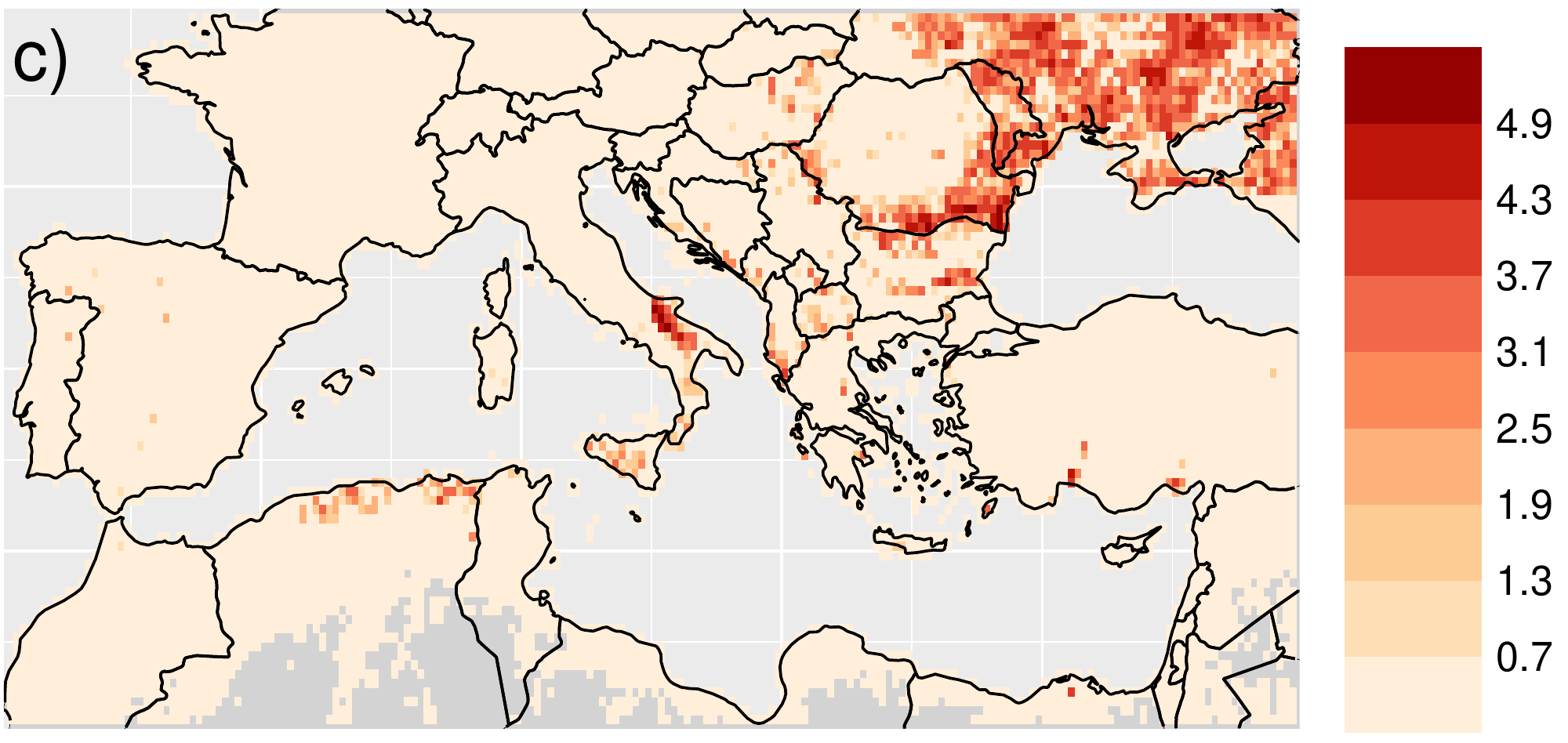} 
\end{minipage}
\hfill
\begin{minipage}{0.49\linewidth}
\centering
\includegraphics[width=\linewidth]{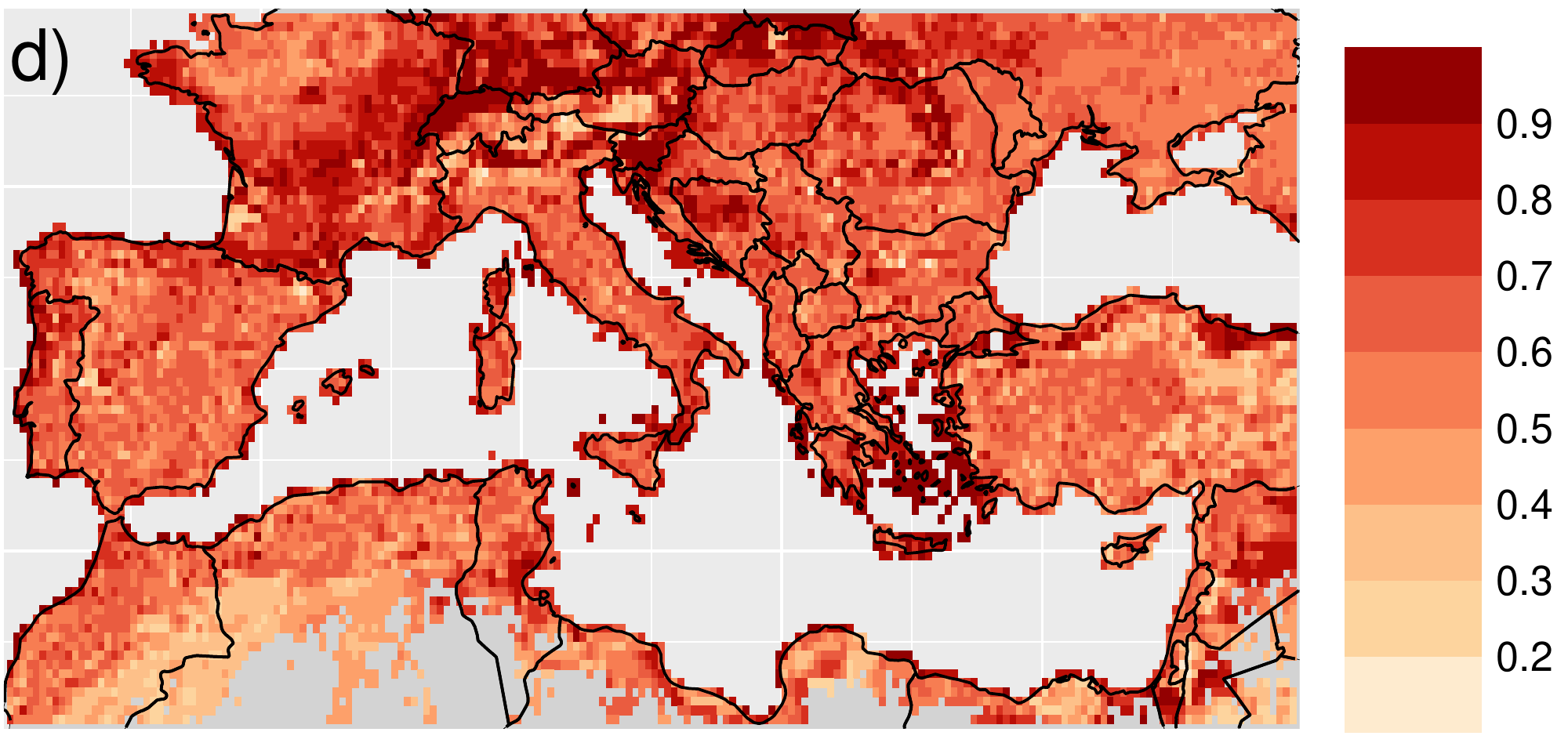} 
\end{minipage}

\caption{Maps of a) observed $\log\{1+{Y}(s,t)\}$ [burnt area; $\log(\mbox{km}^2)$] and corresponding estimates of b) $\log\{1+u(s,t)\}$ [$\log(\mbox{km}^2)$], c) $\log\{1+{Y}(s,t)-u(s,t)\}\;|\;Y(s,t) > u(s,t)$ [exceedances; $\log(\mbox{km}^2)$] and d) $p_u(s,t)$ [unitless] for August 2001. Note that $u(s,t)$ is the $40\%$ quantile of non-zero spread $Y(s,t)\;|\;Y(s,t) > 0$. Estimates are from the first bootstrap sample, see Subsection~\ref{app_overview_sec} of the main text. }
\label{sup_POT_fig}
\end{figure}

\begin{landscape}

\begin{figure}[t!]
\centering
\begin{minipage}{0.32\linewidth}
\centering
\includegraphics[width=\linewidth]{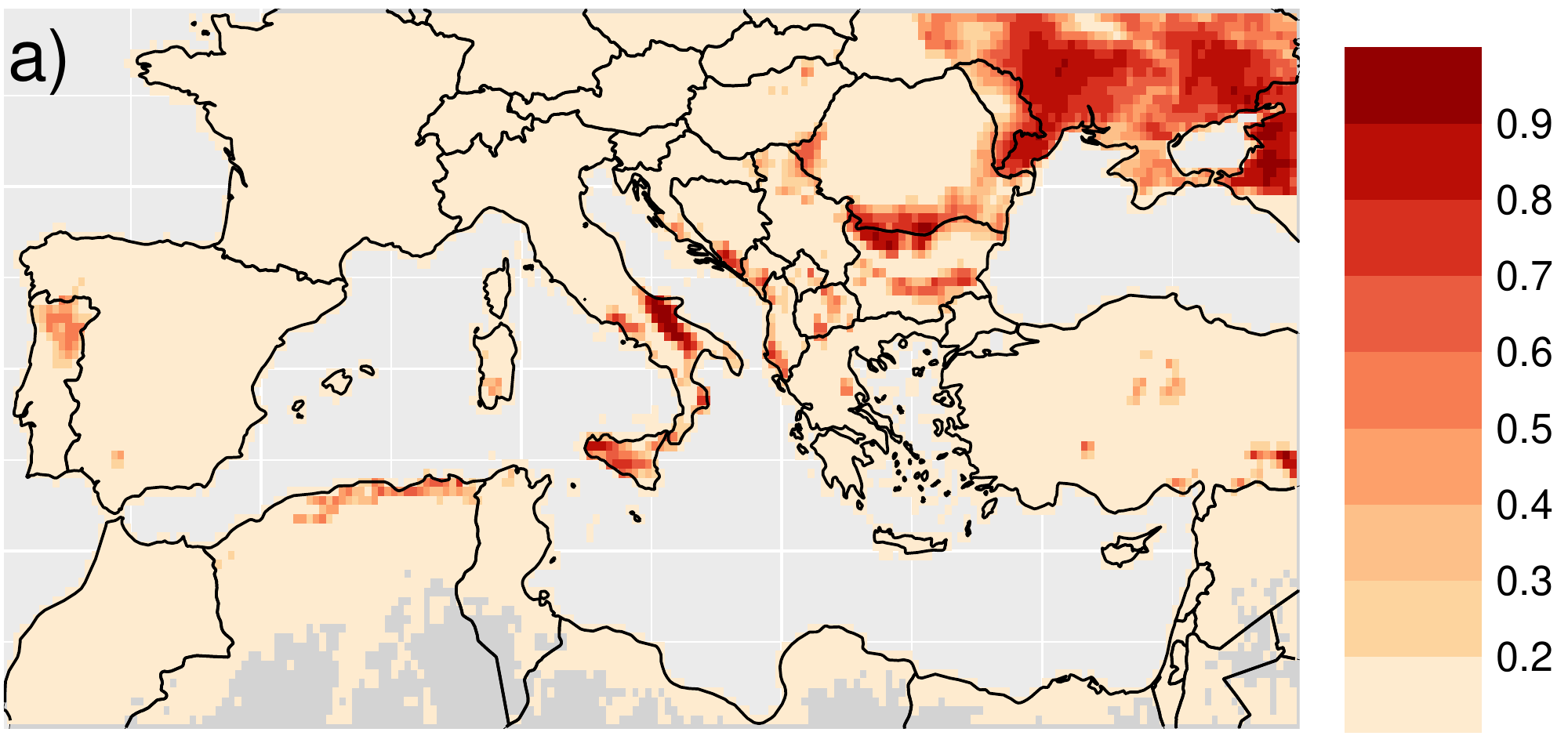} 
\end{minipage}
\begin{minipage}{0.32\linewidth}
\centering
\includegraphics[width=\linewidth]{Images/pZeroMap_med_t3_version5_overlay.pdf} 
\end{minipage}
\begin{minipage}{0.32\linewidth}
\centering
\includegraphics[width=\linewidth]{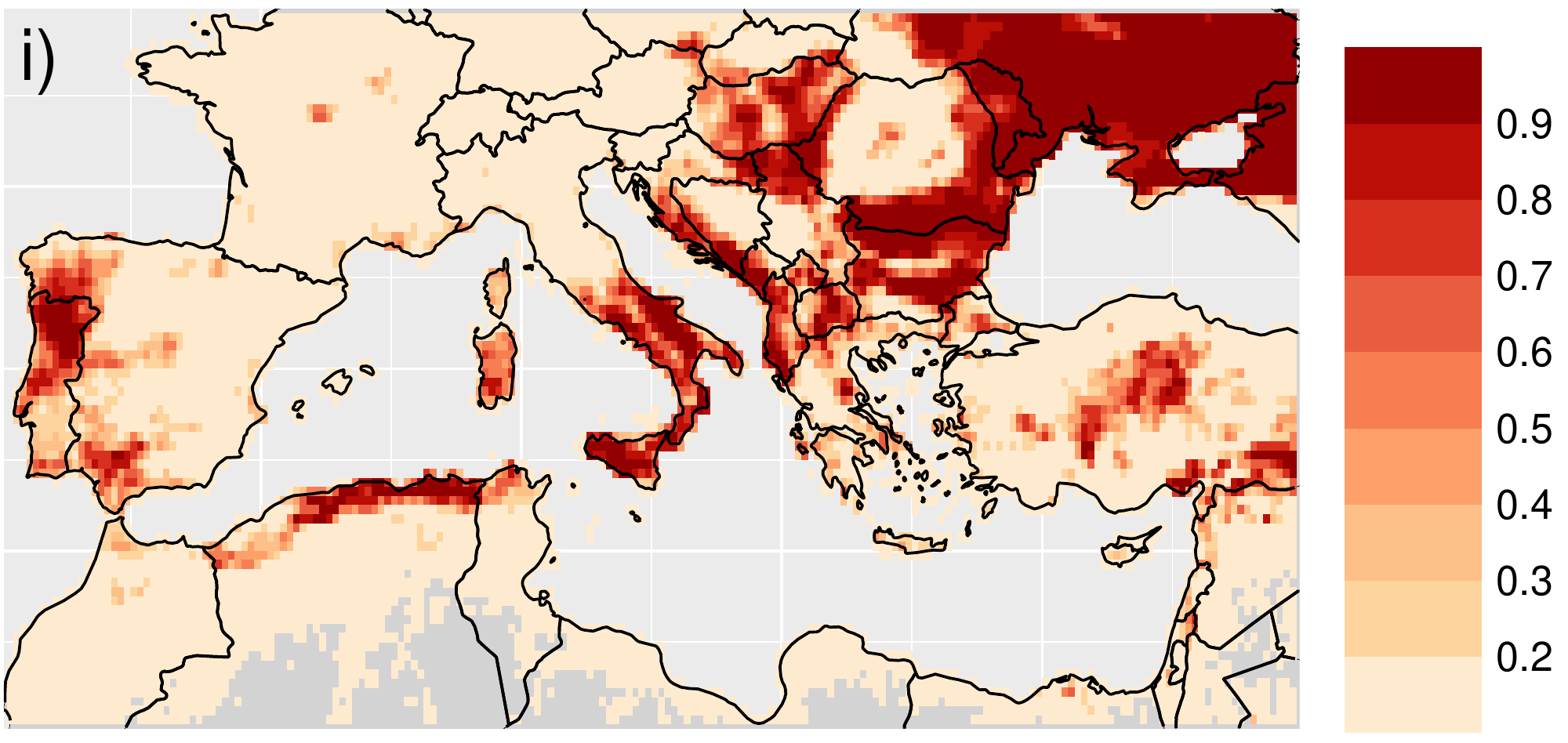} 
\end{minipage}
\begin{minipage}{0.32\linewidth}
\centering
\includegraphics[width=\linewidth]{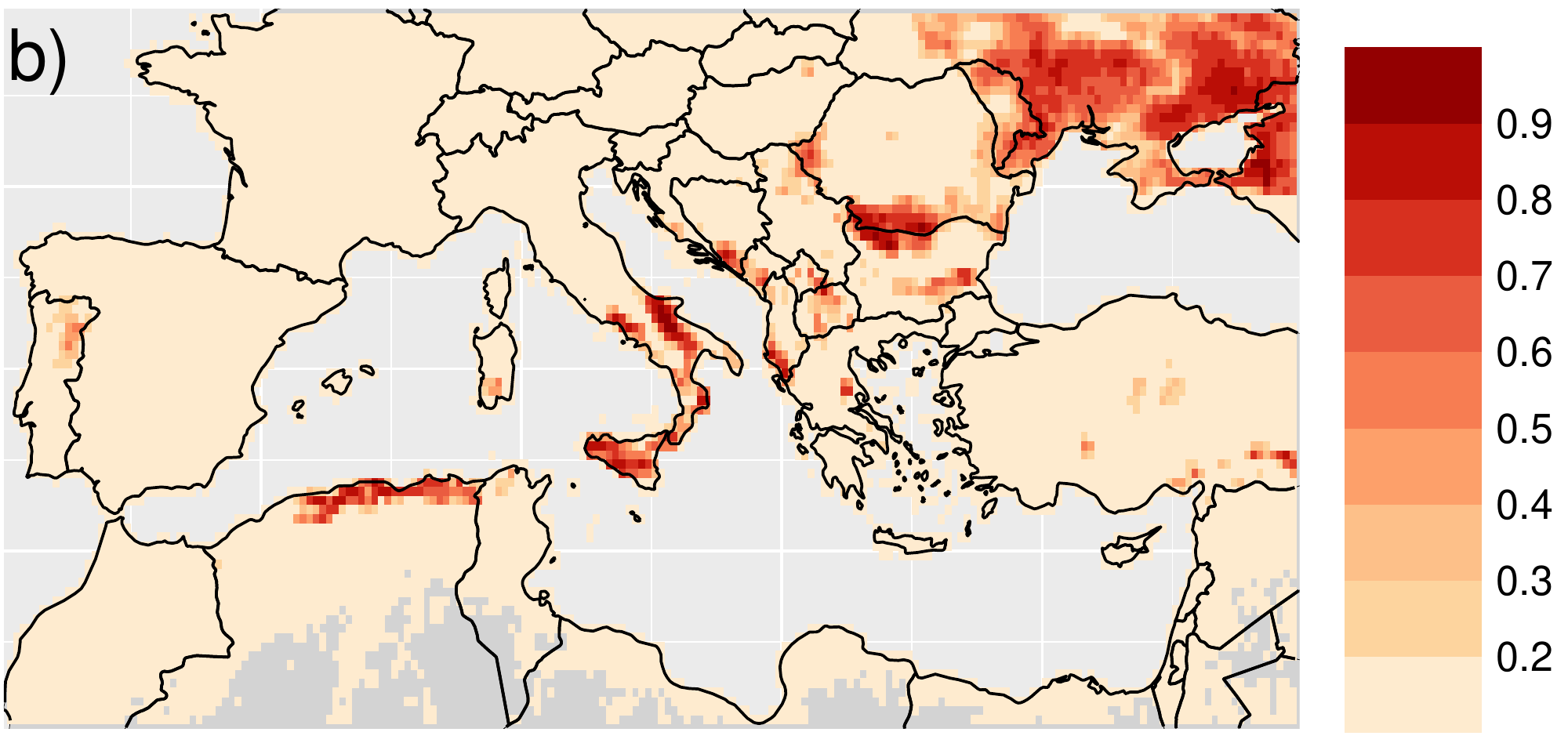} 
\end{minipage}
\begin{minipage}{0.32\linewidth}
\centering
\includegraphics[width=\linewidth]{Images/pZeroMap_med_t45_version5_overlay.pdf} 
\end{minipage}
\begin{minipage}{0.32\linewidth}
\centering
\includegraphics[width=\linewidth]{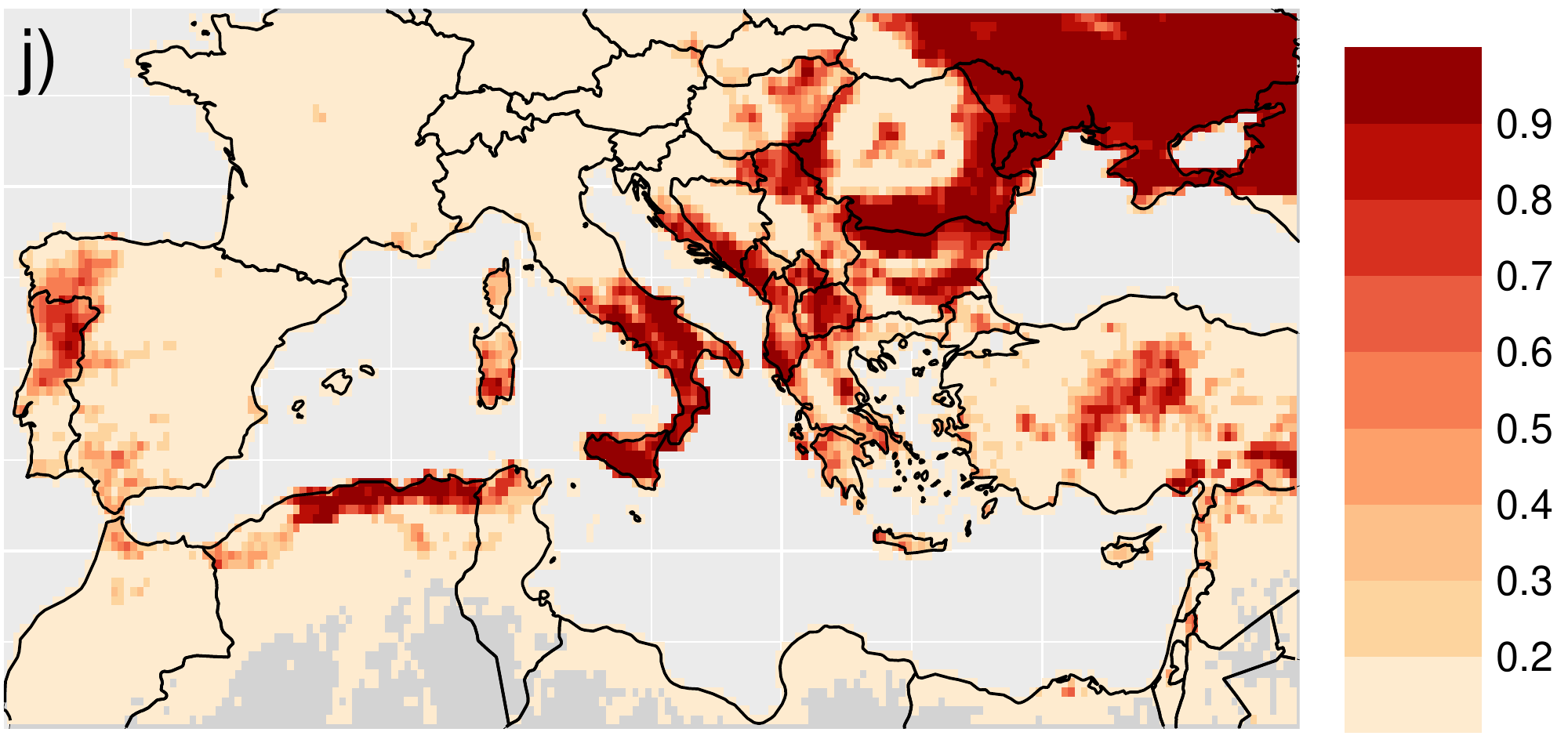} 
\end{minipage}
\begin{minipage}{0.32\linewidth}
\centering
\includegraphics[width=\linewidth]{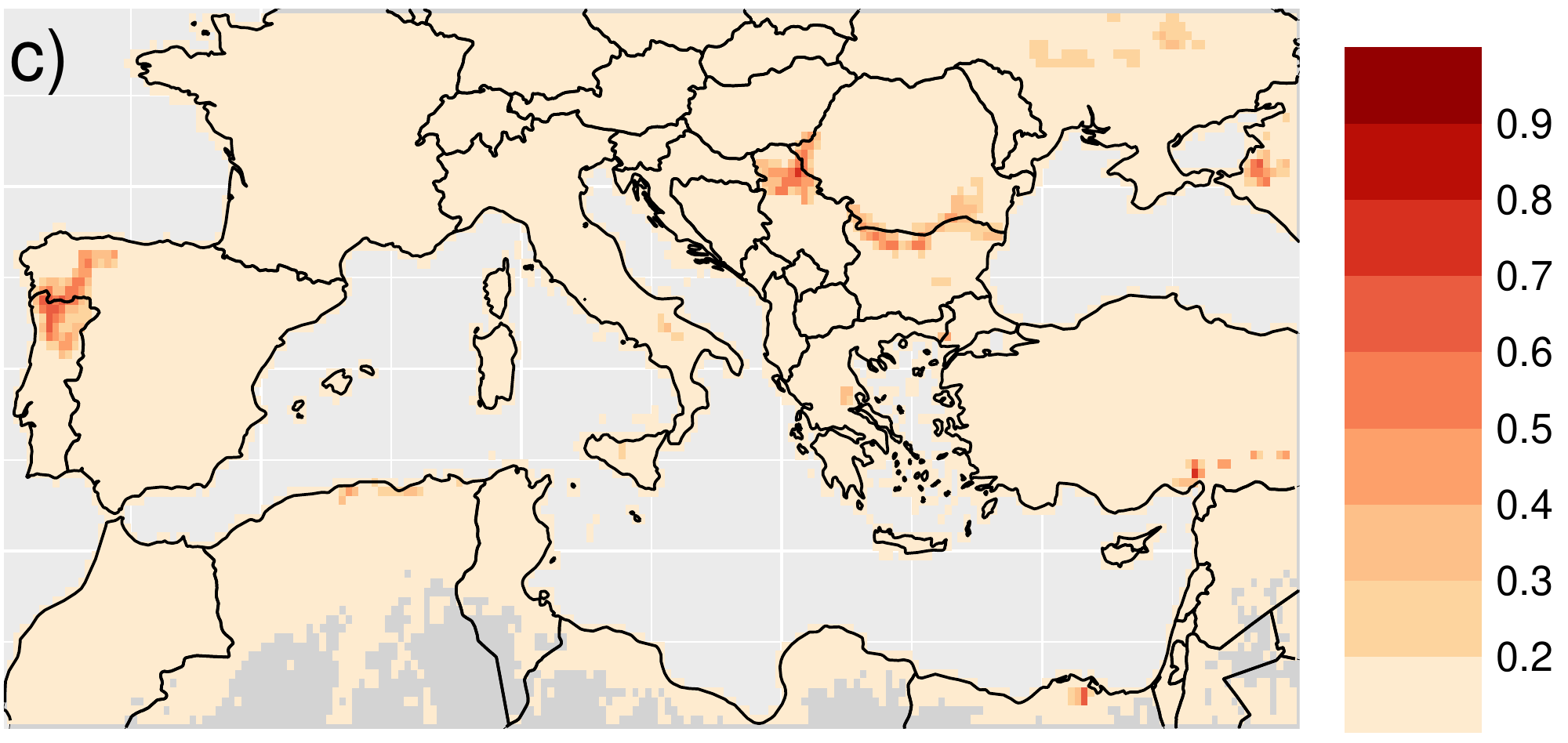} 
\end{minipage}
\begin{minipage}{0.32\linewidth}
\centering
\includegraphics[width=\linewidth]{Images/pZeroMap_med_t101_version5_overlay.pdf} 
\end{minipage}
\begin{minipage}{0.32\linewidth}
\centering
\includegraphics[width=\linewidth]{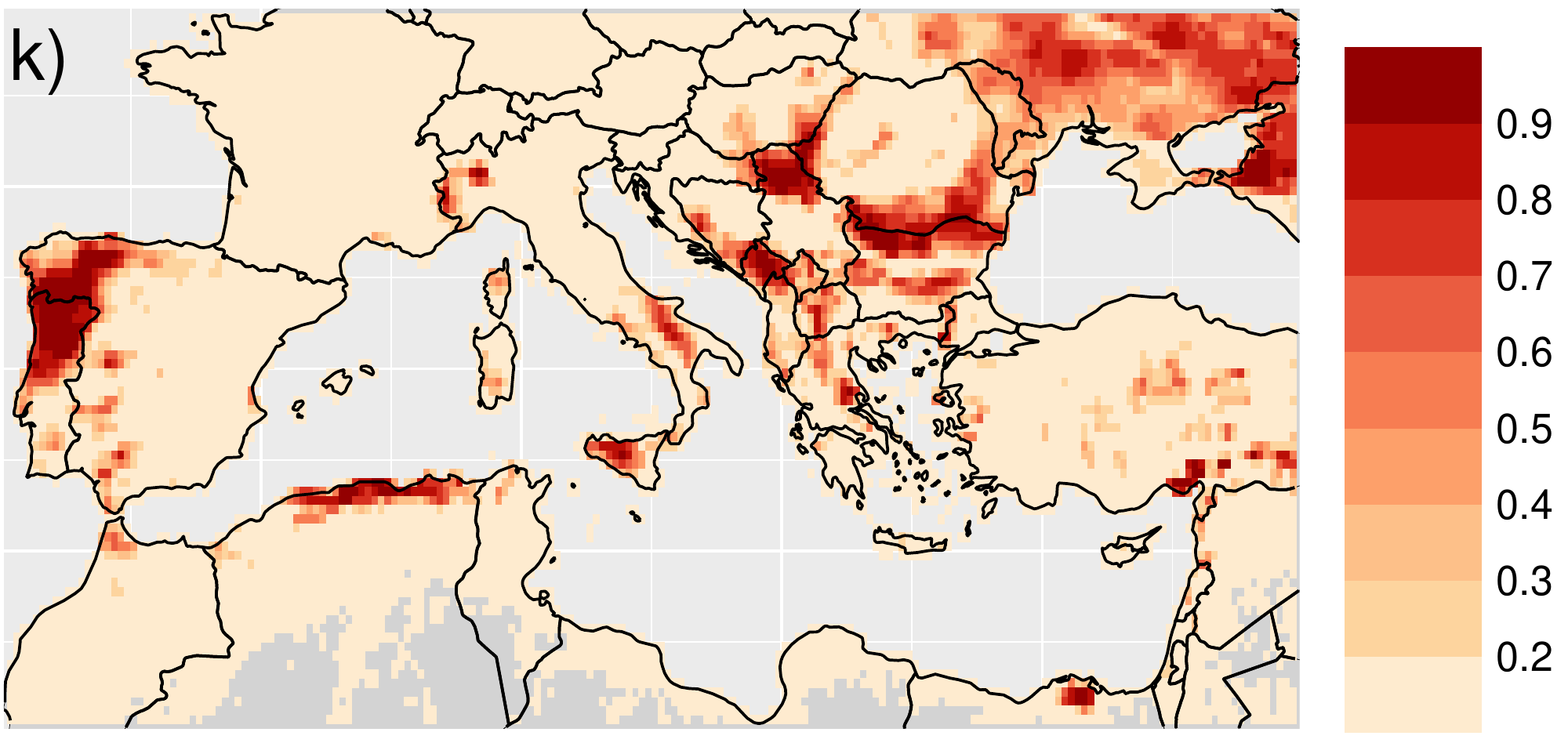} 
\end{minipage}
\begin{minipage}{0.32\linewidth}
\centering
\includegraphics[width=\linewidth]{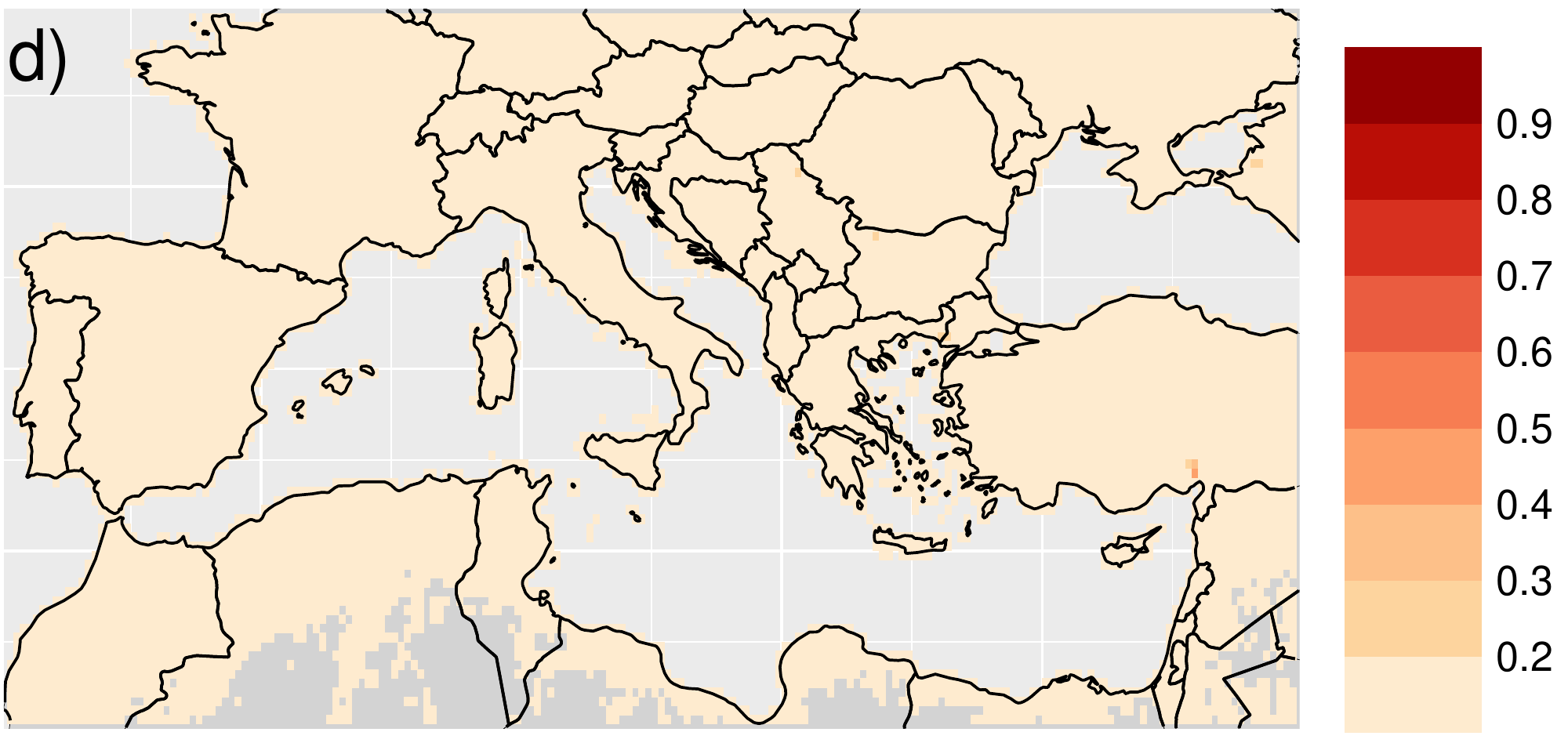} 
\end{minipage}
\begin{minipage}{0.32\linewidth}
\centering
\includegraphics[width=\linewidth]{Images/pZeroMap_med_t120_version5_overlay.pdf} 
\end{minipage}
\begin{minipage}{0.32\linewidth}
\centering
\includegraphics[width=\linewidth]{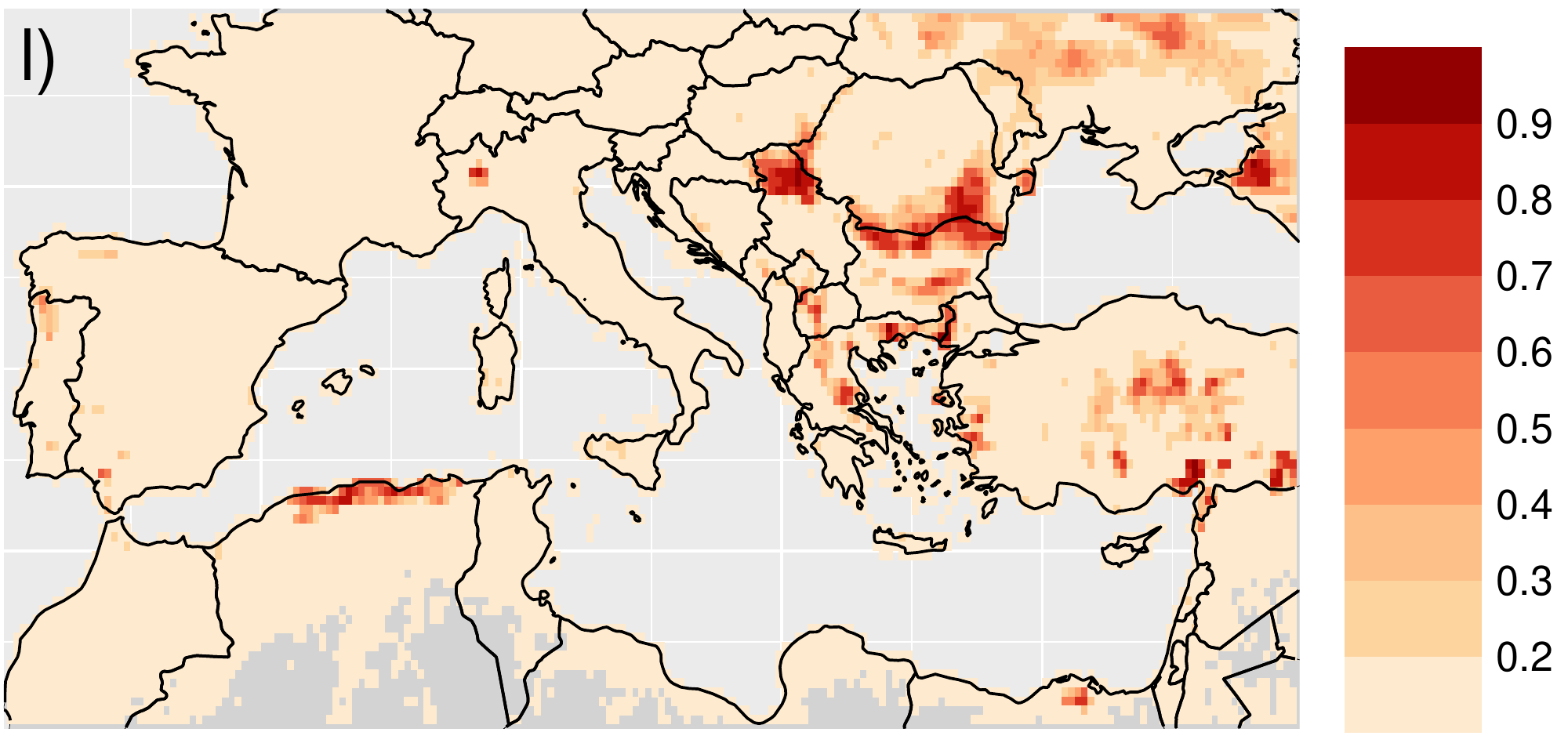} 
\end{minipage}
\caption{
First row, a) $2.5\%$, e) $50\%$, and i) $97.5\%$, bootstrap quantiles of estimated {fire occurence probability} $p_0(s,t)$ [unitless] for August 2001. The second, third, and fourth, rows are as the first row, but for August 2008, October 2017, and  November 2020, respectively.}
\label{p0_uncertain}
\end{figure}
\end{landscape}
\begin{landscape}

\begin{figure}[t!]
\centering
\begin{minipage}{0.32\linewidth}
\centering
\includegraphics[width=\linewidth]{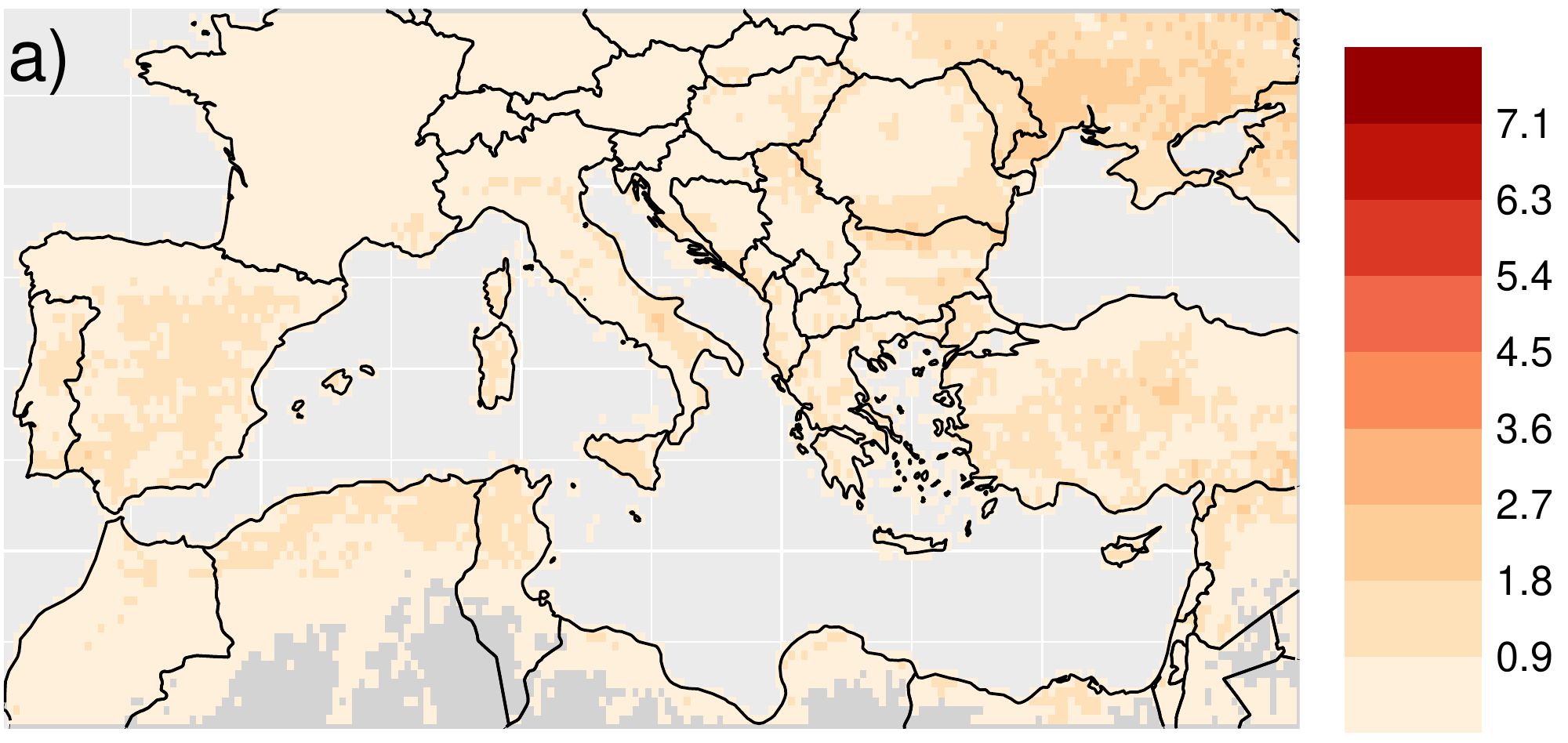} 
\end{minipage}
\begin{minipage}{0.32\linewidth}
\centering
\includegraphics[width=\linewidth]{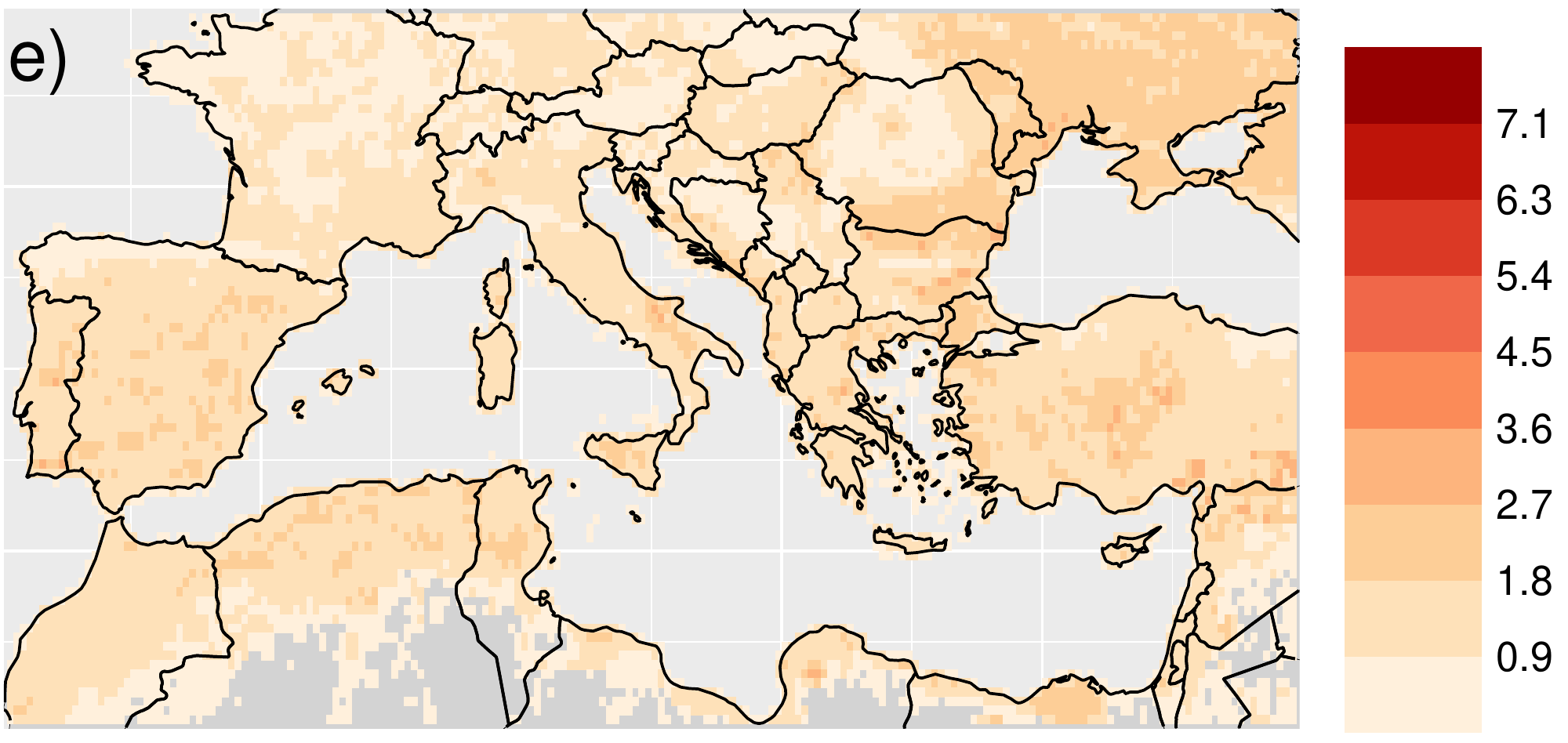} 
\end{minipage}
\begin{minipage}{0.32\linewidth}
\centering
\includegraphics[width=\linewidth]{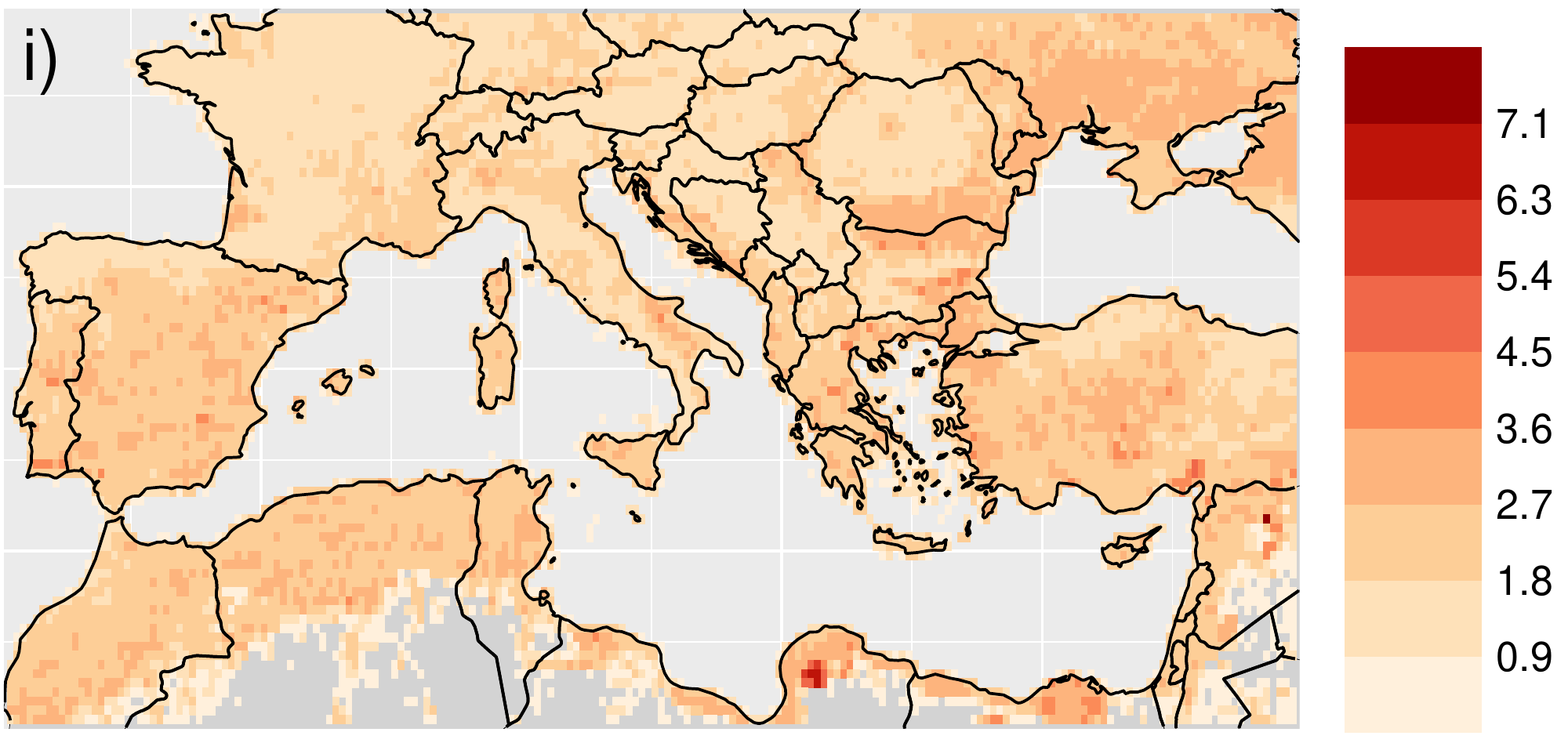} 
\end{minipage}
\begin{minipage}{0.32\linewidth}
\centering
\includegraphics[width=\linewidth]{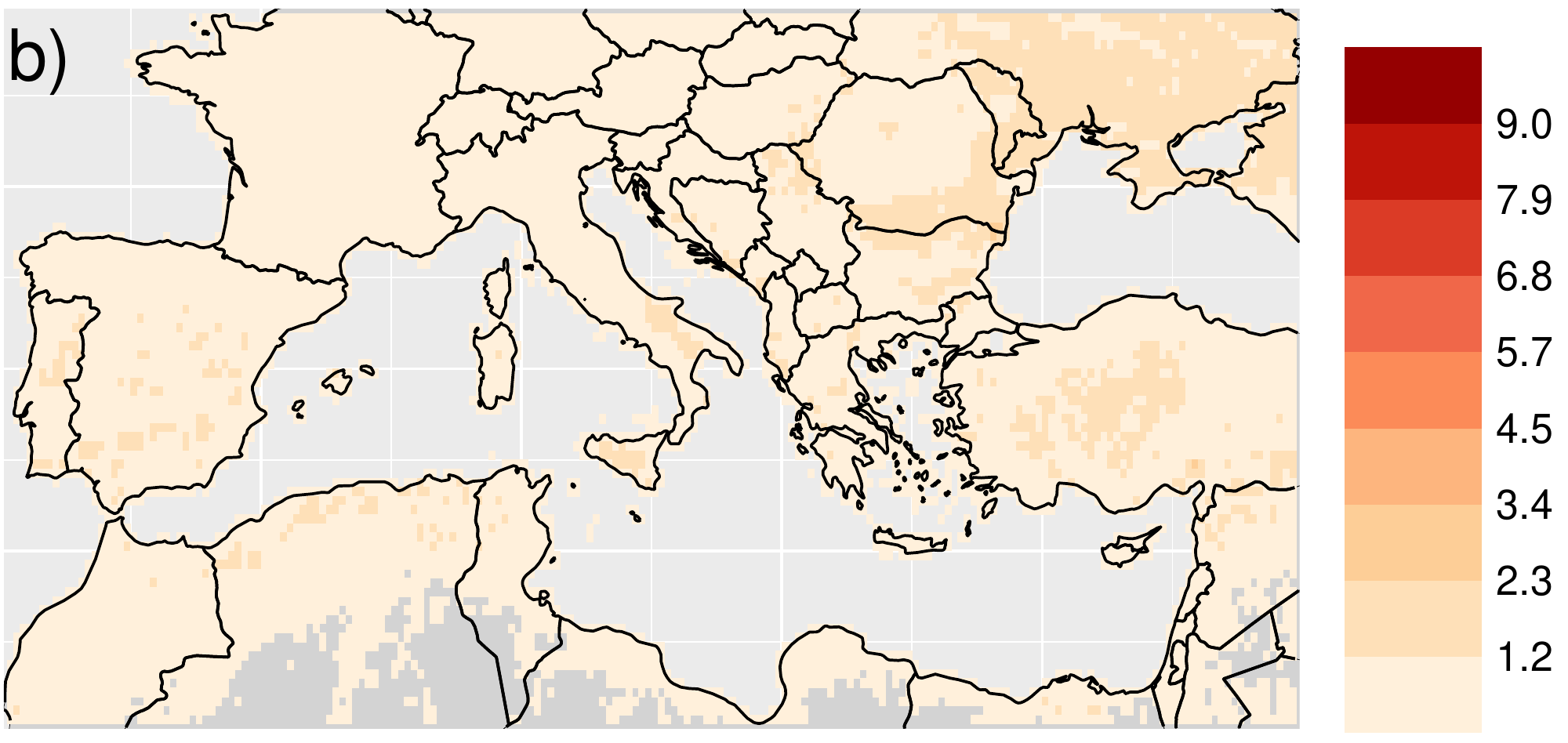} 
\end{minipage}
\begin{minipage}{0.32\linewidth}
\centering
\includegraphics[width=\linewidth]{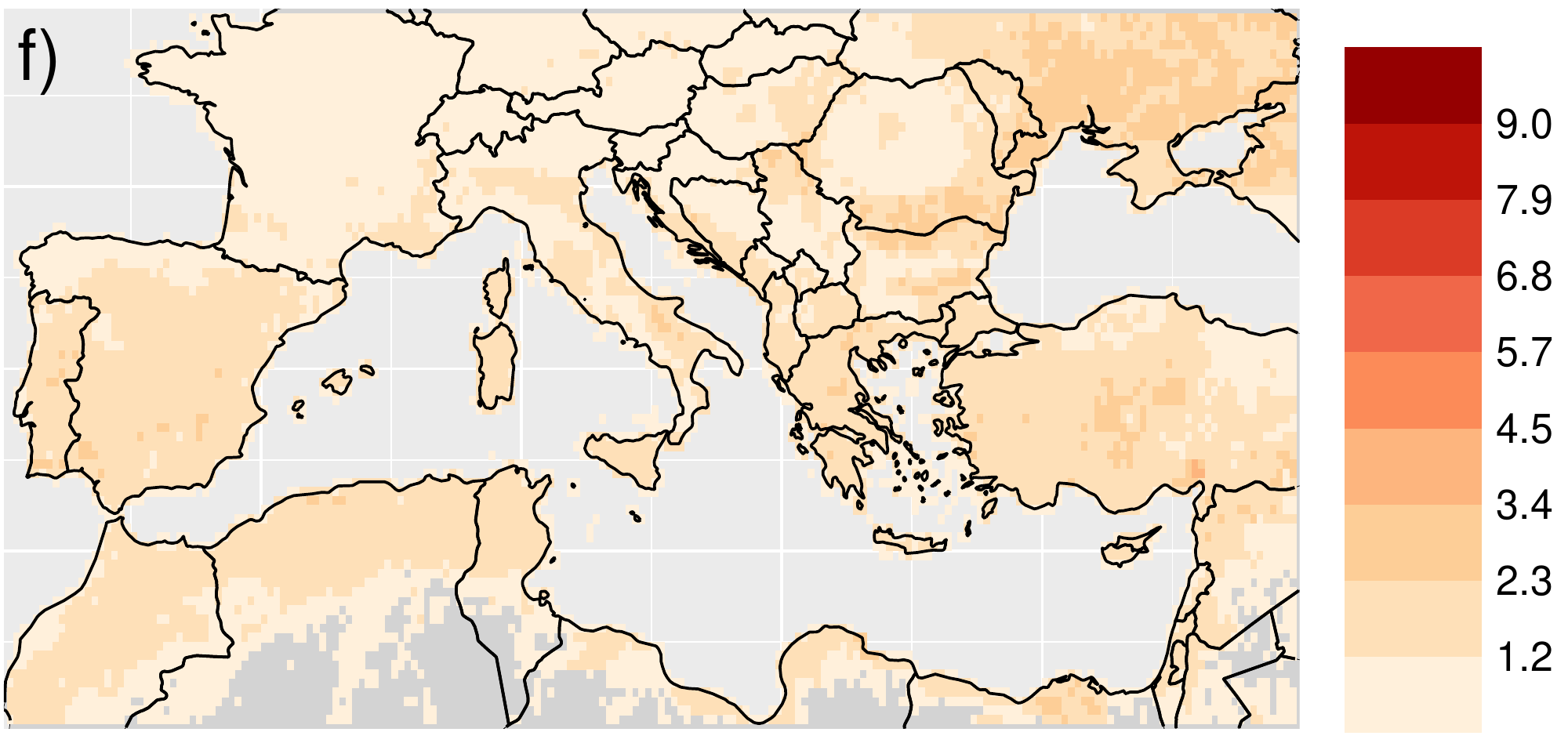} 
\end{minipage}
\begin{minipage}{0.32\linewidth}
\centering
\includegraphics[width=\linewidth]{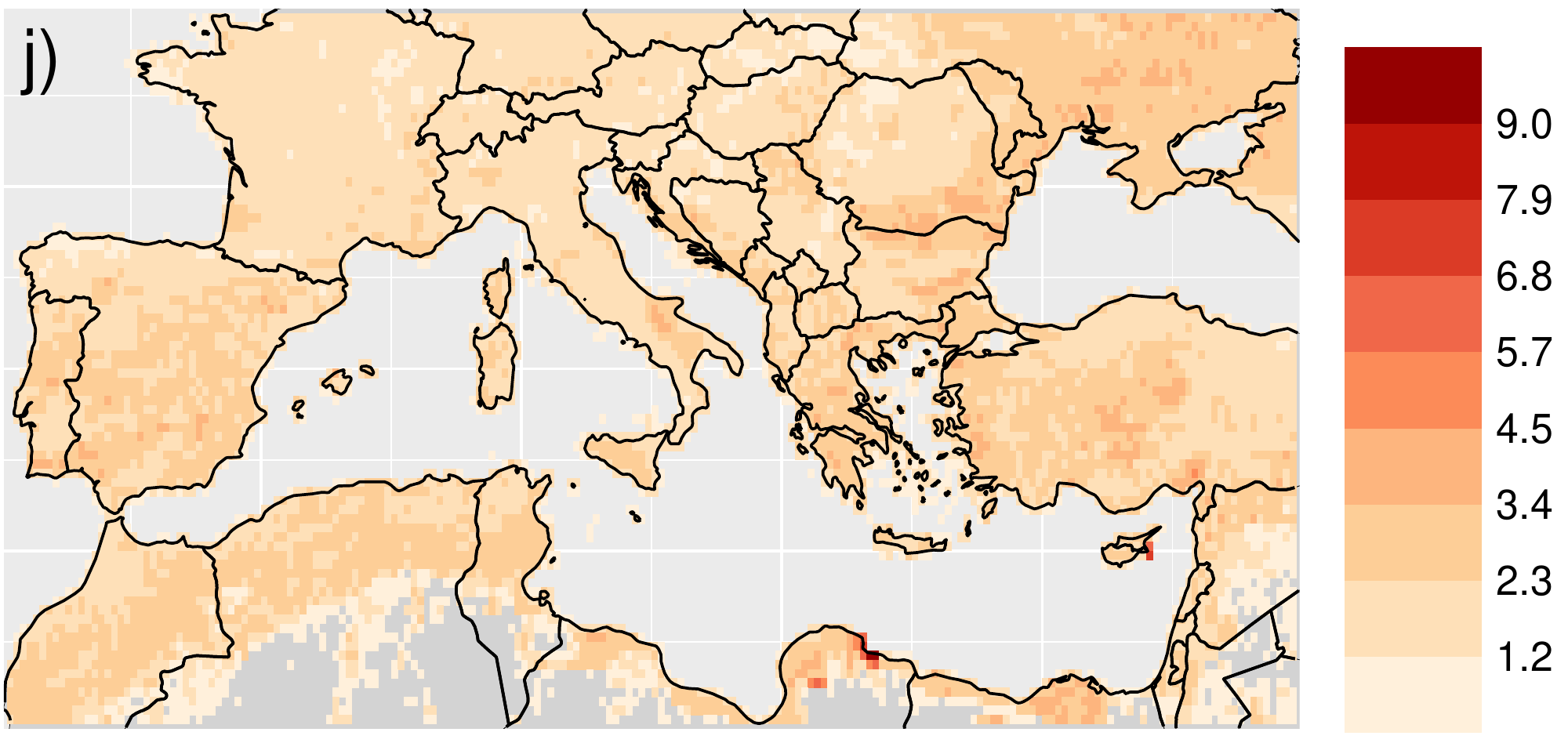} 
\end{minipage}
\begin{minipage}{0.32\linewidth}
\centering
\includegraphics[width=\linewidth]{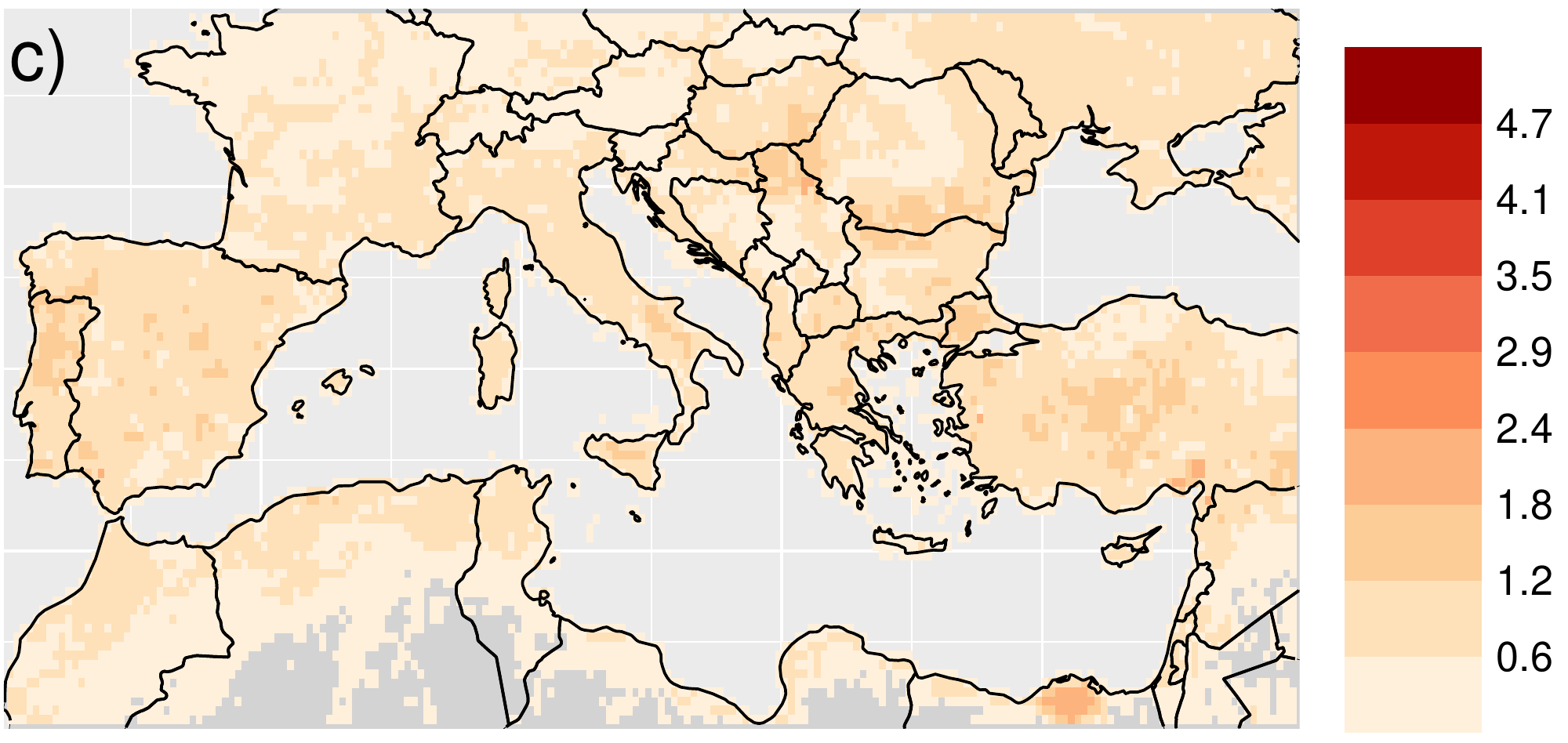} 
\end{minipage}
\begin{minipage}{0.32\linewidth}
\centering
\includegraphics[width=\linewidth]{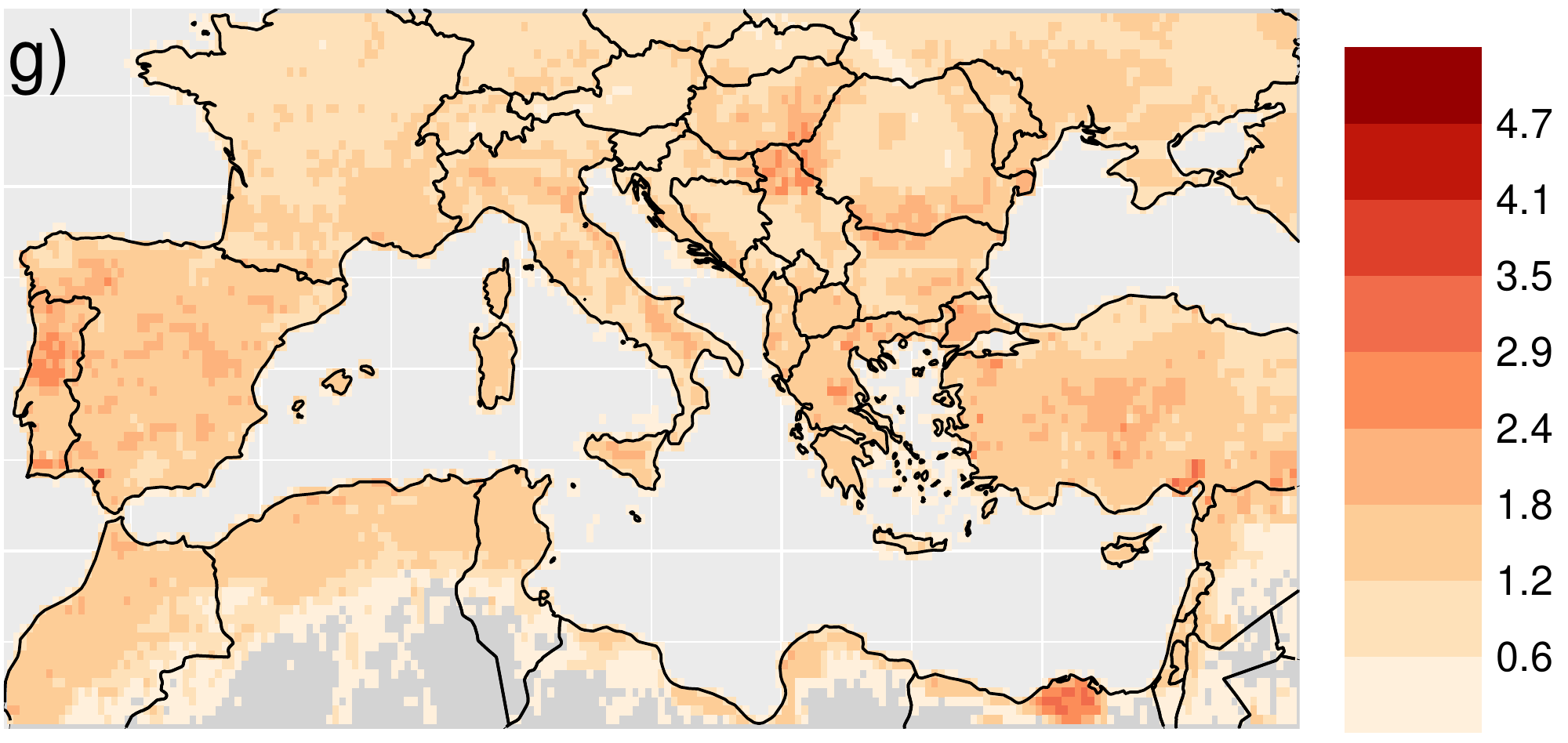} 
\end{minipage}
\begin{minipage}{0.32\linewidth}
\centering
\includegraphics[width=\linewidth]{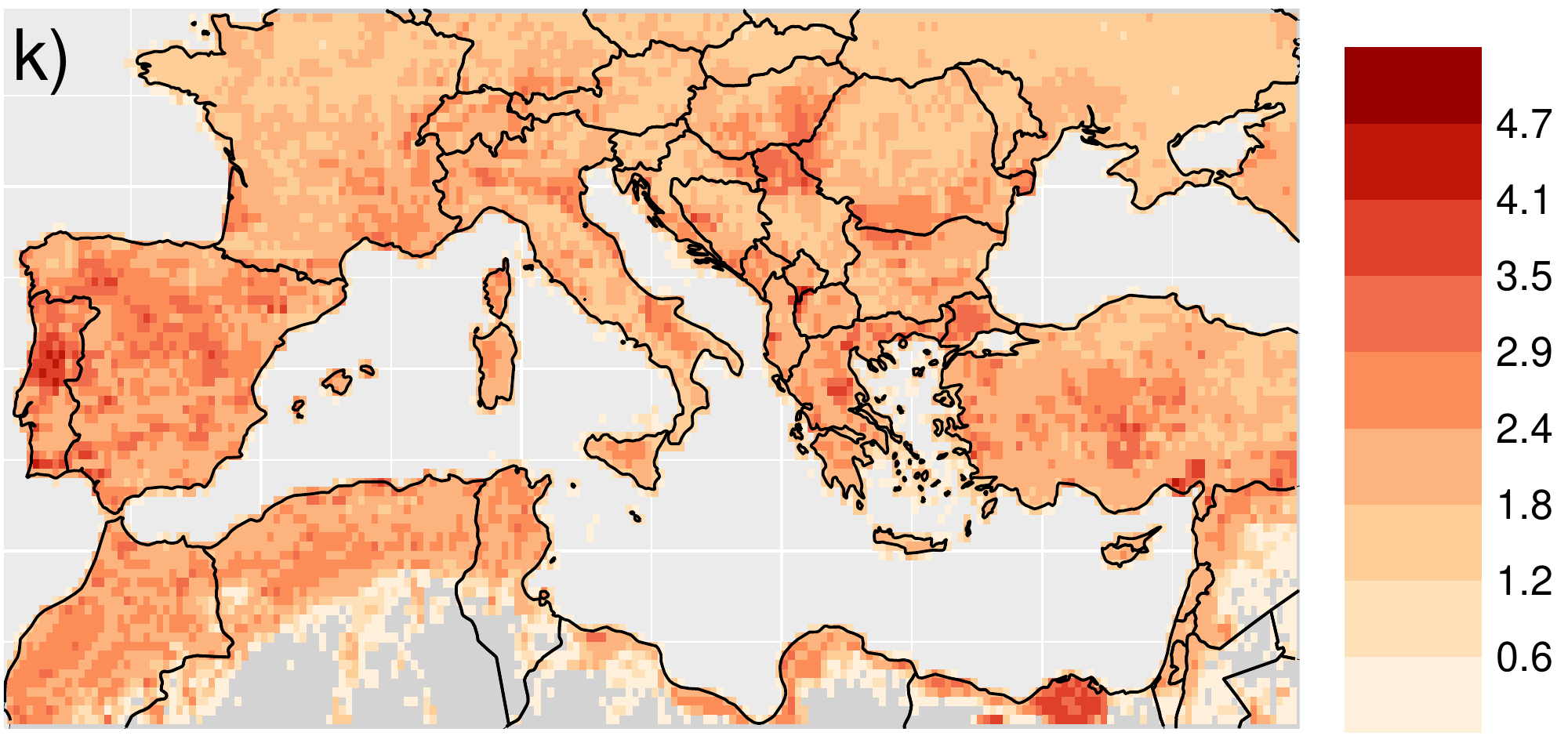} 
\end{minipage}
\begin{minipage}{0.32\linewidth}
\centering
\includegraphics[width=\linewidth]{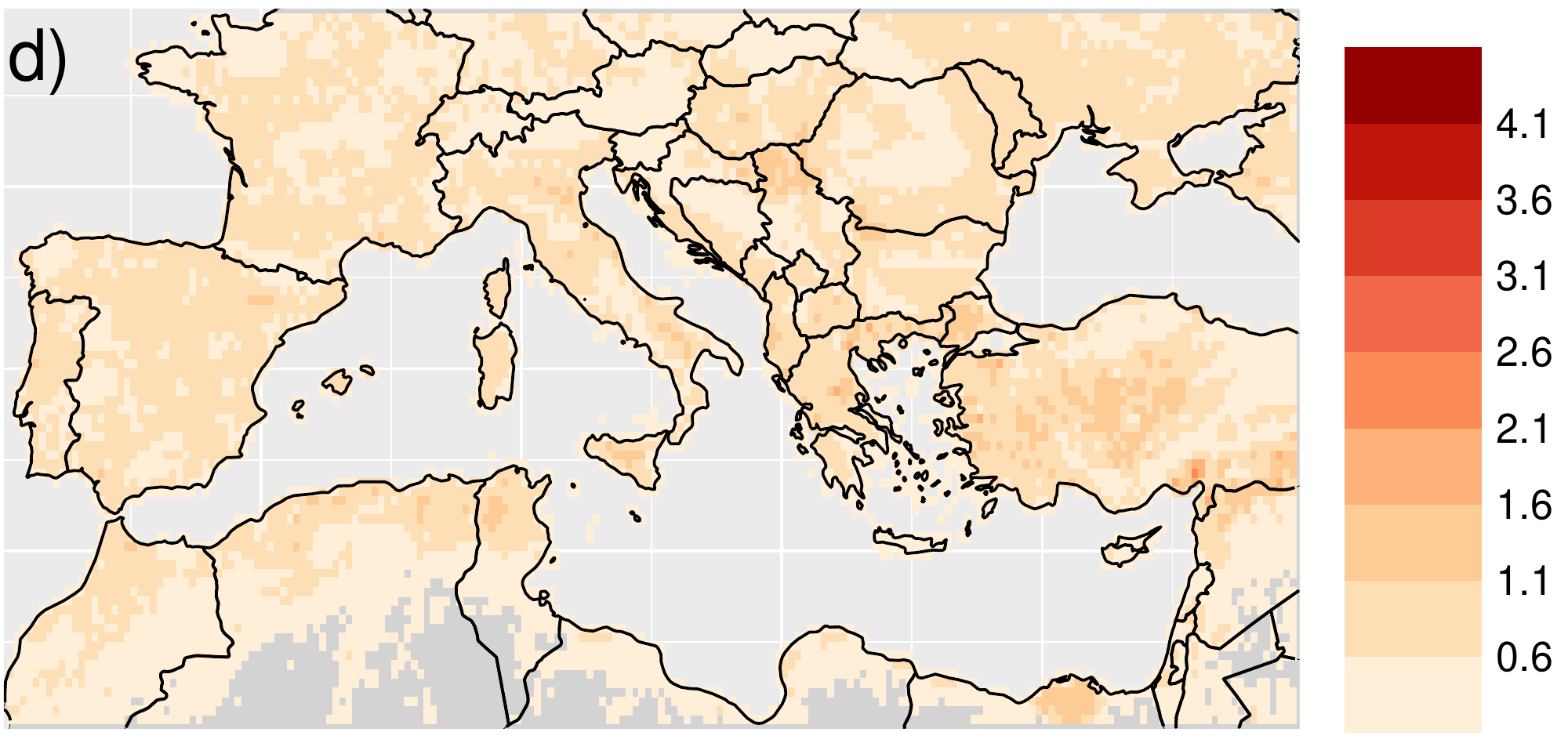} 
\end{minipage}
\begin{minipage}{0.32\linewidth}
\centering
\includegraphics[width=\linewidth]{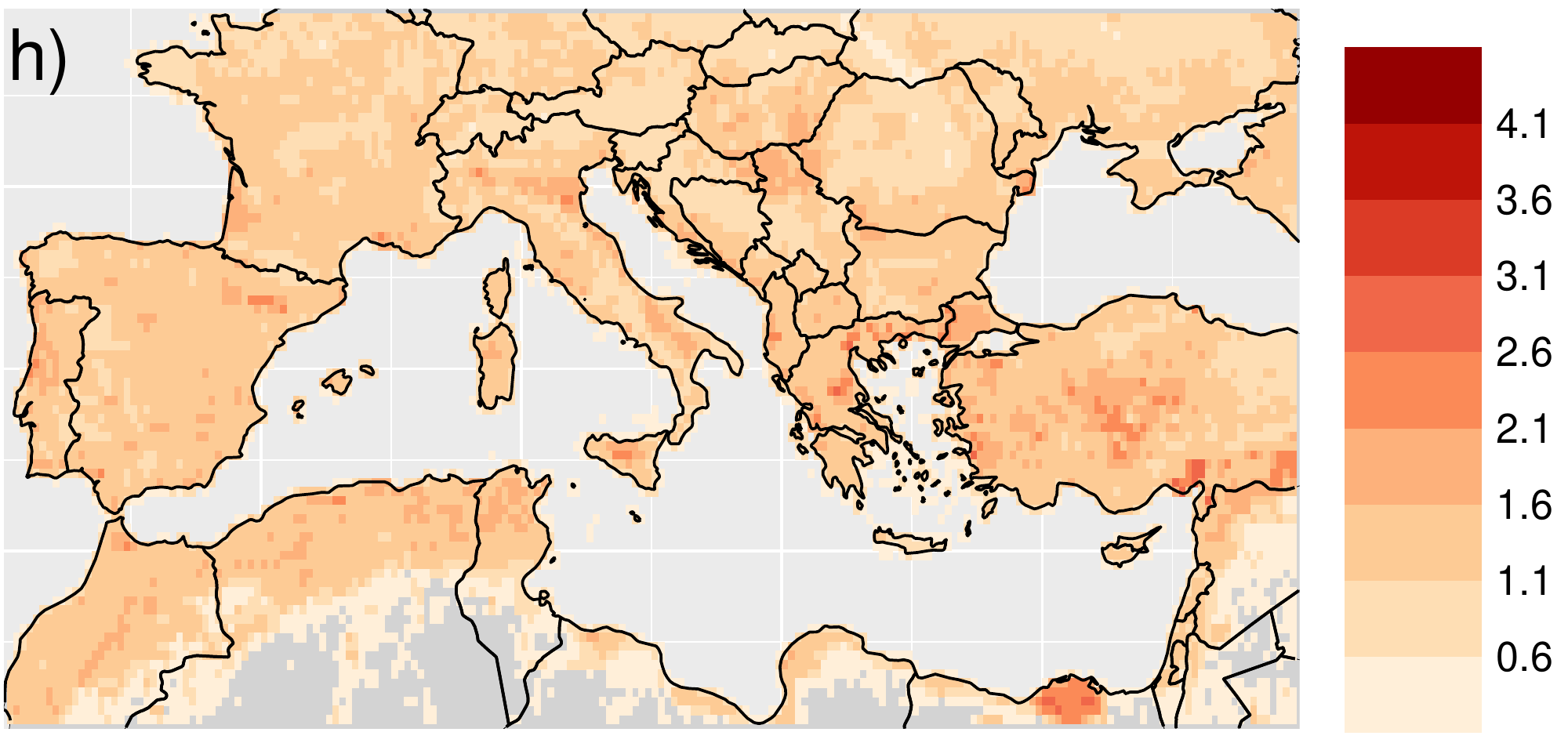} 
\end{minipage}
\begin{minipage}{0.32\linewidth}
\centering
\includegraphics[width=\linewidth]{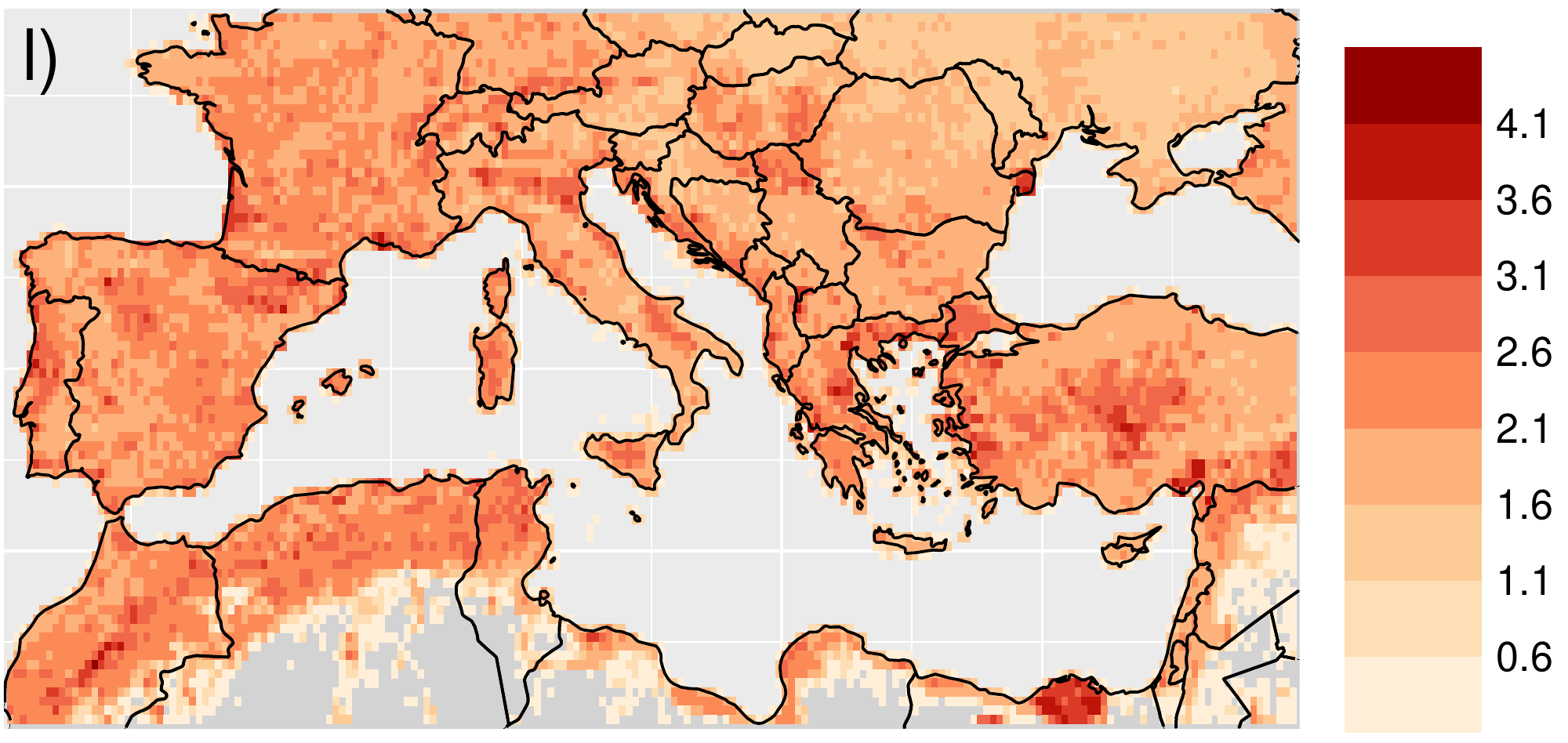} 
\end{minipage}
\caption{First row, a) $2.5\%$, e) $50\%$ and i) $97.5\%$ bootstrap quantiles of estimated $\log\{1+\sigma(s,t)\}$ [conditional spread severity; $\log(\mbox{km}^2)$] for August 2001. The second, third, and fourth, rows are as the first row, but for August 2008, October 2017, and  November 2020, respectively.
}
\label{sig_uncertain}
\end{figure}
\end{landscape}
\begin{landscape}

\begin{figure}[t!]
\centering
\begin{minipage}{0.32\linewidth}
\centering
\includegraphics[width=\linewidth]{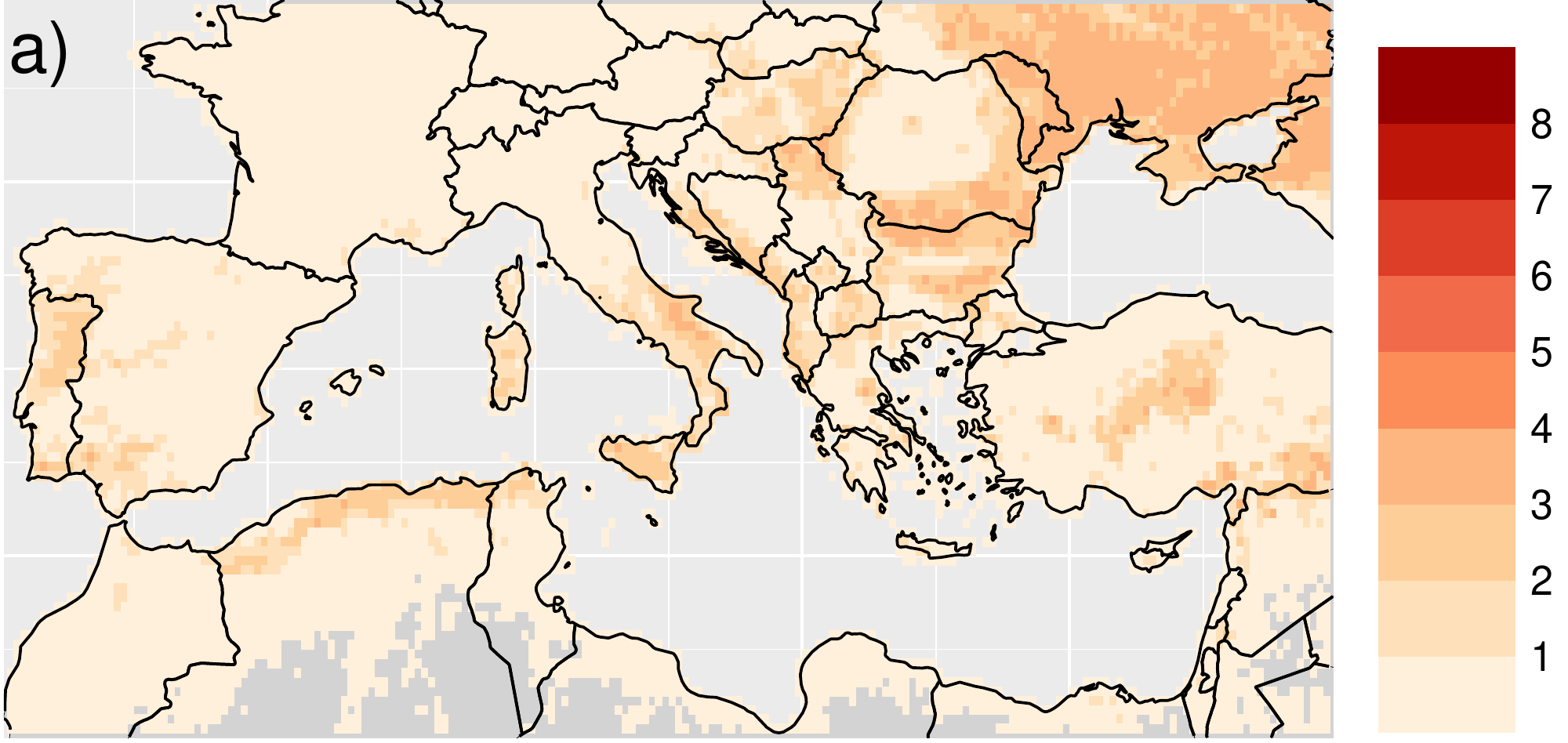} 
\end{minipage}
\begin{minipage}{0.32\linewidth}
\centering
\includegraphics[width=\linewidth]{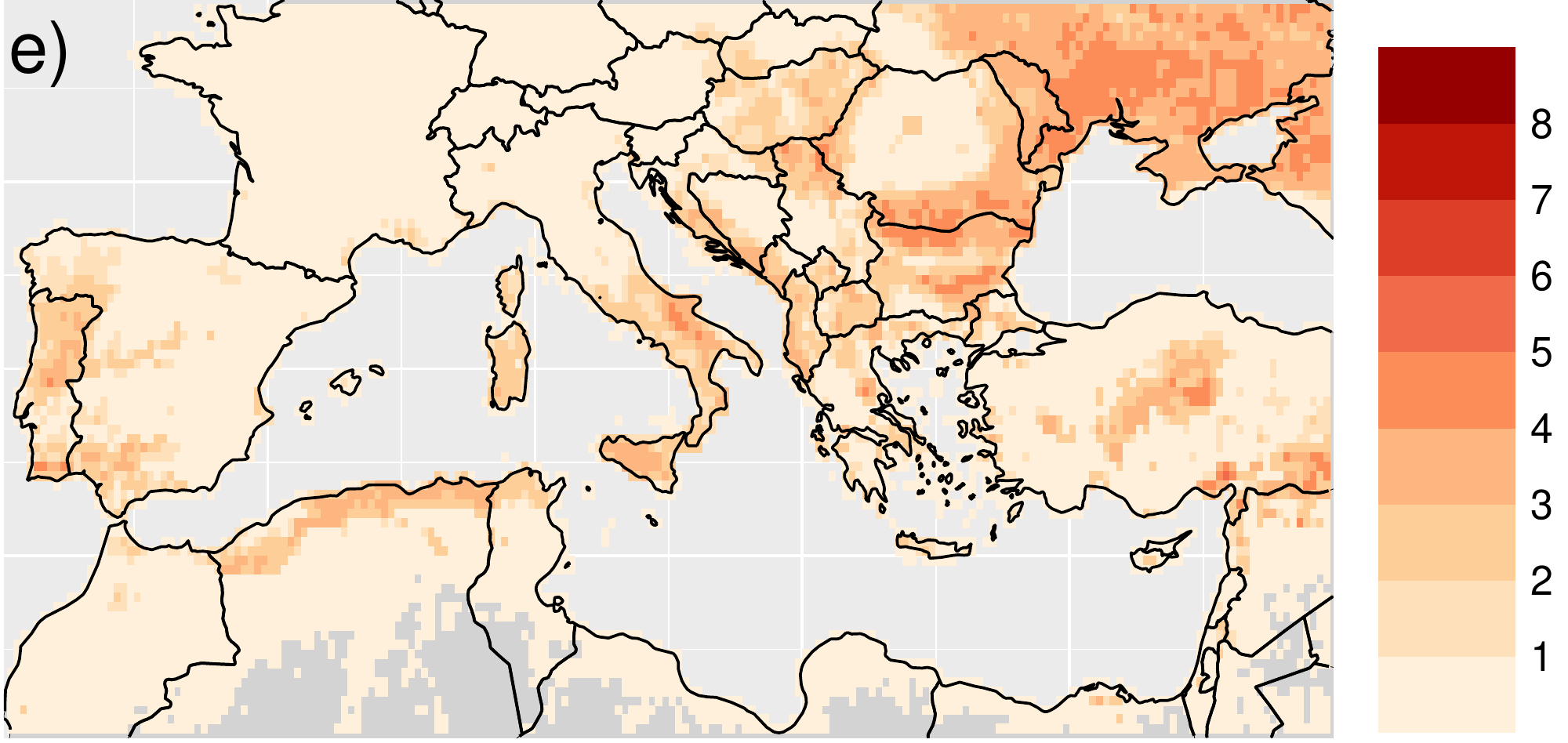} 
\end{minipage}
\begin{minipage}{0.32\linewidth}
\centering
\includegraphics[width=\linewidth]{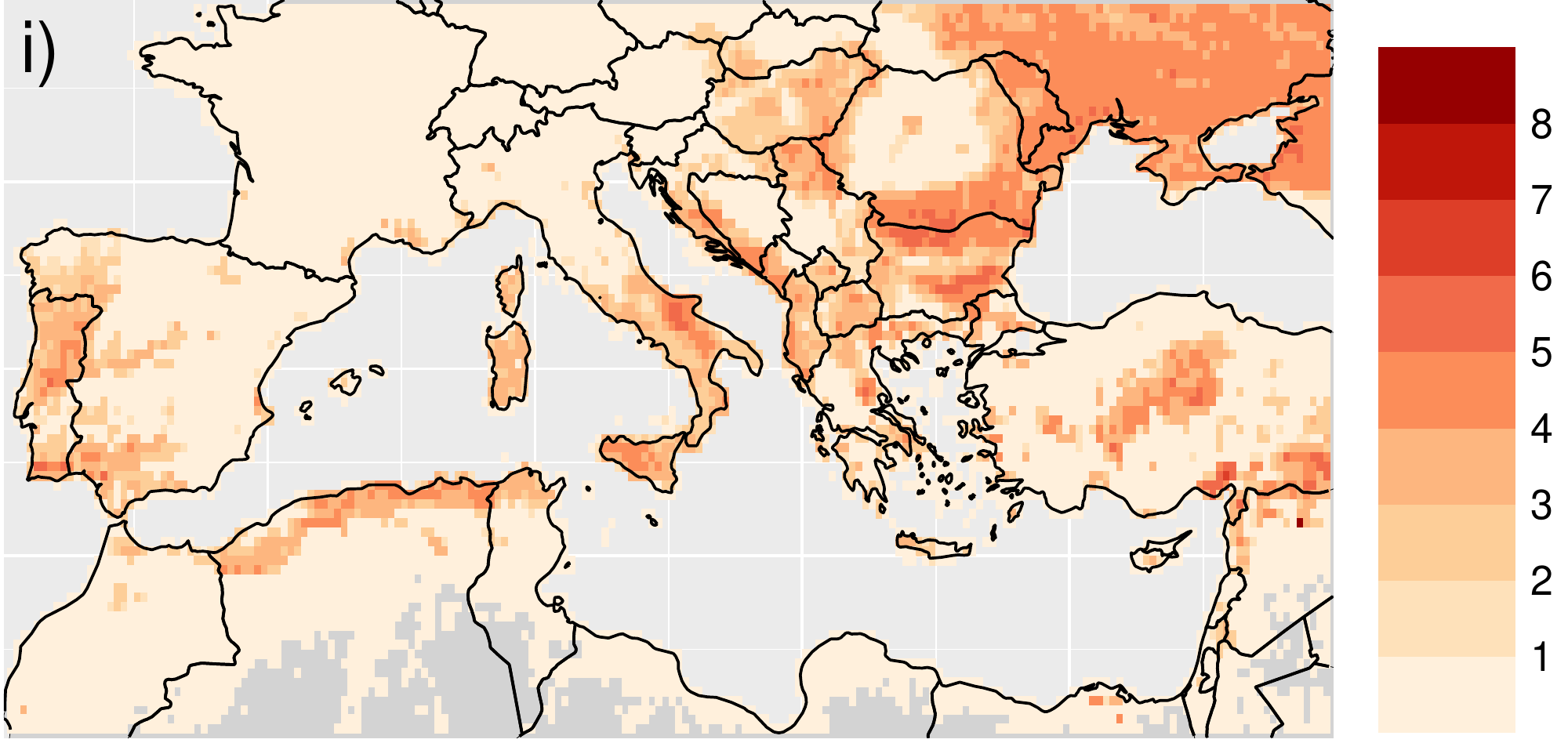} 
\end{minipage}
\begin{minipage}{0.32\linewidth}
\centering
\includegraphics[width=\linewidth]{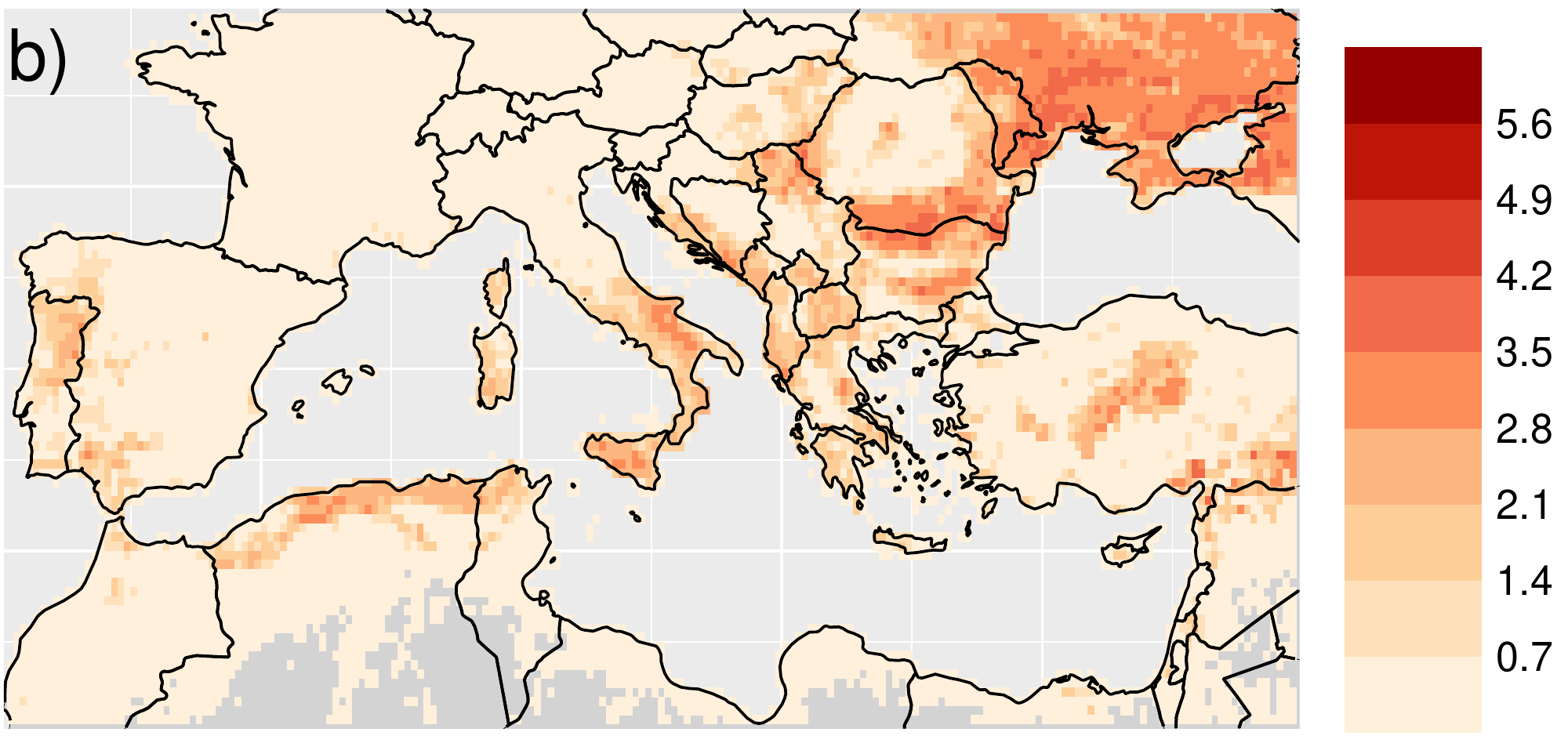} 
\end{minipage}
\begin{minipage}{0.32\linewidth}
\centering
\includegraphics[width=\linewidth]{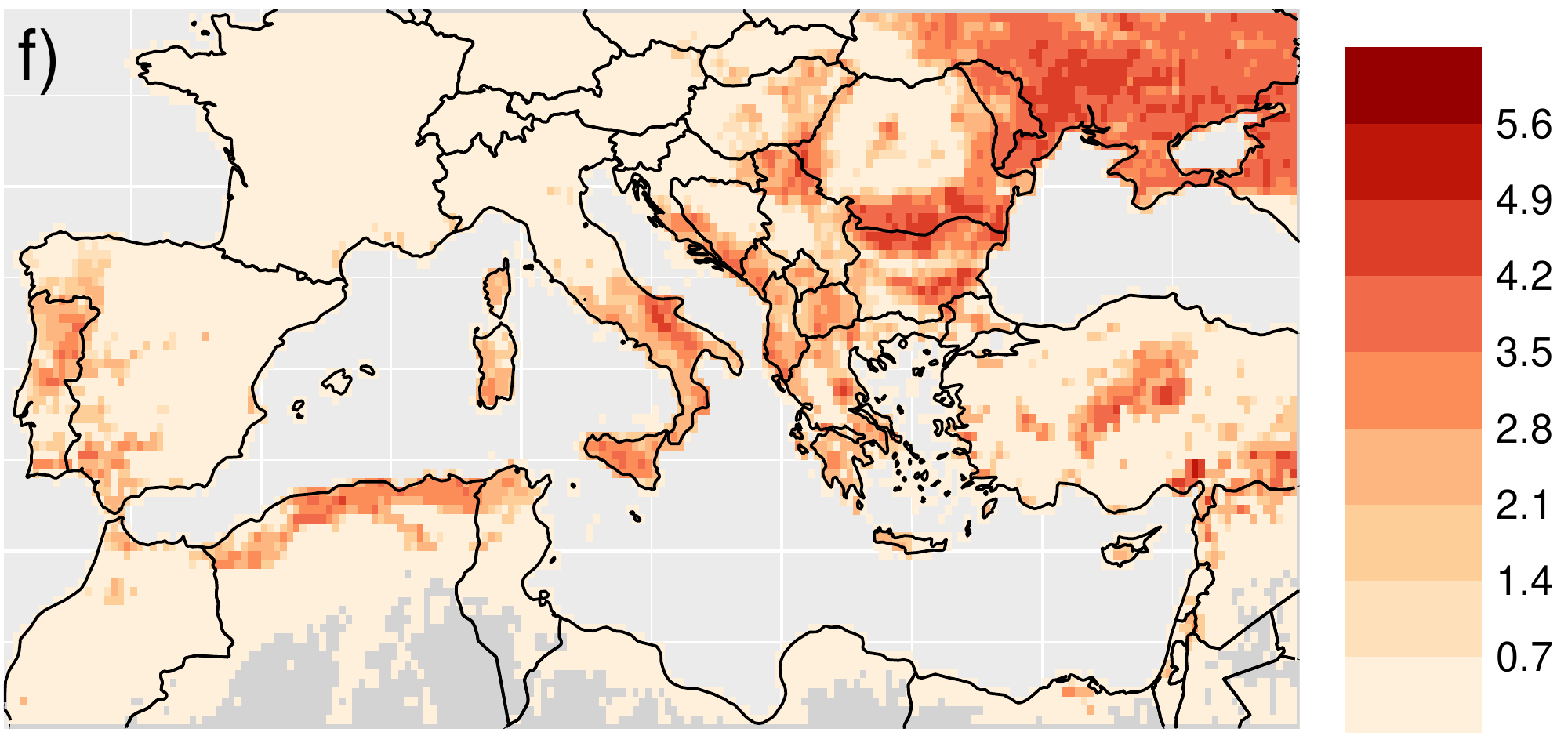} 
\end{minipage}
\begin{minipage}{0.32\linewidth}
\centering
\includegraphics[width=\linewidth]{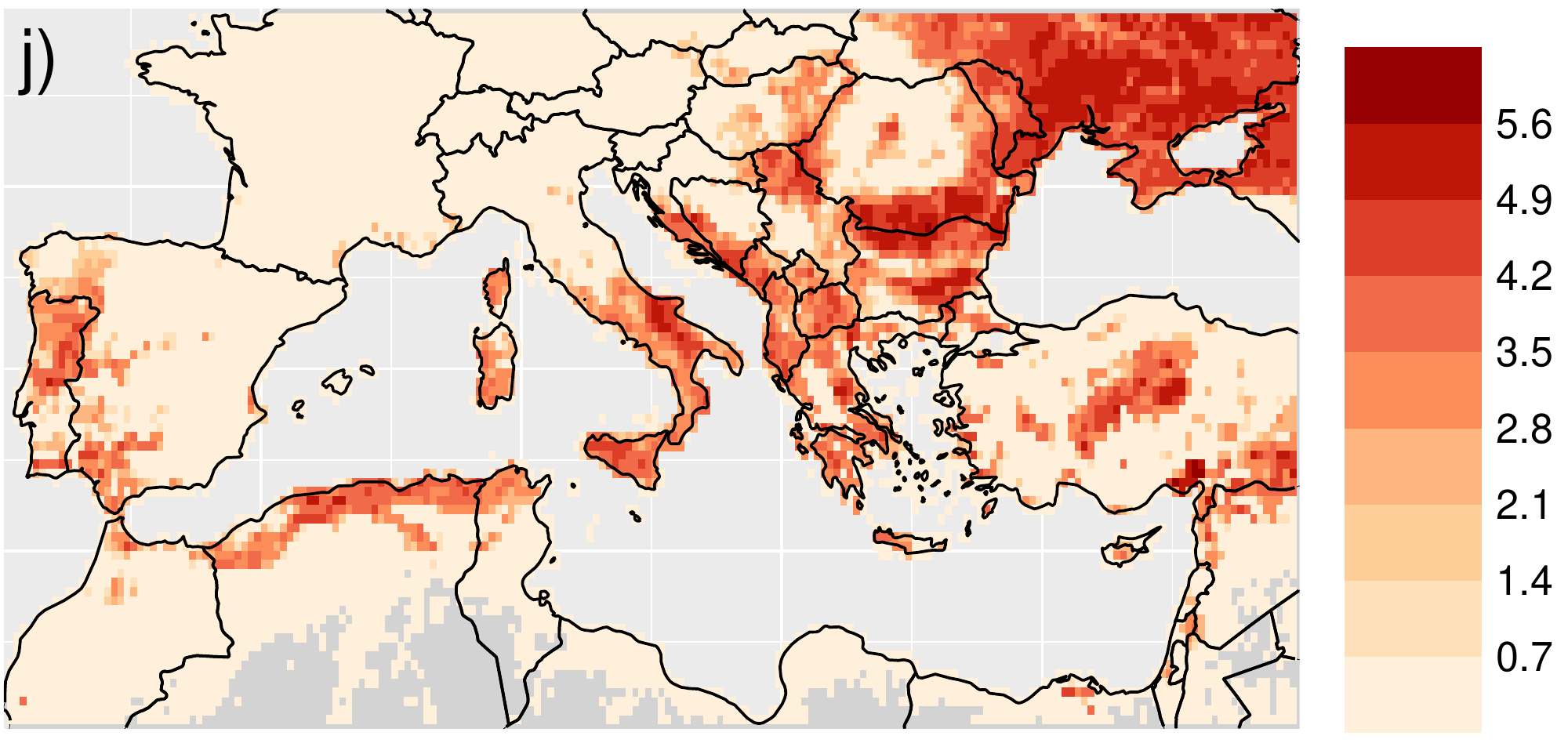} 
\end{minipage}
\begin{minipage}{0.32\linewidth}
\centering
\includegraphics[width=\linewidth]{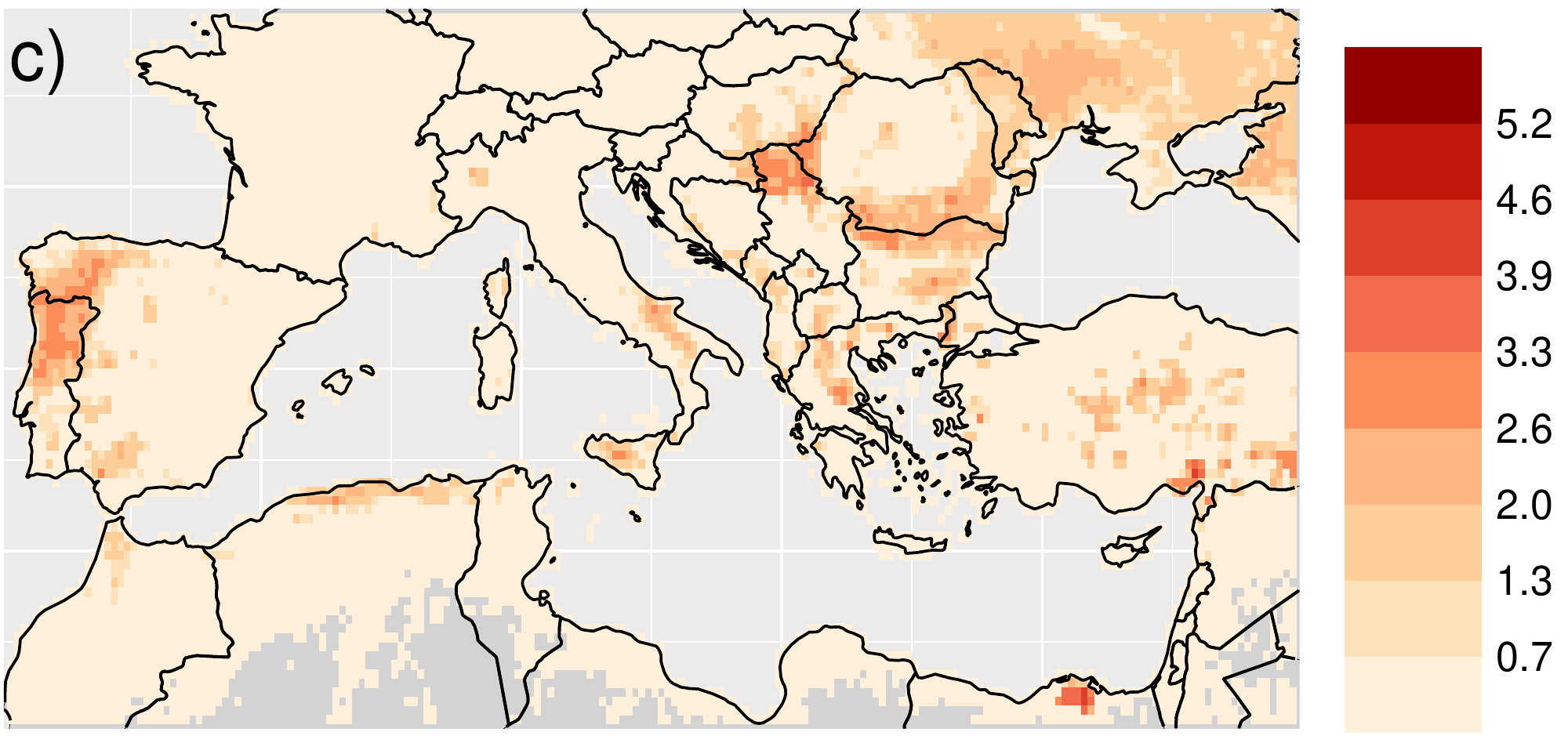} 
\end{minipage}
\begin{minipage}{0.32\linewidth}
\centering
\includegraphics[width=\linewidth]{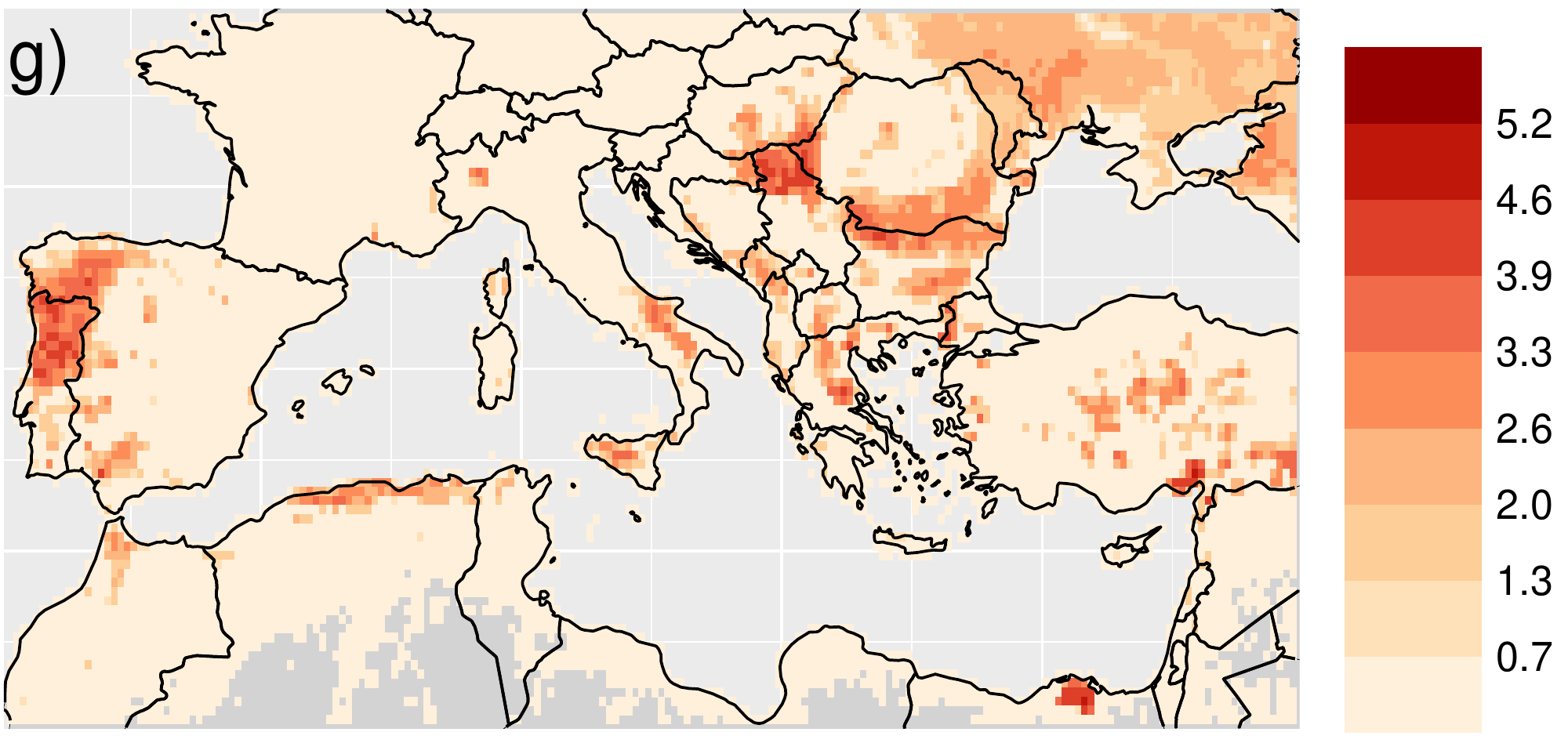} 
\end{minipage}
\begin{minipage}{0.32\linewidth}
\centering
\includegraphics[width=\linewidth]{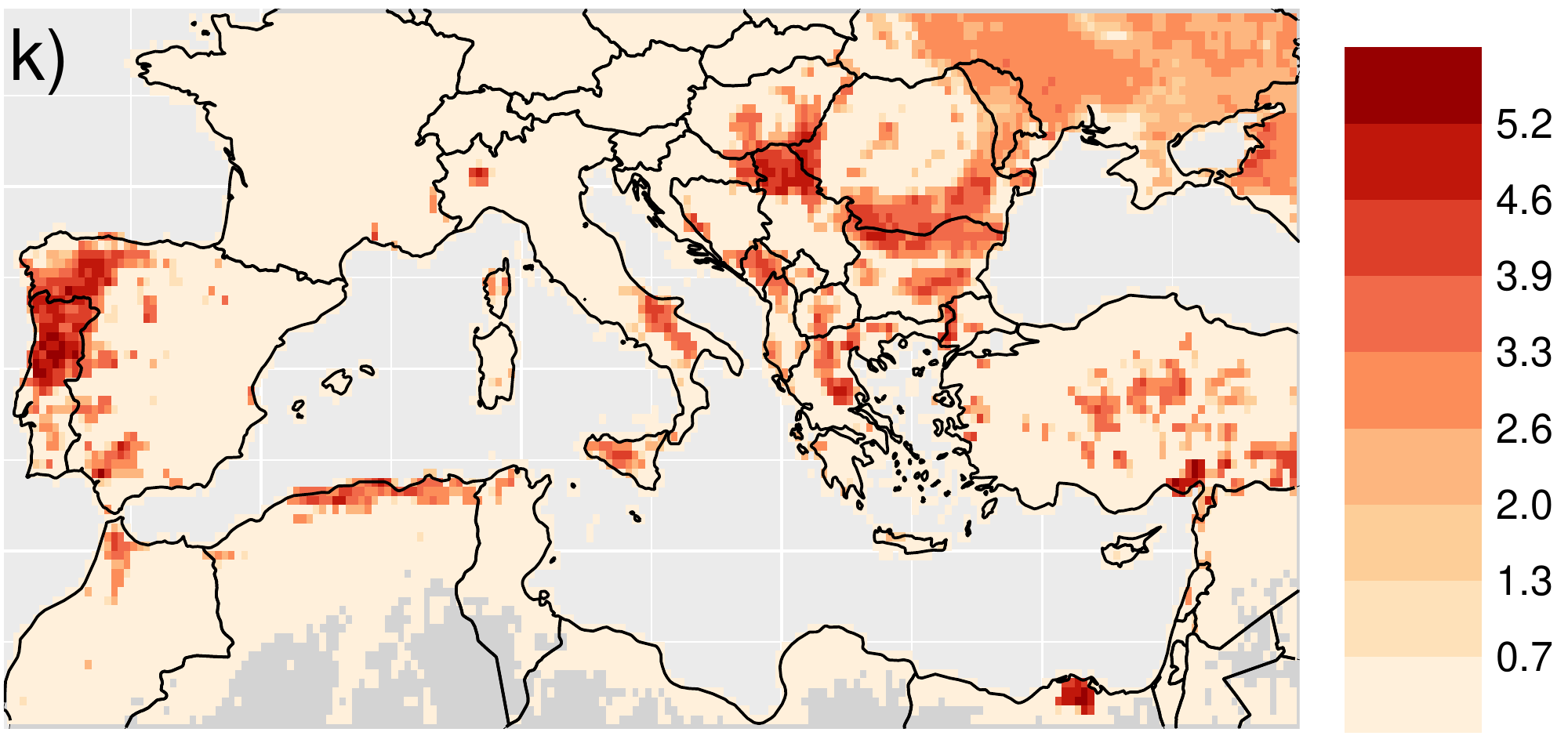} 
\end{minipage}
\begin{minipage}{0.32\linewidth}
\centering
\includegraphics[width=\linewidth]{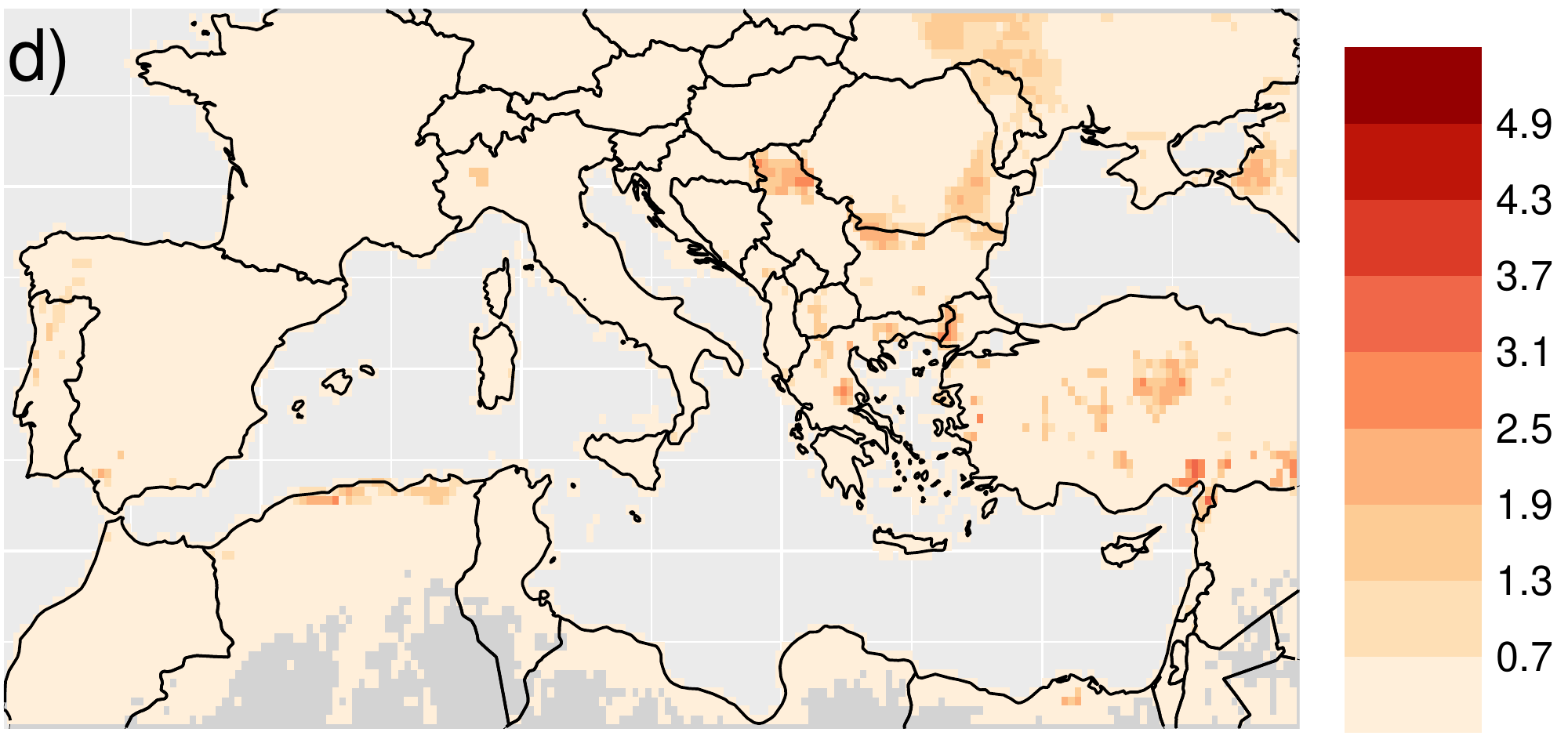} 
\end{minipage}
\begin{minipage}{0.32\linewidth}
\centering
\includegraphics[width=\linewidth]{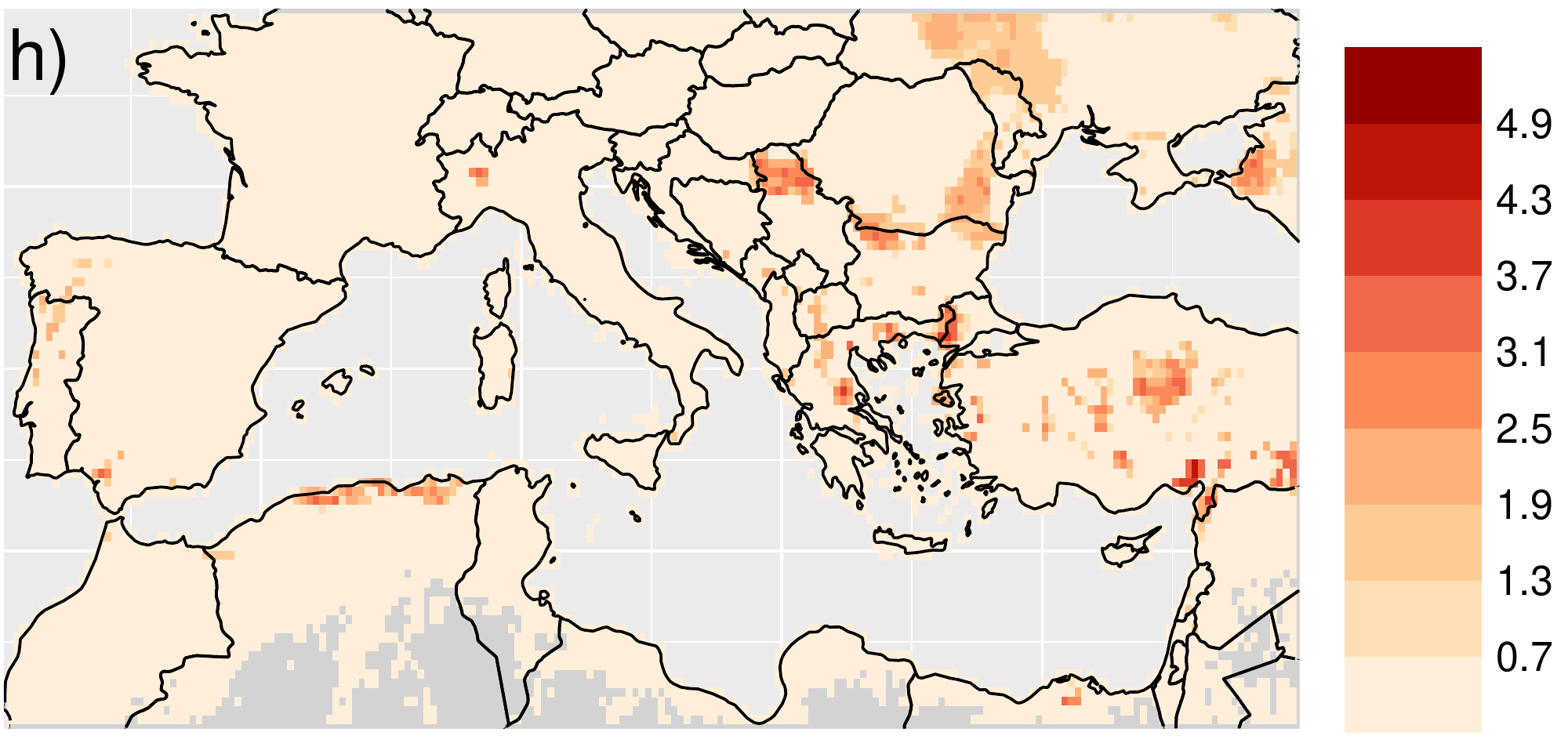} 
\end{minipage}
\begin{minipage}{0.32\linewidth}
\centering
\includegraphics[width=\linewidth]{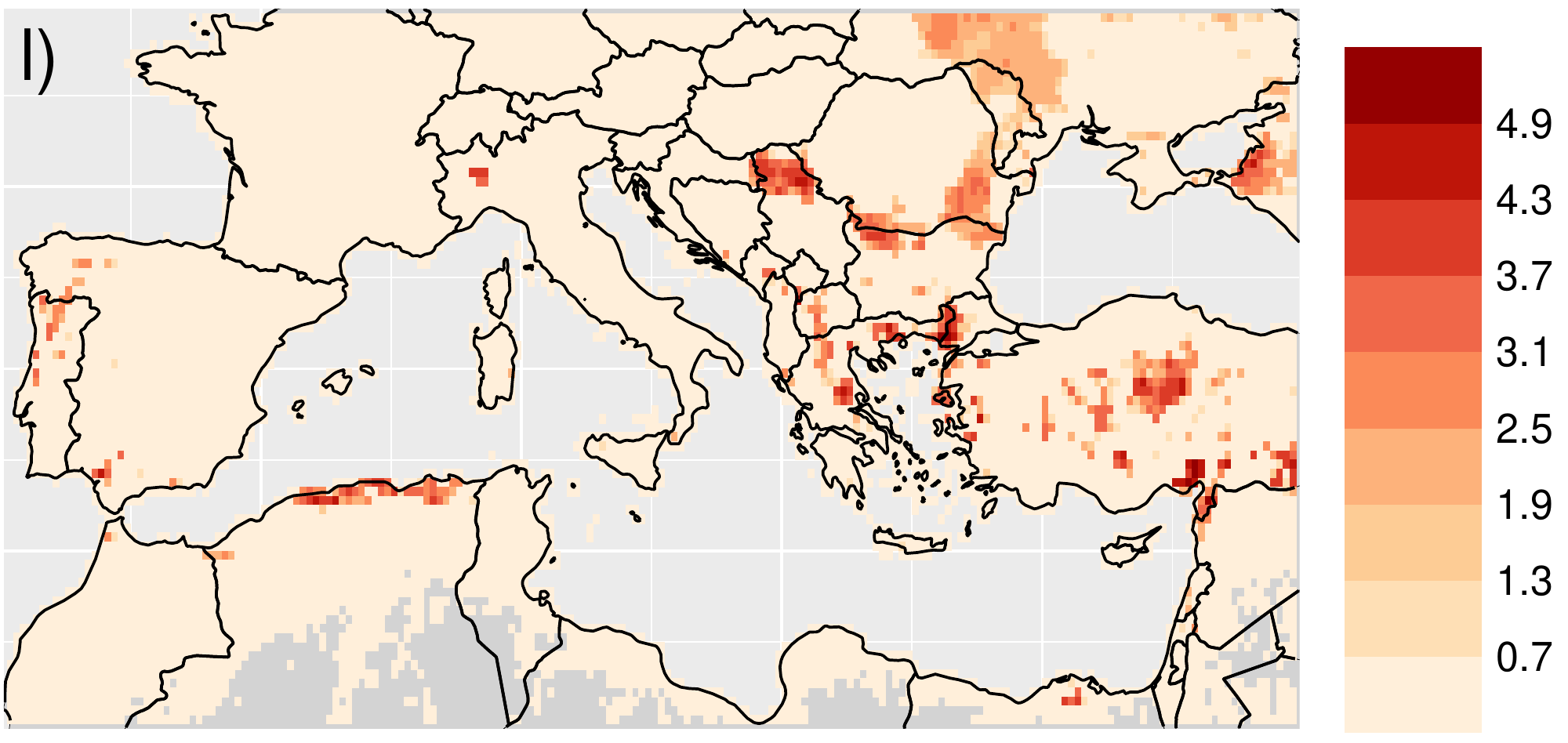} 
\end{minipage}
\caption{First row, a) $2.5\%$, e) $50\%$ and i) $97.5\%$ bootstrap quantiles of estimated $95\%$-quantile for $\log\{1+Y(s,t)\}\;|\;\mathbf{X}(s,t)$ [burnt area; $\log(\mbox{km}^2)$] for August 2001. The second, third, and fourth, rows are as the first row, but for August 2008, October 2017, and  November 2020, respectively.
}
\label{cr_uncertain}
\end{figure}
\end{landscape}

\begin{landscape}

\begin{figure}[t!]
\centering
\begin{minipage}{0.32\linewidth}
\centering
\includegraphics[width=\linewidth]{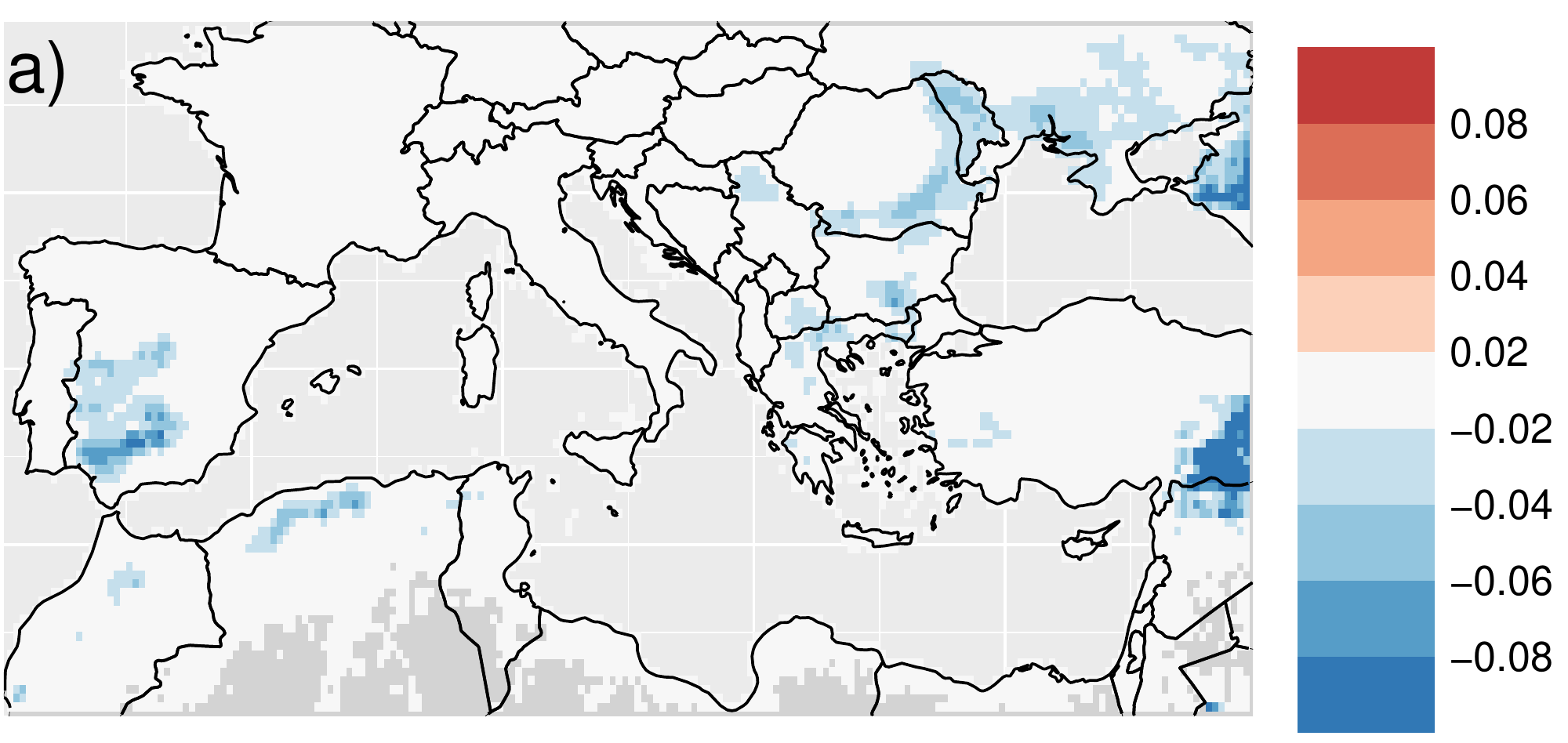} 
\end{minipage}
\begin{minipage}{0.32\linewidth}
\centering
\includegraphics[width=\linewidth]{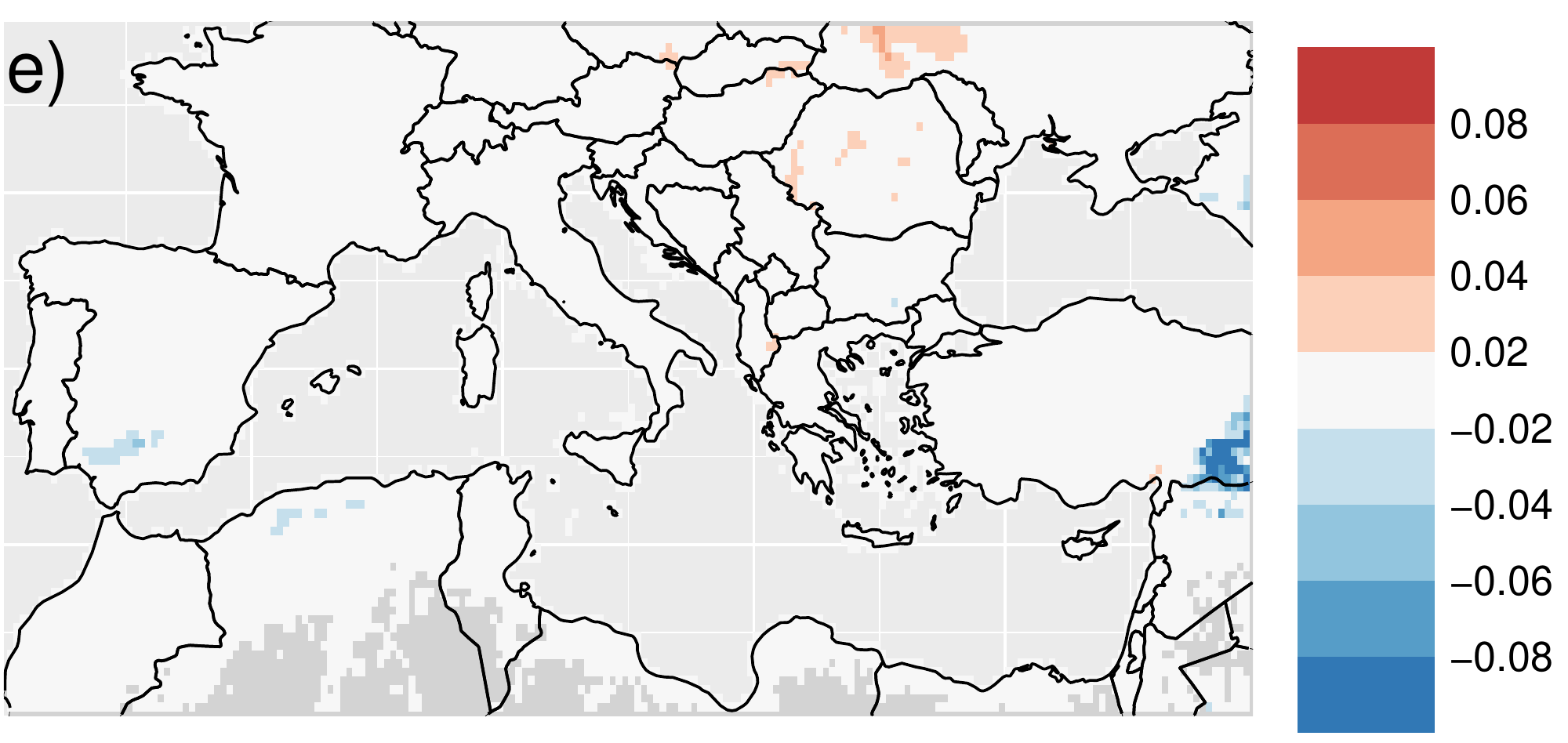} 
\end{minipage}
\begin{minipage}{0.32\linewidth}
\centering
\includegraphics[width=\linewidth]{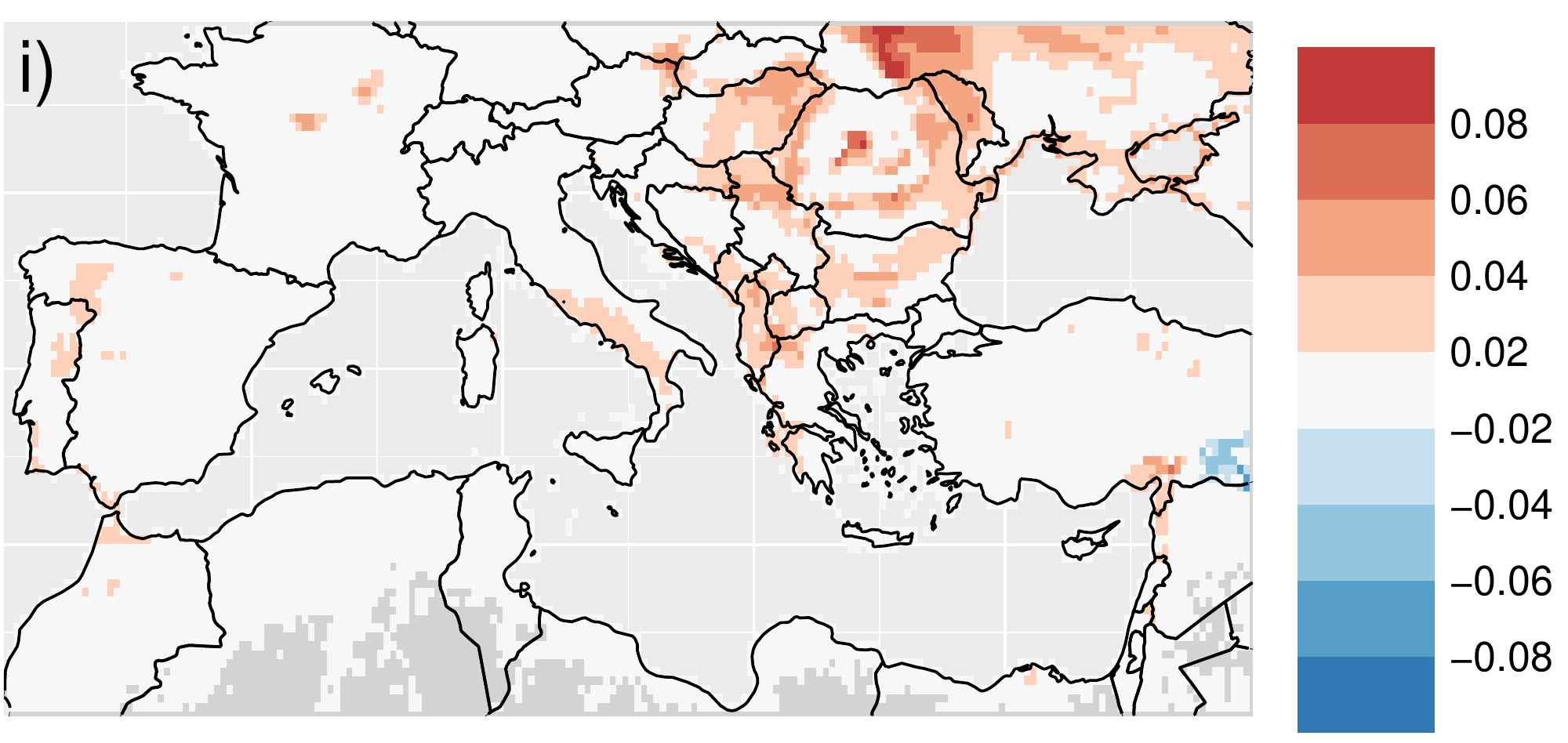} 
\end{minipage}
\begin{minipage}{0.32\linewidth}
\centering
\includegraphics[width=\linewidth]{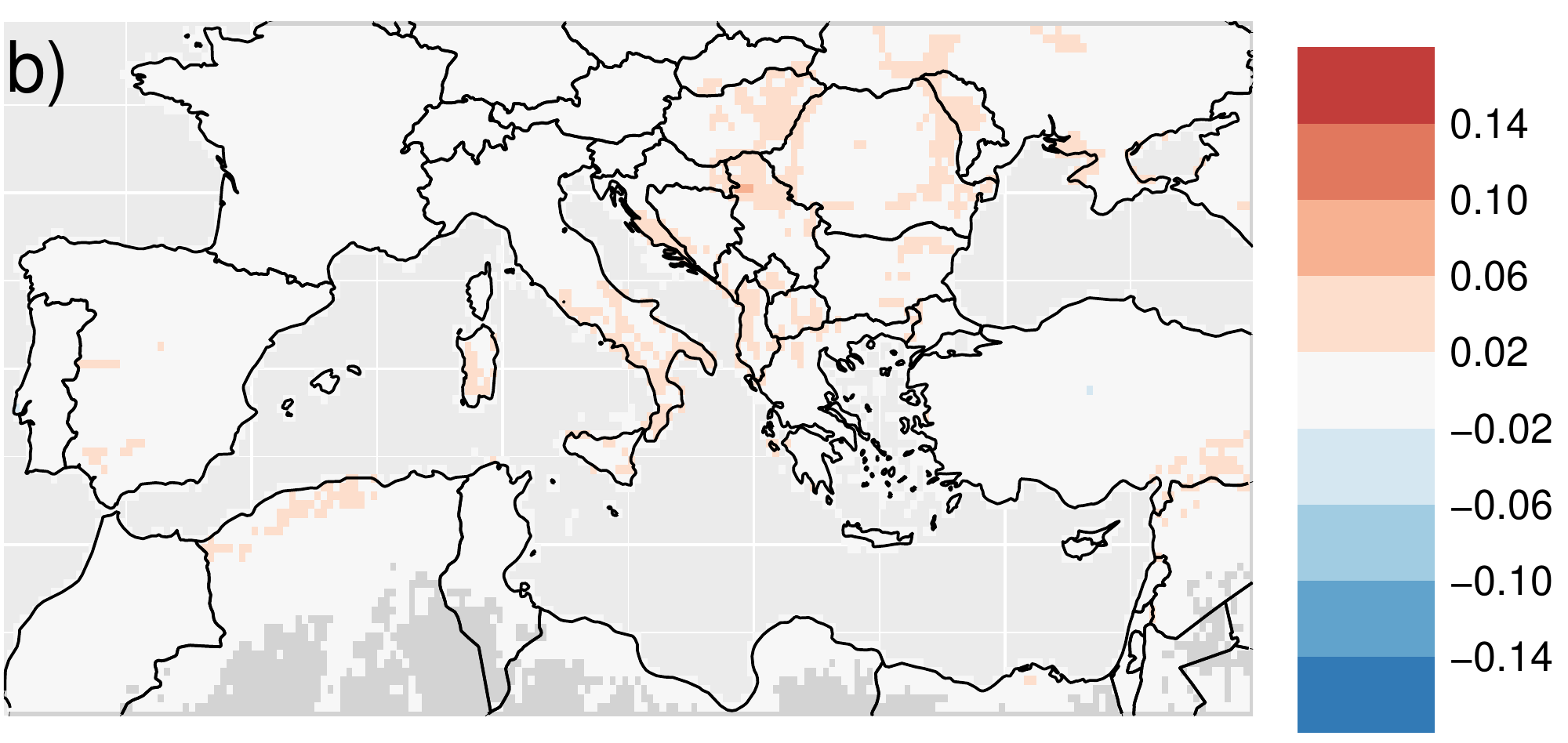} 
\end{minipage}
\begin{minipage}{0.32\linewidth}
\centering
\includegraphics[width=\linewidth]{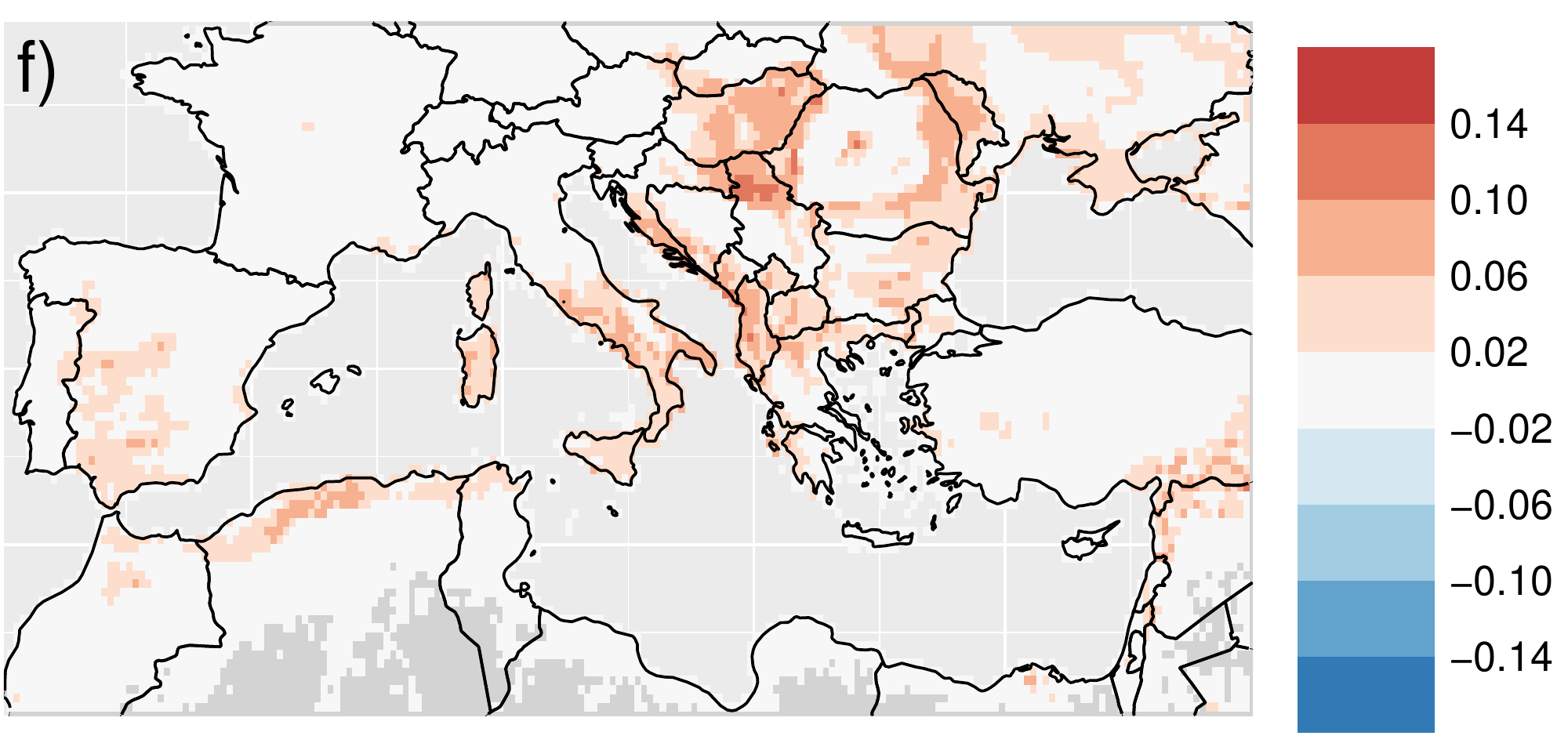} 
\end{minipage}
\begin{minipage}{0.32\linewidth}
\centering
\includegraphics[width=\linewidth]{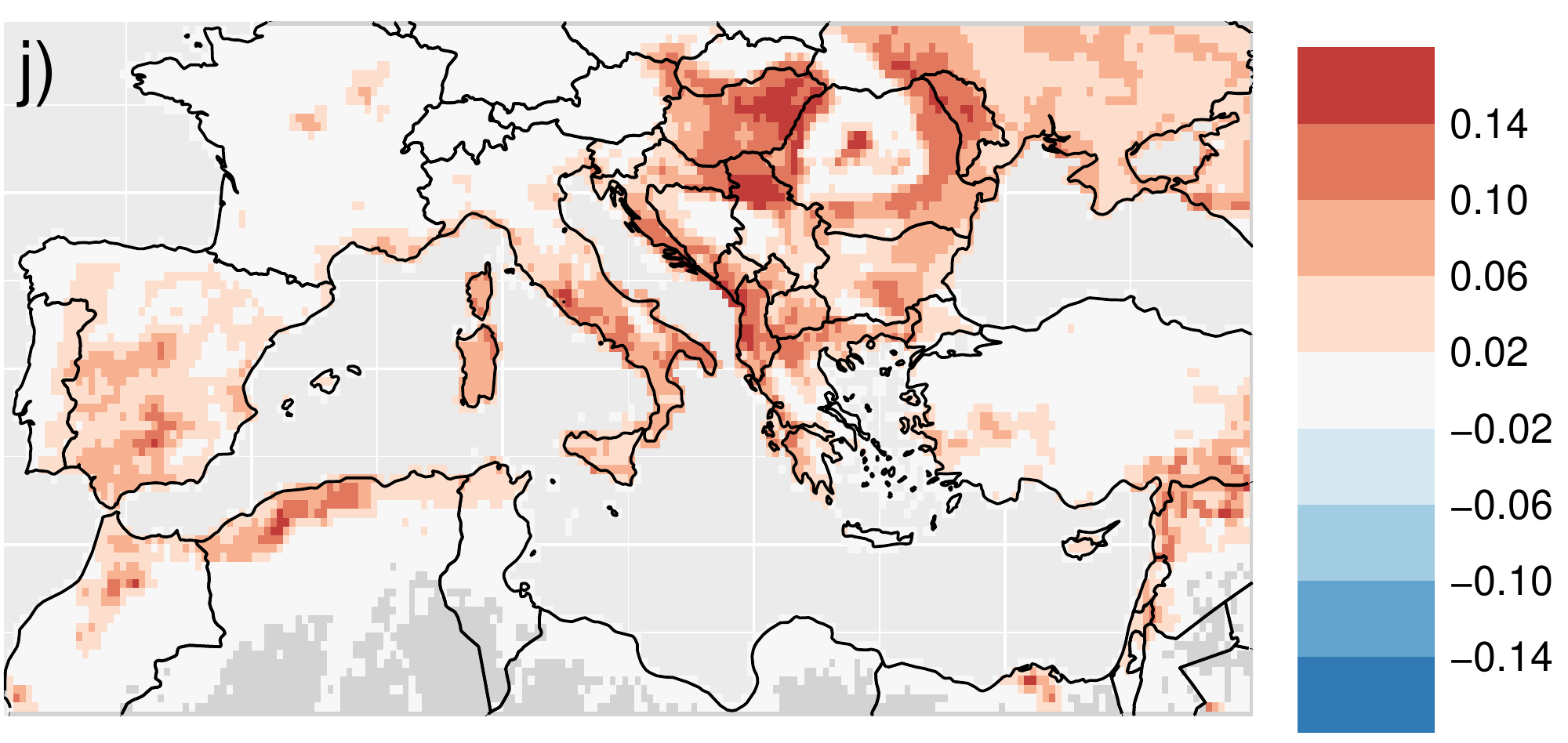} 
\end{minipage}
\begin{minipage}{0.32\linewidth}
\centering
\includegraphics[width=\linewidth]{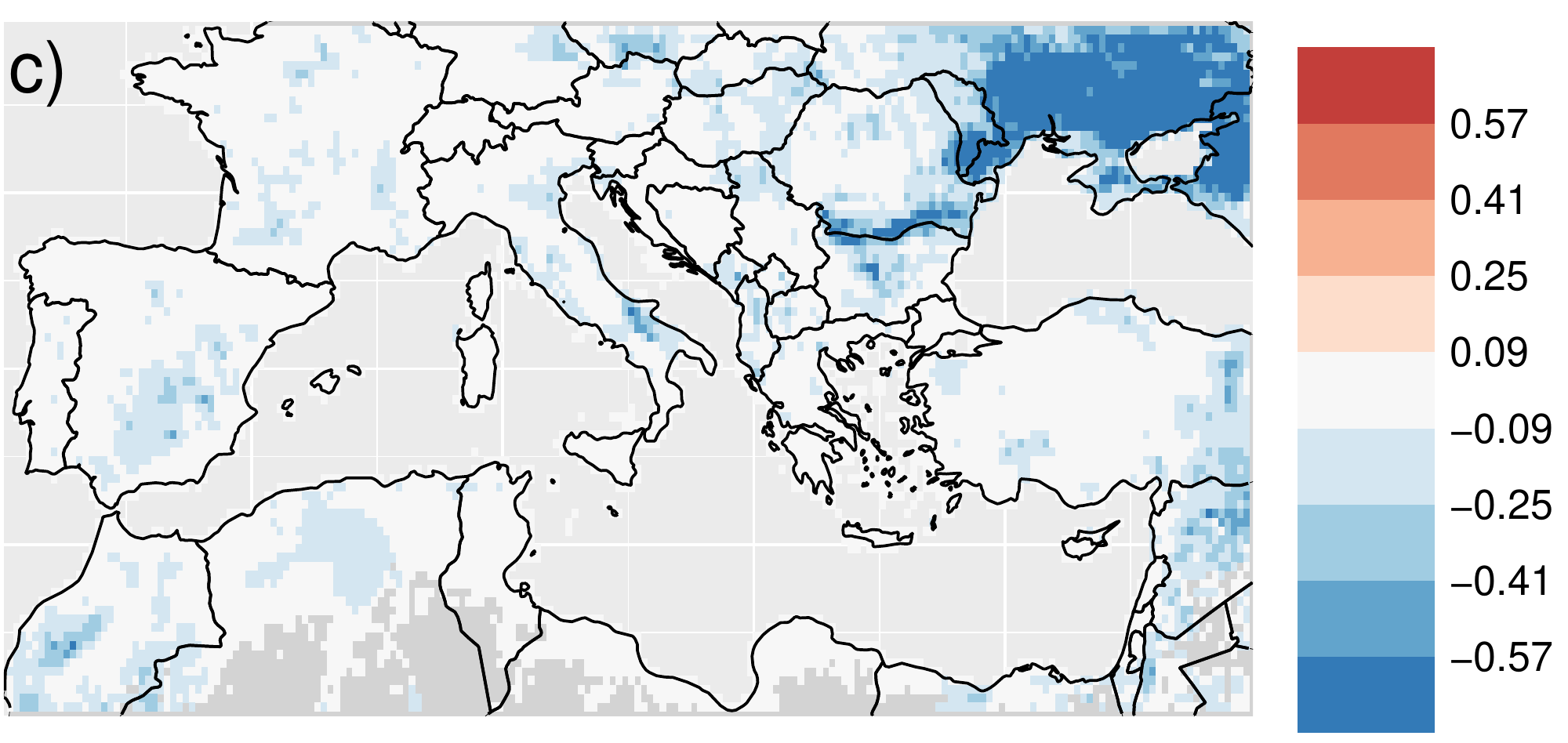} 
\end{minipage}
\begin{minipage}{0.32\linewidth}
\centering
\includegraphics[width=\linewidth]{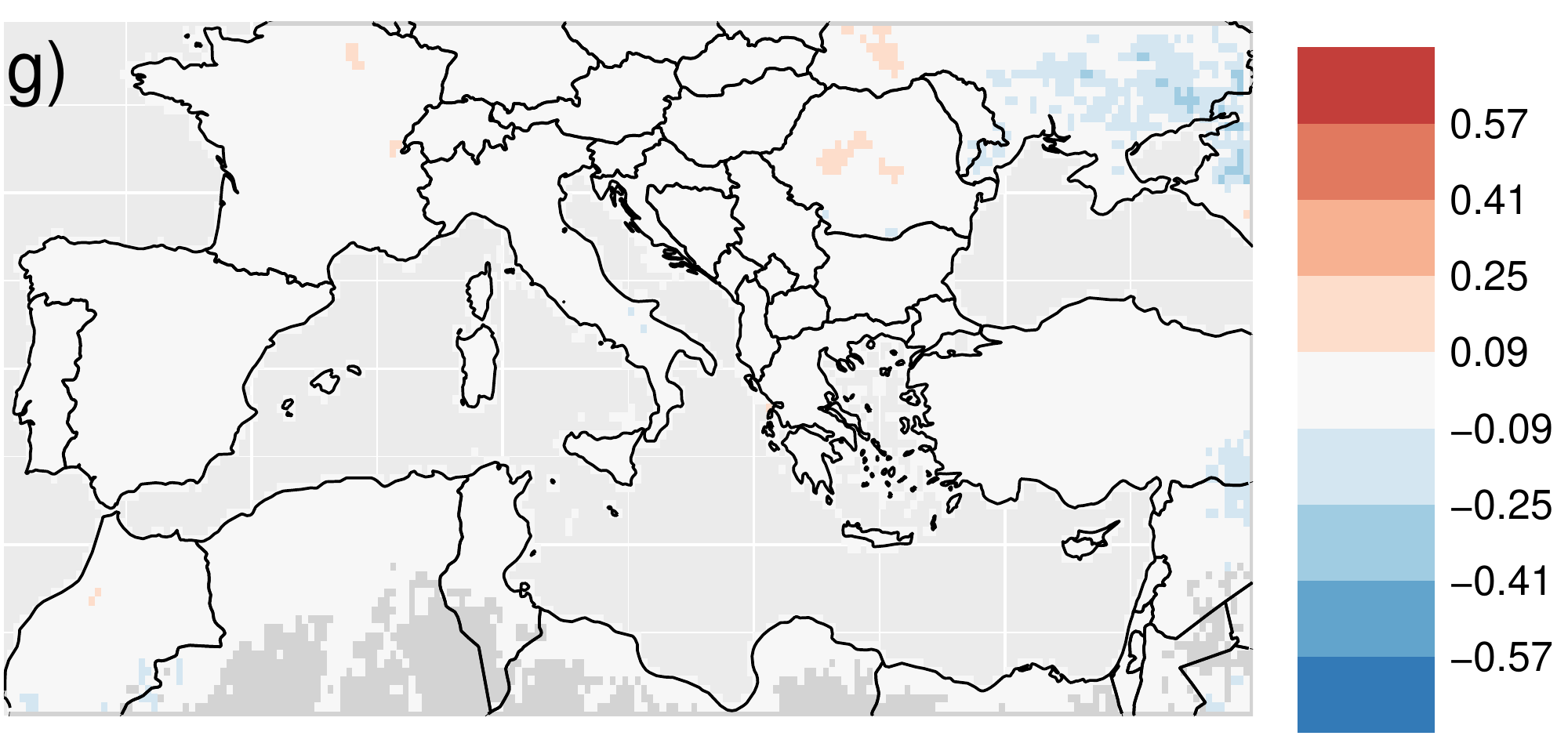} 
\end{minipage}
\begin{minipage}{0.32\linewidth}
\centering
\includegraphics[width=\linewidth]{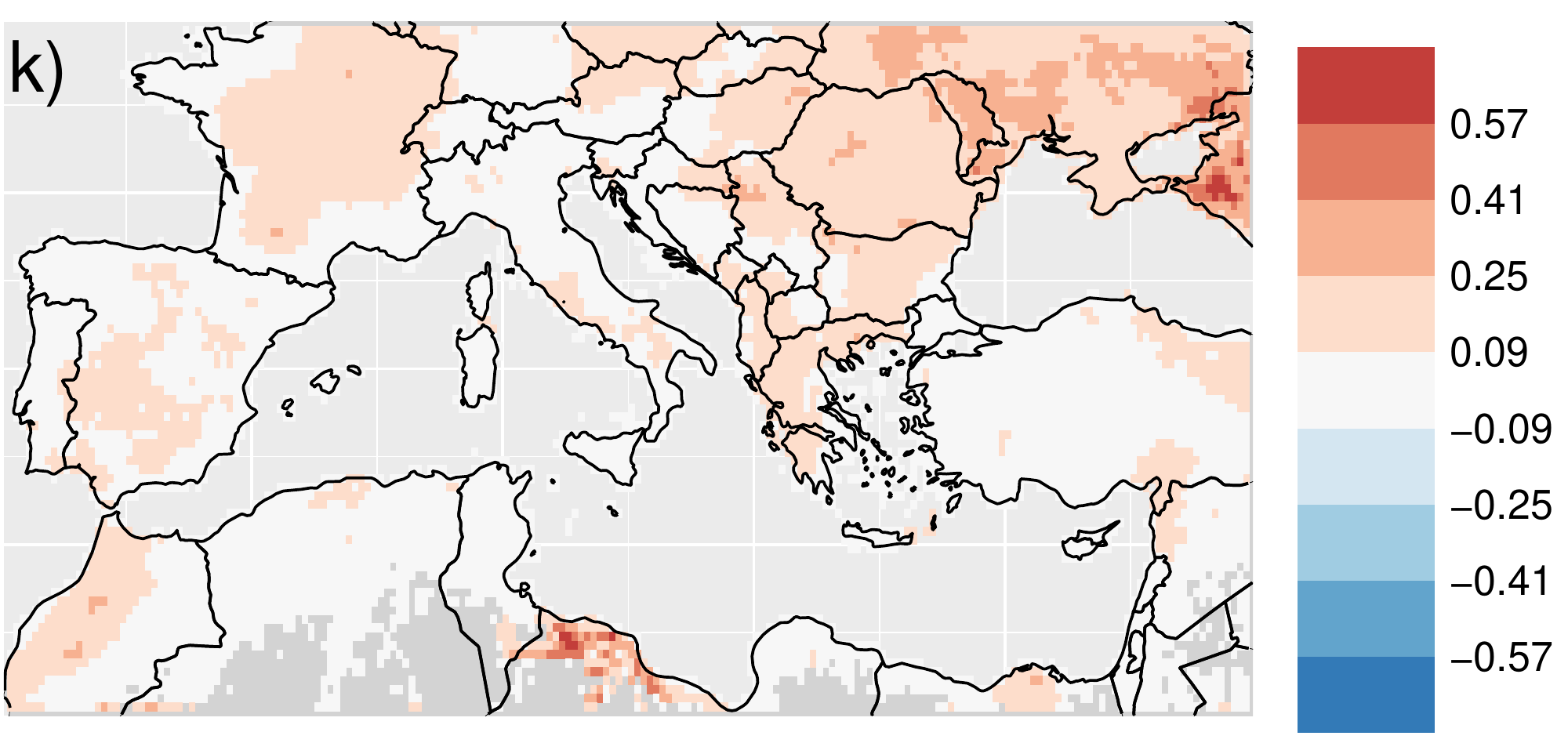} 
\end{minipage}
\begin{minipage}{0.32\linewidth}
\centering
\includegraphics[width=\linewidth]{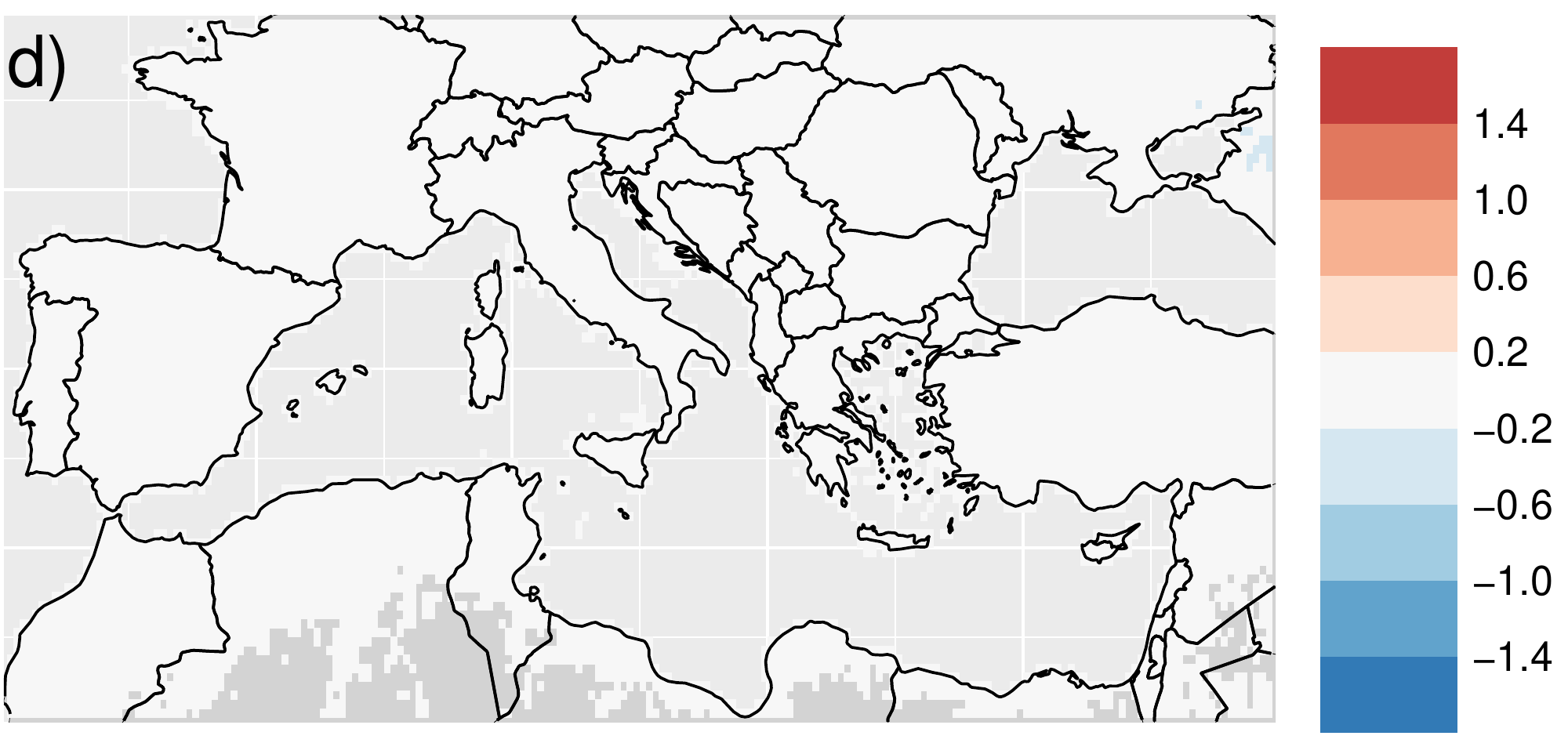} 
\end{minipage}
\begin{minipage}{0.32\linewidth}
\centering
\includegraphics[width=\linewidth]{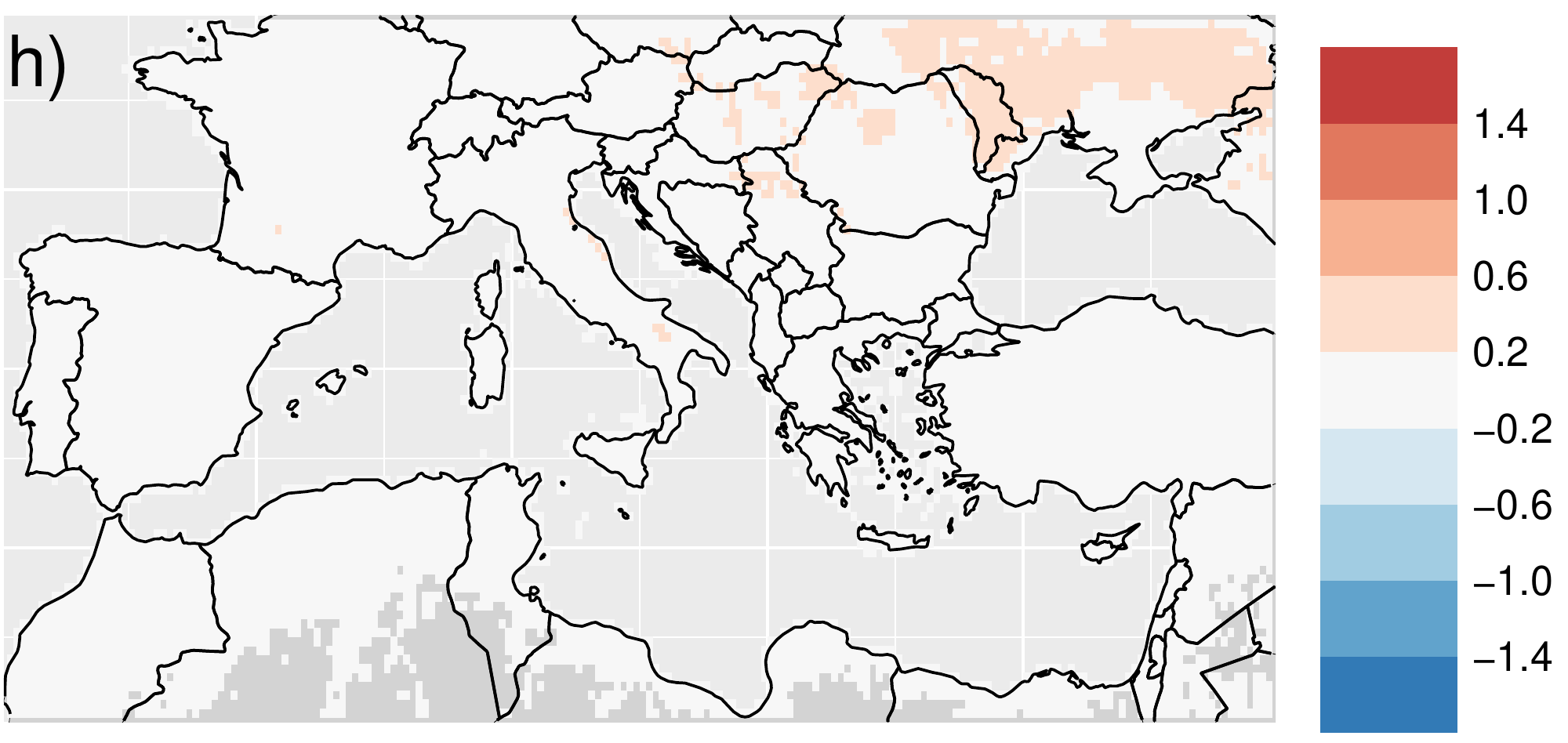} 
\end{minipage}
\begin{minipage}{0.32\linewidth}
\centering
\includegraphics[width=\linewidth]{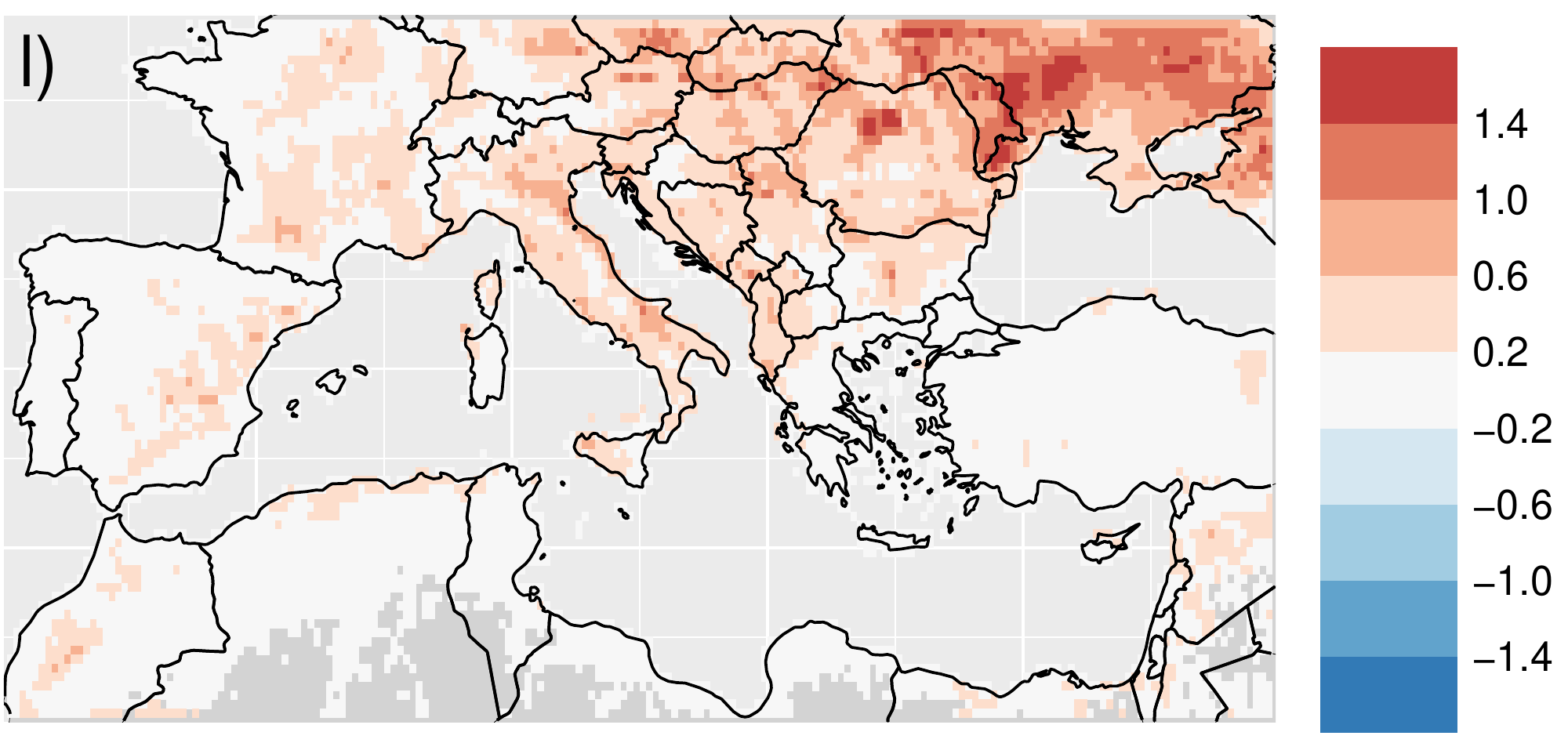} 
\end{minipage}
\caption{First row, a) $2.5\%$, e) $50\%$ and i) $97.5\%$ bootstrap quantiles of site-wise estimated changes, under predicted trends in VPD [Pa] for the period 2001--2020, in the {fire occurrence probability} $p_0(s,t)$ [unitless] for August 2001 (see also Figure~\ref{clim_maps}). The second row is as the first row, but for predicted trends in 2m air temperature [K]; the third and fourth rows are as the first two rows, but instead illustrate changes in  $\log\{1+q^+_{0.9}(s,t)\}$ [intensity; $\log(\mbox{km}^2)$; c], with $q^+_{0.9}(s,t)$ the $90\%$ quantile of spread $Y(s,t)\;|\;\{Y(s,t)>0,\mathbf{X}(s,t)\}$.
}
\label{clim_map_uncertain}
\end{figure}
\end{landscape}
\end{appendix}


\end{document}